\documentclass[doublecolumn]{IEEEtran}
\IEEEoverridecommandlockouts
\usepackage{textcomp}

\usepackage{amsfonts}
\usepackage{graphicx}
\usepackage{amsmath}
\usepackage{bm}
\usepackage[version=4]{mhchem}
\usepackage{siunitx}
\usepackage[ruled,vlined]{algorithm2e}
\usepackage{longtable,tabularx}
\usepackage{amssymb}
\usepackage{setspace}
\usepackage{subcaption}
\usepackage{array}
\usepackage[noadjust]{cite}
\usepackage{xcolor}
\usepackage{tikz}
\usetikzlibrary{positioning, shapes, shadows, arrows}
\usepackage{pdfpages}
\usepackage{lipsum}
\usepackage{wrapfig}
\usepackage{hyperref}
\usepackage{authblk}
\usepackage{stmaryrd}
\usepackage{enumitem}

\allowdisplaybreaks[1]

\begin{document}


\title{Overcoming Beam Squint in Dual-Wideband mmWave MIMO Channel Estimation: A Bayesian Multi-Band Sparsity Approach}

\author{Le~Xu,
        Lei~Cheng,
        Ngai~Wong,
        Yik-Chung~Wu,
        and H. Vincent Poor
\thanks{Le Xu, Ngai Wong and Yik-Chung Wu are with the Department of Electrical and Electronic Engineering, The University of Hong Kong, Hong Kong, P. R. China, (Email: xule@eee.hku.hk, nwong@eee.hku.hk, ycwu@eee.hku.hk).}
\thanks{Lei Cheng is with the College of Information Science and Electronic, Zhejiang University, P. R. China, (Email: lei\_cheng@zju.edu.cn).}
\thanks{H. Vincent Poor is with the Department of Electrical and Computer Engineering, Princeton University, Princeton, NJ 08544 USA (Email: poor@princeton.edu).}}

\markboth{Journal of \LaTeX\ Class Files,~Vol.~14, No.~8, August~2015}%
{Shell \MakeLowercase{\textit{et al.}}: Bare Demo of IEEEtran.cls for IEEE Journals}

\maketitle

\begin{abstract}
The beam squint effect, which manifests in the different steering matrices in different sub-bands, has been widely considered a challenge in millimeter wave (mmWave) multi-input multi-output (MIMO) channel estimation. Existing methods either require specific forms of the precoding/combining matrix, which restrict their general practicality, or simply ignore the beam squint effect by only making use of a single sub-band for channel estimation. Recognizing that different steering matrices are coupled by the same set of unknown channel parameters, this paper proposes to exploit the common sparsity structure of the virtual channel model so that signals from different sub-bands can be jointly utilized to enhance the performance of channel estimation. A probabilistic model is built to induce the common sparsity in the spatial domain, and the first-order Taylor expansion is adopted to get rid of the grid mismatch in the dictionaries. To learn the model parameters, a variational expectation-maximization (EM) algorithm is derived, which automatically obtains the balance between the likelihood function and the common sparsity prior information, and is applicable to arbitrary forms of precoding/combining matrices. Simulation results show the superior estimation accuracy of the proposed algorithm over existing methods under different noise powers and system configurations.
\end{abstract}

\begin{IEEEkeywords}
channel estimation, mmWave MIMO-OFDM, dual-wideband, Bayesian model, beam-squint effect.
\end{IEEEkeywords}

\IEEEpeerreviewmaketitle

\section{Introduction}
With the increasing demands of data transmission amid the overcrowded spectrum in conventional microwave frequencies, it is imperative to utilize the resource of the millimeter (mm) band. As mmWave signals suffer high attenuation and easy blockage \cite{ahmed2018survey}, massive multi-input multi-output (MIMO) systems equipped with hybrid precoder/combiners are often adopted to compensate for the large path loss in mmWave channels. However, in order to maximize the benefits of the massive MIMO systems, accurate channel estimation is an indispensable task in mmWave communications.

While a number of methods have been proposed to estimate the channels in mmWave systems, most of them only apply to narrow-band channel models \cite{bhaskar2013atomic,tang2013compressed,roy1989esprit,ma2020sparse}. Even with sporadic works considering wideband channels, e.g., \cite{venugopal2017channel,gonzalez2018channel,rodriguez2018frequency}, they only consider the frequency-wideband effect \cite{wang2018spatial}, in which only the path loss has different phase shifts under different sub-bands but the steering matrices are the same. 
However, due to the large bandwidth and massive amount of antennas, the transmission delay among different antennas is non-negligible \cite{wang2018spatial}, and the channel steering matrices become disparate in different sub-bands, leading to the spatial-wideband effect (also known as beam-squint effect). Considering both the frequency-wideband effect and spatial wideband effect, i.e., dual-wideband effect, the problem of channel estimation becomes much more difficult.

Existing methods for estimating dual-wideband channels usually require special forms of precoding/combining matrices. For example, the tensor-ESPRIT method \cite{lin2020tensor} estimates the channel parameters by exploiting the shift-invariance relation of the channel. In order to maintain the shift-invariance property of the received signal, the method in \cite{lin2020tensor} requires the precoding and combining matrices to be identity matrices. Another example is the discrete Fourier transform (DFT)-based method in \cite{wang2018spatial1}, which also requires the precoding/combining matrices to be identity matrices; otherwise, the Vandermonde structure of the phase shift matrix does not exist anymore, and the angular-delay signature cannot be recovered by the inverse discrete Fourier transform (IDFT) on the received signal. However, there are practical scenarios where precoding/combining matrices cannot be identity matrices: 1) preliminarily designed precoding/combining matrices must be adopted to ensure an acceptable signal-to-noise ratio (SNR); 2) some sub-bands are used for transmitting pilots for channel estimation of the target user, while other sub-bands are used for transmitting data (which may be for other users and requires beamforming). In the latter case, as the analog precoding/combining matrices are shared among all sub-bands, it might not be possible to simultaneously design exact identity precoding/combining matrices for the pilots, while taking non-identity precoding/combining matrices in effective beamforming for the data.
Therefore, requiring precoding/combining matrices to be identity matrices would severely limit the use cases in practice.



Compressed sensing (CS)-based methods could overcome the challenges mentioned above, with the orthogonal matching pursuit (OMP) being a representative algorithm. However, it can only be performed independently for each sub-band, and this brings up another problem: the channel estimation accuracy cannot be improved by using more training sub-bands. Even though one may propose to combine the estimated channels from different sub-bands in the hope of getting a more accurate one, it still remains a problem on how to combine them, especially when the noise power is large, in which case the estimated channel parameters in different sub-bands may have a different number of components. 
For other methods such as atomic norm minimization \cite{bhaskar2013atomic,yang2015gridless,xi2020gridless,wagner2021gridless}, which are known as continuous CS-based method and have been widely adopted for line spectral estimation, similar problems to those in traditional CS-based methods exist too. 
Although recent works \cite{jian2019angle,wang2019block,wang2018spatial} adopt the common sparsity among different sub-bands in the angular-delay domain so that the information from different sub-bands can be jointly utilized, they only consider the single input multiple outputs (SIMO) case. Since the MIMO case would have one more dimension than SIMO, directly extending them to the MIMO case would require building an over-complete dictionary involving three parameters (delay, AOA, and AOD), which leads to an unacceptable computational burden.


In this paper, to estimate the channel in the general case of dual-wideband mmWave MIMO-OFDM systems, we propose to build a virtual channel model and estimate the channel parameters by exploiting the common sparsity pattern among sub-bands. To avoid building a huge dictionary involving three parameters, we first incorporate the delay-related phase shift into the path loss. This allows the virtual channel model only
explicitly depends on the angular parameters, so the sizes of the dictionaries are greatly reduced, compared to those in \cite{jian2019angle,wang2019block,wang2018spatial}, where dictionaries are built on both angular and delay domains. 
In addition, to get rid of grid mismatch, the dictionaries are augmented by a correction obtained from first-order Taylor approximations. In this way, the received signals from different sub-bands can be accurately expressed.

While the proposed dictionaries are different in different sub-bands, it is found that they share the same sparsity pattern, i.e., a non-zero basis (corresponding to an AOD-AOA pair) would appear in the same position in all dictionaries. To exploit this property, a probabilistic model is proposed by assigning the channel responses of the same path but in different sub-channels a common sparsity-promoting prior. Considering the noise power is commonly unknown in practice, it is also treated as a random variable. Then a variational EM algorithm is derived to learn the model parameters. After convergence of the EM algorithm, the transmission delay and path loss are decoupled in an extra step and the channel is recovered. 
The advantages of the proposed method are: 
\begin{itemize}[leftmargin=*]
    \item The precoding and combining matrices can take arbitrary forms;
    \item The received signals from different sub-channels can be jointly utilized so that a more accurate channel estimation could be obtained;
    \item The algorithm could adapt itself to different noise power and no parameter tuning is required.
\end{itemize}
These advantages are demonstrated in the simulation results, in which the proposed method achieves superior performance over tensor-ESPRIT \cite{lin2020tensor} and OMP, under different noise levels and system configurations including different precoding/combining matrix designs.

The rest of this paper is organized as follows. Section \ref{sec:systemandchannel} introduces the dual-wideband channel model in mmWave MIMO-OFDM systems. Section \ref{sec:problem} presents the virtual channel model and problem formulation. The probabilistic model which exploits the common sparsity among sub-channels is built in Section \ref{sec:model}. The proposed inference algorithm is derived in Section \ref{sec:algorithm}. Simulation results are presented in Section \ref{sec:experiments}. The conclusion is drawn in Section \ref{sec:conclusion}.

\textit{Notations:} Boldface lowercase and uppercase letters are used to denote vectors and matrices, respectively, e.g., $\bm a$ and $\bm A$. Subscripts are adopted to denote specific elements of a vector or matrix, e.g., $\bm{a}_i$ and $\bm{A}_{i,j}$ denote the $i$-th element of $\bm{a}$ and the $(i,j)$-th element of $\bm{A}$, respectively. $\bm{A}^T$, $\bm{A}^H$ and $\bm{A}^*$ represent the transpose, hermitian transpose, and element-wise conjugate of $\bm A$, respectively. $\text{vec}(\bm A)$ denotes the vectorization of $\bm A$, and $\text{diag}(\bm a)$ denotes a diagonal matrix with $\bm a $ on its diagonal.
The operators $\odot$, $\otimes$ and $\ast$ denote the Khatri-Rao product, Kronecker product, and Hadamard product, respectively. $\mathbb{CN}(\bm\mu,\bm\Sigma)$ denotes the complex Gaussian distribution with mean $\bm \mu$ and covariance $\bm \Sigma$. $\text{Gamma}(\gamma,\beta)$ denotes the Gamma distribution with shape $\gamma$ and rate $\beta$. $\mathbb E\llbracket.\rrbracket$ denotes the expectation of a variable.

\section{Channel and System Models}
\label{sec:systemandchannel}
\subsection{Channel Model with Dual-Wideband Effects}
Consider the downlink channel with $N_\text{t}$ and $N_\text{r}$ antennas at the transmitter and receiver, respectively, both configured as uniform linear arrays (ULA). Suppose there are $L$ resolvable paths in the channel, and for the $\ell$-th path, ${\alpha}_\ell$, ${\tau}_\ell$, $\varphi_\ell$ and $\vartheta_\ell$ denote the complex channel gain, the real-valued time delay from the $1$st transmitting antenna to the $1$st receiving antenna, the angle-of departure and arrival (AOD/AOA), respectively. Then the channel response between the $n$-th transmit antenna and the $m$-th receive antenna at time $\tau$ is \cite{wang2018spatial,wang2019beam,gao2021wideband}
\begin{align}
    h_{m,n}(\tau) &= \sum_{\ell=1}^{L} {\alpha}_\ell e^{-j2\pi f_{\text{c}} ({\tau}_\ell + \frac{(n-1){\phi}_\ell}{f_{\text{c}}} + \frac{(m-1){\theta}_\ell}{f_{\text{c}}})} \nonumber \\ 
    &\times \delta\bigg(\tau-\big({\tau}_\ell + \frac{(n-1){\phi}_\ell}{f_{\text{c}}} + \frac{(m-1){\theta}_\ell}{f_{\text{c}}}\big)\bigg),
\end{align}
where ${\phi}_\ell = d \sin(\varphi_\ell) / \lambda_c$ and ${\theta}_\ell = d \sin(\vartheta_\ell) / \lambda_c$ are the normalized AOD and AOA of the $\ell$-th path, respectively, with $d$ denoting the distance between two adjacent antennas, and $f_{\text{c}}$ and $\lambda_c$ denoting the carrier frequency and wavelength, respectively.

After Fourier transformation, the channel response in the frequency domain is obtained as
\begin{align}
    {H}_{m,n}(f) = \sum_{\ell=1}^{L} {\alpha}_\ell e^{-j2\pi (f_{\text{c}}+f) ({\tau}_\ell + \frac{(n-1){\phi}_\ell}{f_{\text{c}}} + \frac{(m-1){\theta}_\ell}{f_{\text{c}}})}.
\end{align}
Consider evenly spaced channels, of which the $k$-th channel is located at frequency $f_\text{c}+f_k$, with $f_k = (k-1)f_\text{s}/K_0$ and $f_\text{s}$ denoting the total bandwidth. Then the frequency domain representation of the $k$-th channel is
\begin{align}
    H_{m,n,k} &= \sum_{\ell=1}^{L} {\alpha}_\ell e^{-j2\pi (f_{\text{c}}+f_k){\tau}_\ell} e^{-j2\pi  (1 + \frac{f_k}{f_{\text{c}}}) \big((n-1){\phi}_\ell + (m-1){\theta}_\ell\big)}.
    \label{eqn:Hf_mnk}
\end{align}

\begin{figure*}[!t]
    \centering
    \includegraphics[width=1\linewidth]{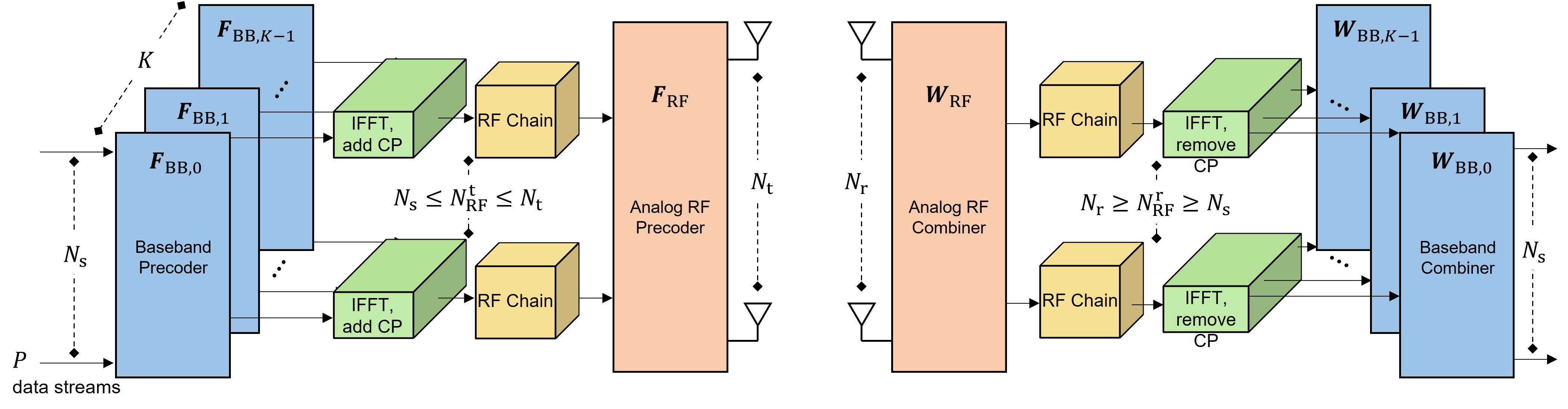}
    \caption{A MIMO-OFDM system with hybrid precoder and combiner.}
    \label{fig:system_model}
\end{figure*}

By defining
\begin{align}
    \bm{a}_{\text{bs},k}({\phi}_\ell) &= [1, e^{-j2\pi (1+\frac{f_k}{f_{\text{c}}}) {\phi}_\ell}, \ldots, e^{-j2\pi (1+\frac{f_k}{f_{\text{c}}})(N_\text{t}-1){\phi}_\ell}]^T, \nonumber \\
    \bm{a}_{\text{ms},k}({\theta}_\ell) &= [1, e^{-j2\pi (1+\frac{f_k}{f_{\text{c}}}) {\theta}_\ell}, \ldots, e^{-j2\pi (1+\frac{f_k}{f_{\text{c}}})(N_\text{r}-1){\theta}_\ell}]^T, \nonumber \\
    \bm{g}_k &= [{\bar\alpha}_1 e^{-j2\pi f_k{\tau}_1},\ldots,{\bar\alpha}_L e^{-j2\pi f_k{\tau}_L}]^T,
    \label{eqn:AbsAmsGk}
\end{align}
for $\ell = 1$ to $L$ and $k=1$ to $K_0$, in which $\{{\bar\alpha}_\ell = \alpha_\ell e^{-j2\pi f_{\text{c}}{\tau}_\ell}\}_{\ell = 1}^{L}$ are the equivalent path gain, the channel response in (\ref{eqn:Hf_mnk}) can be written in matrix notation $\bm{H}_k \in \mathbb{C}^{N_\text{r} \times N_{\text{t}}}$, with
\begin{align}
    \bm{H}_{k} &= \sum_{\ell=1}^{L}  [\bm{g}_k]_\ell \bm{a}_{\text{ms},k}({\theta}_\ell) \bm{a}_{\text{bs},k}^T({\phi}_\ell) \nonumber \\
    &= \bm{A}_{\text{ms},k}(\{\bm{\theta}_\ell\}_{\ell=1}^L) \text{diag} (\bm{g}_k) \bm{A}_{\text{bs},k}^T(\{\bm{\phi}_\ell\}_{\ell=1}^L),
\label{eqn:dual_wideband_channel_model}
\end{align}
where $\bm{A}_{\text{bs},k}(\{\bm{\phi}_\ell\}_{\ell=1}^L) = [\bm{a}_{\text{bs},k}(\phi_1),\ldots,\bm{a}_{\text{bs},k}(\phi_L)]$, $\bm{A}_{\text{ms},k}(\{\bm{\theta}_\ell\}_{\ell=1}^L) = [\bm{a}_{\text{ms},k}(\theta_1),\ldots,\bm{a}_{\text{ms},k}(\theta_L)]$, and $\text{diag}(\bm{g}_k)$ denotes a diagonal matrix with $\bm{g}_k$ being its diagonal.

Compared with traditional channel models like \cite{rodriguez2018frequency} or \cite{ertel1998overview}, (\ref{eqn:dual_wideband_channel_model}) is more comprehensive as it takes into consideration the wideband effects in both the augmented path loss (i.e., frequency wideband) and the steering matrices (i.e., spatial wideband or beam squint), and therefore is called the dual-wideband model \cite{wang2018spatial}. Specifically, in traditional wideband channel models like \cite{venugopal2017channel}, the phase shifts $2\pi (f_k/f_\text{c})(n-1)\phi_\ell$ and $2\pi (f_k/f_\text{c})(m-1)\theta_\ell$ are often neglected from the steering matrices because of the small value of $f_k$ compared with $f_\text{c}$, and therefore $\bm{A}_{\text{bs},k}$ and $\bm{A}_{\text{ms},k}$ would take the same form for different $k$.

However, in the mmWave systems where bandwidth and number of antennas are typically large, the above-mentioned phase shifts in $\bm{A}_{\text{bs},k}$ and $\bm{A}_{\text{ms},k}$ cannot be neglected anymore. For example, consider a MIMO configuration with $64$ ULA antennas at both sides, the carrier frequency $f_{\text{c}} = 60\text{GHz}$, and $K_0 = 64$ sub-channels with the total bandwidth $f_s=1\text{GHz}$. For the path with AOD $30^\circ$ and AOA $60^\circ$, there will be a phase shift around $1.46\pi$ between $H_{N_{\text{r}},N_{\text{t}},1}$ and $H_{N_{\text{r}},N_{\text{t}},K_0}$, and the relative square difference $\| \bm{H}_{K_0} - \bm{H}_{1}\|_F^2/\| \bm{H}_{1}\|_F^2$ is $2.744$. Therefore, the dual-wideband channel model (\ref{eqn:dual_wideband_channel_model}) is vital in multi-carrier mmWave massive MIMO systems.

\subsection{Received Signal in MIMO-OFDM System with Hybrid Beamforming}
\label{subsec:systemmodel}
As shown in Fig. \ref{fig:system_model}, we consider a MIMO-OFDM mm-Wave system with $N_\text{t}$ antennas at the transmitter, $N_\text{r}$ antennas at the receiver and $K_0$ subcarriers. In the $p$-th time slot, $\bm{S}(p)\in \mathbb{C}^{N_\text{s}\times K_0}$ is input into the transmitter, with its $k$-th column $\bm{s}_{k}(p) \in \mathbb{C}^{N_\text{s}}$ denoting $N_\text{s}$ modulated symbols to transmit on the $k$-th sub-carrier.
Out of the $K_0$ sub-carriers, $K$ of them are allocated for pilot symbols, and the rest are for data.

The input signals are firstly precoded by the baseband precoders $\{\bm{F}_{\text{BB},k} \in \mathbb{C}^{N_\text{s} \times N_{\text{RF}}^\text{t}}\}_{k=1}^{K_0}$ where $N_\text{RF}^\text{t}$ denotes the number of RF-chains with $N_\text{s} \leq N_{\text{RF}}^\text{t} \leq N_\text{t}$ \cite{el2014spatially}, and then transformed into time domain signals by a $K_0$-point inverse fast Fourier transform (IFFT) denoted by $\bm{Q}$. After adding the cyclic prefix (CP), the baseband signals go through an analog RF precoder $\bm{F}_{\text{RF}}\in \mathbb{C}^{N_{\text{RF}}^{\text{t}} \times N_\text{t}}$, and transmitted through the $N_\text{t}$ antennas.

Considering the dual-wideband channel (\ref{eqn:dual_wideband_channel_model}) and suppose one common baseband precoder $\bm{F}_{\text{BB}}\in \mathbb{C}^{N_\text{s} \times N_{\text{RF}}^\text{t}}$ is adopted for all the sub-bands.
then the received signal at the $p$-th time slot (after CP removal) at the receiver is
\begin{align}
    \bm{X}_\text{r}(p) = \sum_{k=1}^{K_0}\bigg( \bm{H}_k \bm{F}_{\text{RF}} \bm{F}_{\text{BB}} \bm{s}_{k}(p) \bm{q}_{k}^T \bigg) + \bm{N}_{\text{r}}(p),
\end{align}
where $\bm{q}_k$ denotes the $k$-th column of $\bm{Q}$, and $\bm{N}_{\text{r}}(p) \in \mathbb{C}^{N_{\text{r}}\times K_0}$ is additive white Gaussian noise (AWGN) with each element having zero mean and variance $\sigma^2$. At the receiver side, the inverse operations are performed, as shown in the right-hand side of Fig. \ref{fig:system_model}. Specifically, the number of RF chains on the user side follows $N_\text{r} \geq N_{\text{RF}}^{\text{r}} \geq N_{\text{s}}$. Going through the RF combiner $\bm{W}_{\text{RF}} \in \mathbb{C}^{N_\text{r}\times N_\text{RF}^\text{r}}$, the signal is transformed back into the frequency domain via FFT. Assume one common baseband combiner $\bm{W}_{\text{BB}} \in \mathbb{C}^{N_\text{RF}^\text{t} \times N_\text{d}}$ is adopted at the receiver, then the received signal on the $k$-th subcarrier is
\begin{align}
    \bm{y}_{k}(p) &= \bm{W}_{\text{BB}}^T\bm{W}_{\text{RF}}^T\bm{X}_\text{r}(p) \bm{q}_k^* \nonumber \\
    & = \bm{W}^T\bm{H}_k\bm{F} \bm{s}_k(p) + \bm{W}^T\bm{n}_k(p),
\label{eqn:Y_t_received}
\end{align}
in which the second line is due to $\bm{Q}^T \bm{Q}^* = \bm{I}_{K_0}$, and $\bm{F} = \bm{F}_{\text{RF}} \bm{F}_{\text{BB}}$ and $\bm{W} = \bm{W}_{\text{RF}}\bm{W}_{\text{BB}}$ are used for a more concise expression. Noise $\bm{n}_k(p)$ is the $k$-th column of the FFT of $\bm{N}_{\text{r}}(p)$ and remains as AWGN with zero mean and covariance $\sigma^2 \bm{I}_{N_{\text{r}}}$.

Collect the transmitted signal, received signal and noise at the $k$-th sub-band along time $p = [1,\ldots,P]$, and denote $\bm{S}_k = [\bm{s}_{k}(1),\ldots,\bm{s}_{k}(P)]$, $\bm{Y}_k =[\bm{y}_{k}(1),\ldots,\bm{y}_{k}(P)]$ and $\bm{N}_k =[\bm{n}_{k}(1),\ldots,\bm{n}_{k}(P)]$. Then using (\ref{eqn:Y_t_received}) and (\ref{eqn:dual_wideband_channel_model}), the following input-output relationship can be derived,
\begin{align}
    \bm{Y}_k & = \bm{W}^T{\bm{A}}_{\text{ms},k}(\{\bm{\theta}_\ell\}_{\ell=1}^L) \text{diag}(\bm{g}_k) {\bm{A}}_{\text{bs},k}^T(\{\bm{\phi}_\ell\}_{\ell=1}^L) \bm{F} \bm{S}_k \nonumber \\ 
    & + \bm{W}^T \bm{N}_k, \text{ for } k=1 \text{ to } K_0.
    \label{eqn:Y_k_received}
\end{align}

\begin{figure*}[!t]
    \centering
    \begin{subfigure}[b]{0.25\textwidth}
        \includegraphics[width=1\linewidth]{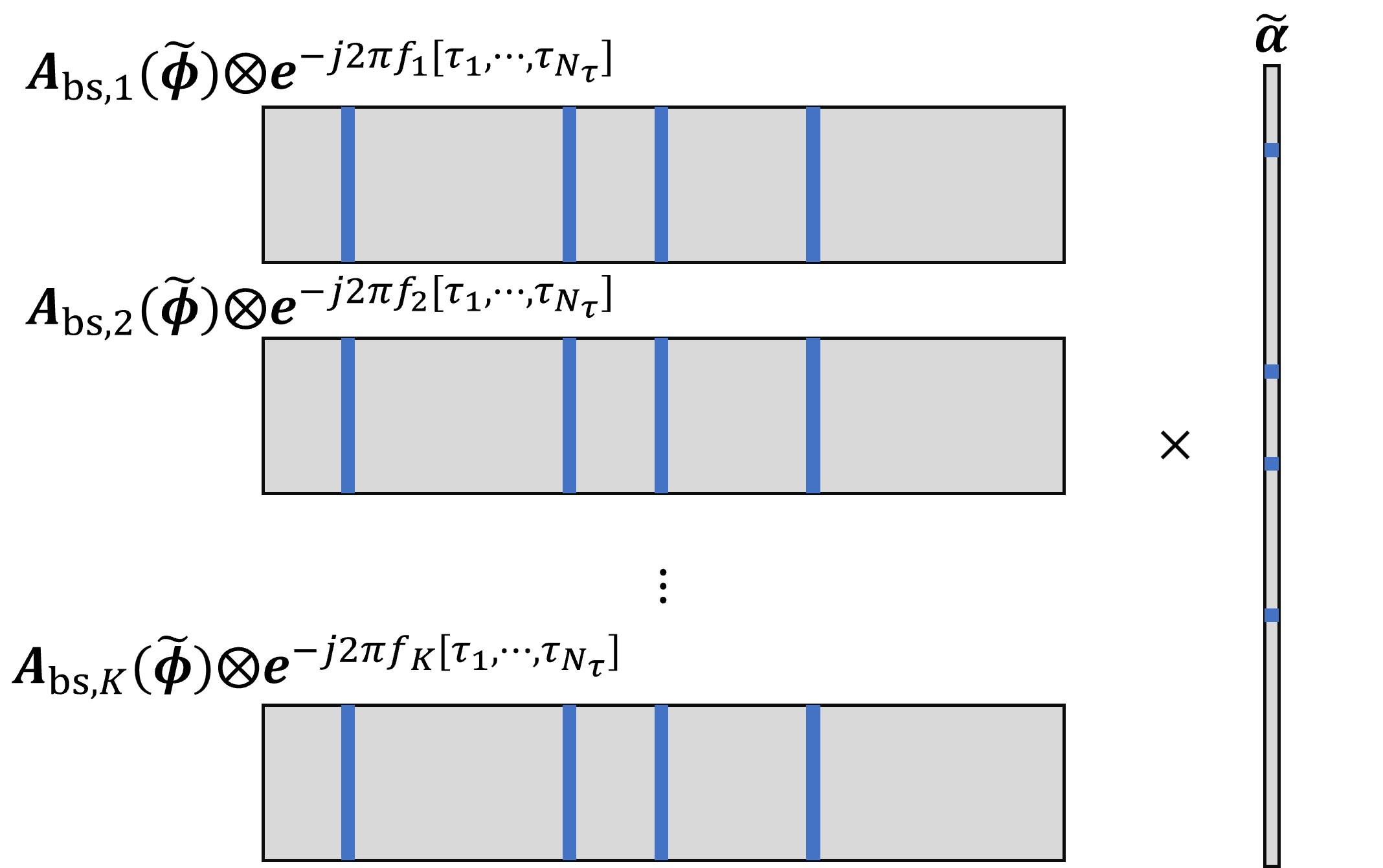}
        \caption{Wideband SIMO}
        \label{subfig:widebandsimovirtual}
    \end{subfigure}
    \begin{subfigure}[b]{0.19\textwidth}
        \includegraphics[width=1\linewidth]{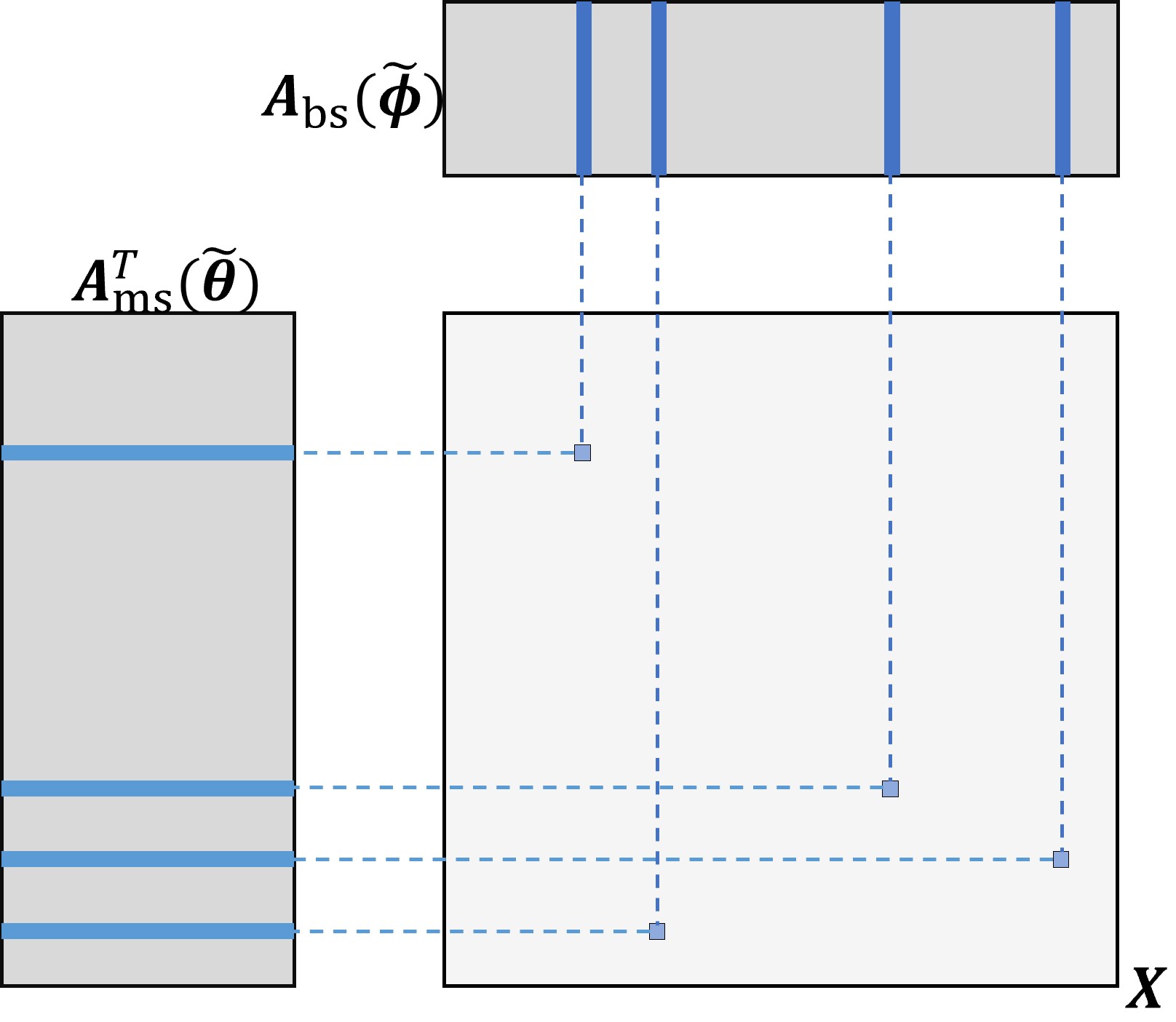}
        \caption{Narrowband MIMO}
        \label{subfig:narrowbandvirtual}
    \end{subfigure}
    \begin{subfigure}[b]{0.24\textwidth}
        \includegraphics[width=1\linewidth]{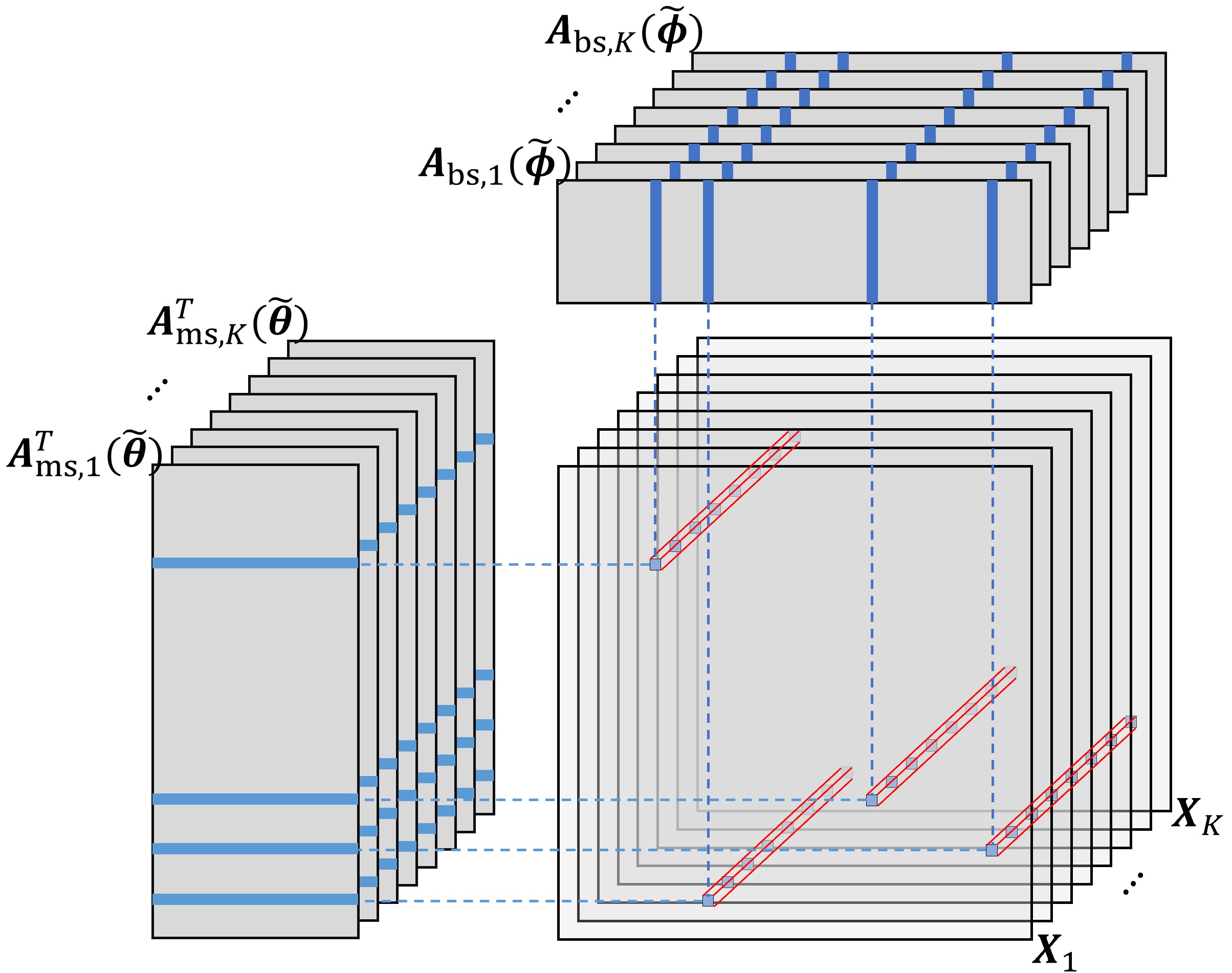}
        \caption{Dual-Wideband MIMO}
        \label{subfig:dualwidebandvirtual}
    \end{subfigure}
    \begin{subfigure}[b]{0.20\textwidth}
        \includegraphics[width=1\linewidth]{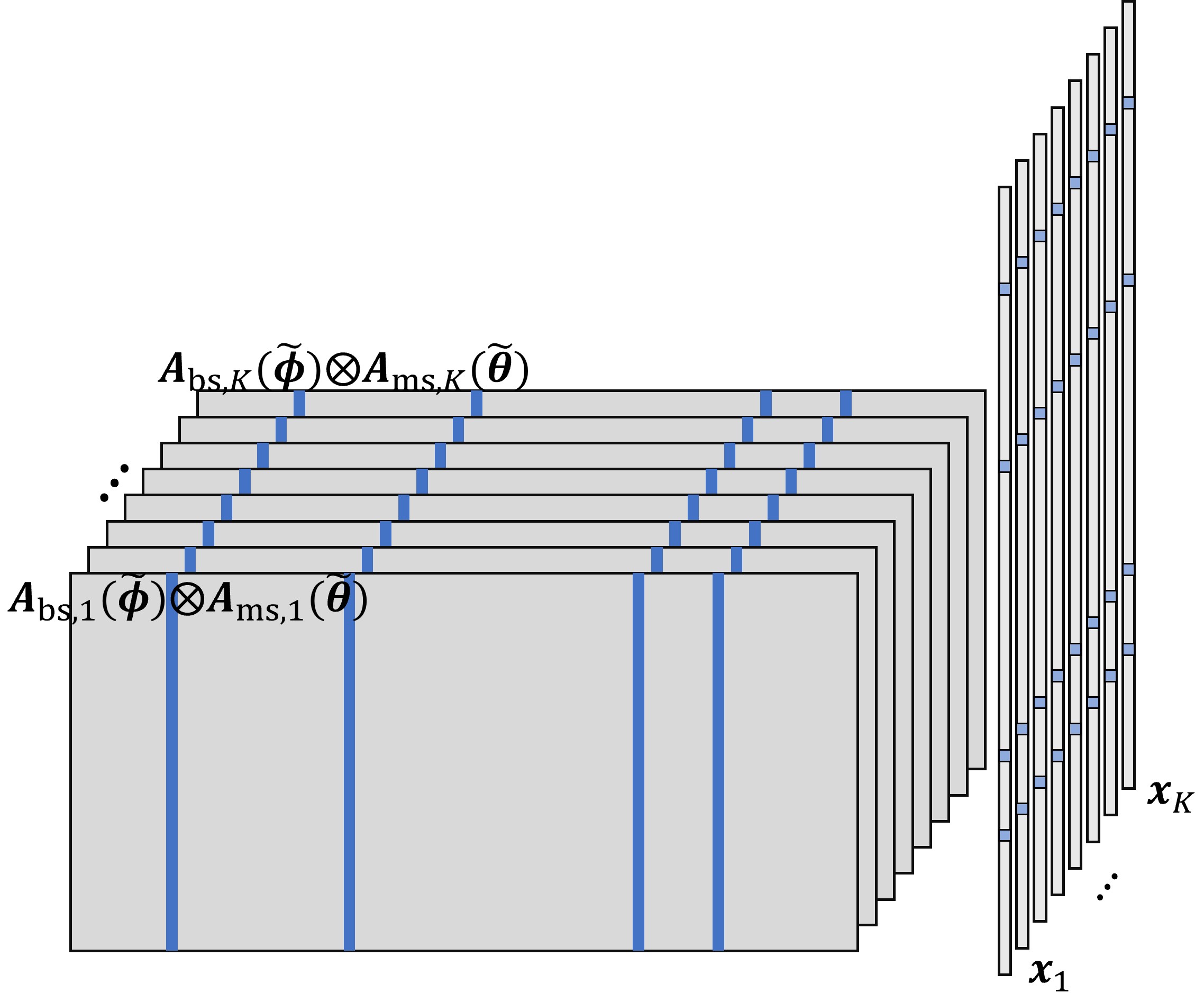}
        \caption{Dual-Wideband MIMO,\\ vectorized format}
        \label{subfig:dualwidebandvirtualvec}
    \end{subfigure}
    \caption{Illustrations of the virtual channel models.}
    \label{fig:common_sparsity}
\end{figure*}

Many existing works on mmWave channel estimation assume $\bm{W}$ to be $\bm{I}_{N_\text{r}}$ or $[\bm{I}_{N_\text{s}},\bm{0}]^T$ \cite{lin2020tensor,wang2018spatial1}, such that the column space of ${\bm{A}}_{\text{ms},k}$ is preserved and the noise is maintained as AWGN. While it is valid if all the sub-carriers contain training signals and this simplifies the subsequent analysis, $\bm{W}$ may not be an identity matrix when some of the sub-carriers contain user data and non-identity combining is needed. This paper considers the general case where $\bm{W}$ is not restricted to any particular form. In this case, a whitening step is required because the noise in (\ref{eqn:Y_k_received}) is not AWGN anymore. Noise whitening can be achieved by multiplying $\bm{\Sigma}^{-\frac{1}{2}}\bm{U}^H$ to $\bm{Y}_k$ from the left, where $\bm{W}^T\bm{W}^* = \bm{U}\bm{\Sigma}\bm{U}^H$.

\section{Problem formulation and challenges ahead}
\label{sec:problem}
Given the received training signals in $K$ subcarriers (out of totally $K_0$ subcarriers) and making use of (\ref{eqn:Y_k_received}), the problem of channel estimation is stated as
\begin{align}
    &\min_{\{{\phi}_\ell,{\theta}_\ell,{\bar\alpha}_\ell,{\tau}_\ell\}_{\ell=1}^L}  \sum_{k=1}^K \bigg\| \bm{Y}_{\text{w},k} - \bm{W}_{\text{w}}^T{\bm{A}}_{\text{ms},k}(\{\bm{\phi}_\ell\}_{\ell=1}^L) \nonumber \\ 
    & \quad \quad \times \text{diag}(\bm{g}_k) {\bm{A}}_{\text{bs},k}^T(\{\bm{\theta}_\ell\}_{\ell=1}^L) \bm{F} \bm{S}_k \bigg\|_F^2,
    \label{eqn:problemformulation}
\end{align}
where $\bm{Y}_{\text{w},k} = \bm{\Sigma}^{-\frac{1}{2}}\bm{U}^H \bm{Y}_{\text{w},k}$ and $\bm{W}_{\text{w}}^T = \bm{\Sigma}^{-\frac{1}{2}}\bm{U}^H\bm{W}^T$ are the received signal and the normalized combining matrix after noise whitening. There are several challenges in solving (\ref{eqn:problemformulation}): 
\begin{itemize}[leftmargin=*]
    \item \textit{The objective of (\ref{eqn:problemformulation}) is non-linear and non-convex with respect to variables ${\{{\phi}_\ell,{\theta}_\ell,{\tau}_\ell\}_{\ell=1}^L}$.} Searching a combination of these variables to achieve the minimum error is NP-hard;
    \item \textit{Different terms in (\ref{eqn:problemformulation}) are coupled by the dual-wideband effects,} since $\bm{A}_{\text{bs},k}$, $\bm{A}_{\text{ms},k}$ and $\bm{g}_k$ all take different expressions in different sub-bands but share one common set of channel parameters, i.e., ${\{{\phi}_\ell,{\theta}_\ell,{\bar\alpha}_\ell,{\tau}_\ell\}_{\ell=1}^L}$. This makes (\ref{eqn:problemformulation}) more complicated than traditional channel models without the beam squint effect, in which case $\bm{A}_{\text{ms},k}$ and $\bm{A}_{\text{bs},k}$ do not change with $k$ \cite{rodriguez2018frequency}. Even if one may estimate the channel parameters separately in each sub-channel, it is not clear how to combine the results;
    \item \textit{The number of channel paths is generally unknown in practice.} An under-estimated number of paths would lead to a large estimation error, while an over-estimated one would lead to overfitting of noise.
\end{itemize}

When there is no beamforming matrices, special structure of $\bm{A}_{\text{bs},k}$ and $\bm{A}_{\text{ms},k}$ can be exploited to facilitate a solution. For example, a tensor ESPRIT method is proposed in \cite{lin2020tensor} to solve (\ref{eqn:problemformulation}). However, in order for the ESPRIT to work, the training sub-carriers need to be evenly spaced, and the effective precoding and combining matrix $\bm{F}$ and $\bm{W}_{\text{w}}$ also have to take special forms. This limits the applicability of the tensor ESPRIT algorithm in general settings.

\subsection{Dual-Wideband Virtual Channel Reformulation}
To solve (\ref{eqn:problemformulation}) in its generic form, a virtual channel model shall be adopted. In the SIMO case, \cite{jian2019angle,wang2019block,wang2018spatial} solve a simpler version of (\ref{eqn:problemformulation}) by first building a sensing matrix based on grids of the AOD $\phi$ and the time delay $\tau$. Then by stacking the training signals in all $K$ sub-channels, the problem becomes a sparse vector recovery problem under an over-completed dictionary, as illustrated in Fig. \ref{subfig:widebandsimovirtual}. However, if we directly extend this idea to the MIMO case, there will be three parameters in the grids, i.e., $\{\phi,\theta,\tau\}$. Suppose the number of each parameter grid is set to $N_\text{G}$, then the resulting sensing matrix will have $KN_\text{G}^3$ bases, which brings an unacceptable model complexity. For example, consider $K=16$, $P = 16$ and $N_s = 6$ in the MIMO configuration and setting the grid size to $128$ for each parameter, the number of bases is over $33$ millions. If 64 bits representation is used, it would lead to a sensing matrix taking $48$GB memory.

Instead, we propose to adopt the extended virtual channel model as in \cite{heath2016overview} by treating the variable $\tau$ as the phase of the parameter $\bar\alpha$. By assuming the true equivalent AODs and AOAs lie in the $N_\theta$- and $N_\phi$-length angle grids $\tilde{\bm{\theta}} = [\tilde{{\theta}}_1,{\tilde{{\theta}}}_2,\ldots,\tilde{{\theta}}_{N_\theta}]$ and $\tilde{\bm{\phi}} = [\tilde{\phi}_1,\tilde{\phi}_2,\ldots,\tilde{\phi}_{N_\phi}]$, respectively, both of which are uniformly spaced between $[-1/2,1/2]$, then the dual-wideband channel in the $k$-th sub-band (\ref{eqn:dual_wideband_channel_model}) can be expressed as
\begin{align}
    \bm{H}_k \simeq {\bm A}_{\text{ms},k}(\tilde{\bm{\theta}}) {\bm{X}}_k {\bm A}_{\text{bs},k}^T(\tilde{\bm{\phi}}),
    \label{eqn:Hkvirtual}
\end{align}
in which ${\bm A}_{\text{ms},k}(\tilde{\bm{\theta}})$ and ${\bm A}_{\text{bs},k}(\tilde{\bm{\phi}})$ are defined similarly to those in (\ref{eqn:AbsAmsGk}) but with different channel parameter inputs $\tilde{\bm{\theta}}$ and $\tilde{\bm{\phi}}$. Since ${\bm A}_{\text{ms},k}(\tilde{\bm{\theta}})$ and ${\bm A}_{\text{bs},k}(\tilde{\bm{\phi}})$ are over-completed but there are actually only $L$ paths in the channel, $\bm{X}_k$ is a sparse matrix with only $L$ non-zero elements, each corresponds to a value from ${\bm{g}_k}$ in (\ref{eqn:AbsAmsGk}). 
On the other hand, compared to directly extending the virtual channel model of wideband SIMO \cite{jian2019angle} to the MIMO case, the number of bases of (\ref{eqn:Hkvirtual}) is only $KN_\text{G}^2$. With the same configuration as assumed above, the sensing matrix only takes $384$MB memory. Compared with the virtual channel model in the narrowband case \cite{heath2016overview} (shown in Fig \ref{subfig:narrowbandvirtual}), (\ref{eqn:Hkvirtual}) shows more complicated coupling among the $K$ training sub-carriers because of the dual-wideband effect (shown in Fig. \ref{subfig:dualwidebandvirtual}). The coupling not only occurs in $\tilde{\bm{\phi}}$ and $\tilde{\bm{\theta}}$ among $K$ steering matrices, but also in the elements of $\{\bm{X}_k\}_{k=1}^K$, as
\begin{align}
    [\bm{X}_k]_{i,j} = [\bm{X}_1]_{i,j} e^{-j2\pi (f_k-f_1) \bm{\Psi}_{i,j}},
    \label{eqn:Xkrelation}
\end{align}
where
$\bm{\Psi}_{i,j}$ is the time delay in the virtual path with AOA $\tilde{\bm{\theta}}_i$ and AOD $\tilde{\bm{\phi}}_j$.

While seems to be a complication, the coupling in (\ref{eqn:Xkrelation}) inspires us to take a two-step approach in channel estimation. Firstly, we estimate $\tilde{\bm{\phi}}$, $\tilde{\bm{\theta}}$ and $\{\bm{X}_k\}_{k=1}^K$ in which the relationship (\ref{eqn:Xkrelation}) is temporarily relaxed. In the second step, (\ref{eqn:Xkrelation}) is adopted to recover the corresponding path loss and time delay from the estimated $\{\bm{X}_k\}_{k=1}^K$.

According to (\ref{eqn:Xkrelation}), as there is common sparsity of elements of $\bm{X}_k$ across $k$ (shown in Fig. \ref{subfig:dualwidebandvirtual}), this sparsity also exists in $\bm{x}_k = \text{vec}(\bm{X}_k)$, as shown in Fig. \ref{subfig:dualwidebandvirtualvec}. Therefore, the first step of the channel estimation problem can be written as
\begin{align}
    &\min_{ \{\bm{x}_k\}_{k=1}^K} \Big\| [\bm{x}_1,\bm{x}_2,\ldots,\bm{x}_K] \Big\|_{1,0},\nonumber \\
    & \text{ s.t. } \sum_{k=1}^K \Big\| \bm{y}_{\text{w},k}-\Big({\bm D}_{\text{bs},k}(\tilde{\bm{\phi}}) \otimes \bm{D}_{\text{ms},k}(\tilde{\bm{\theta}})\Big)\bm{x}_k \Big\|_2^2 < \epsilon,
    \label{eqn:optproblem}
\end{align}
where $\|.\|_{1,0}$ denotes the number of rows that are not with all-zero elements,
$\otimes$ denotes the Kronecker product, $\bm{y}_{\text{w},k}$ is the vectorization of $\bm{Y}_{\text{w},k}$, $\bm{D}_{\text{ms},k}(\tilde{\bm{\theta}}) = \bm{W}_{\text{w}}^T {\bm A}_{\text{ms},k}(\tilde{\bm{\theta}})$ and $\bm{D}_{\text{bs},k}(\tilde{\bm{\phi}}) = \bm{S}_k^T \bm{F}^T {\bm A}_{\text{bs},k}(\tilde{\bm{\phi}})$. When $K=1$, (\ref{eqn:optproblem}) reduces to the narrowband case, which can be re-formulated as a standard sparse signal recovery problem and solved using methods such as OMP. However, when $K\ge2$, OMP is not applicable, as the non-zero elements of $\bm{x}_k$ for different $k$ occur in the same locations rather than independently. One may suggest solving (\ref{eqn:optproblem}) for each $k$ and then averaging the results, but the estimation results from different sub-channels contain multiple AOD/AOA's, and it is not known in prior which AOD/AOA's from each $k$ should be combined. Therefore, it is important to exploit and enforce the common sparsity pattern of $\{\bm{x}_k\}_{k=1}^K$ together with channel estimation. We will introduce a probabilistic model in Section \ref{sec:model} to achieve this.

\subsection{Off-Grid Steering Matrices Modeling}
The problem formulation (\ref{eqn:optproblem}) would be adequate if the true $\{\phi_\ell\}_{\ell=1}^L$ and $\{\theta_\ell\}_{\ell=1}^L$ lie on the grid points of $\tilde{\bm{\phi}}$ and $\tilde{\bm{\theta}}$, respectively. However, in practice, it is not quite realistic to make such an assumption even if we have a very fine grid \cite{yang2012off, hu2012compressed}. Therefore, we add a small shift $\bm{\delta}_\phi \in \mathbb{R}^{N_{\phi}}$ and $\bm{\delta}_\theta \in \mathbb{R}^{N_\theta}$ to $\tilde{\bm{\phi}}$ and $\tilde{\bm{\theta}}$, respectively, and allow for their update in (\ref{eqn:optproblem}). In particular, the dictionary ${\bm D}_{\text{bs},k}(\tilde{\bm{\phi}}) \otimes \bm{D}_{\text{ms},k}(\tilde{\bm{\theta}})$ should be replaced by the off-grid one $\bm{D}_k(\bm{\delta}_\phi,\bm{\delta}_\theta) = \bm{D}_{\text{bs},k}(\tilde{\bm{\phi}}+\bm{\delta}_\phi)\otimes \bm{D}_{\text{ms},k}(\tilde{\bm{\theta}}+\bm{\delta}_\theta)$. However, as $\bm{D}_k(\bm{\delta}_\phi,\bm{\delta}_\theta)$ is non-linear w.r.t. $\bm{\delta}_\phi$ and $\bm{\delta}_\theta$, their update would not be straightforward (in fact it is this non-linearity that we introduce grids in the first place). Fortunately, as long as the grid is dense enough, $\bm{D}_k(\bm{\delta}_\phi,\bm{\delta}_\theta)$ can be approximated by
\begin{align}
    \bm{D}_k(\bm{\delta}_\phi,\bm{\delta}_\theta)  \approx & \bm{D}_{\text{bs},k}(\tilde{\bm{\phi}})\otimes \bm{D}_{\text{ms},k}(\tilde{\bm{\theta}}) \nonumber \\
    & + \Big(\frac{\partial \bm{D}_{\text{bs},k}(\tilde{\bm{\phi}})}{\partial \tilde{\bm{\phi}}} \text{diag}(\bm{\delta}_\phi) \Big)\otimes \bm{D}_{\text{ms},k}(\tilde{\bm{\theta}}) \nonumber \\
    & + \bm{D}_{\text{bs},k}(\tilde{\bm{\phi}})\otimes \Big(\frac{\partial \bm{D}_{\text{ms},k}(\tilde{\bm{\theta}})}{\partial \tilde{\bm{\theta}}} \text{diag}(\bm{\delta}_\theta)\Big),
    \label{eqn:taylorapprox}
\end{align}
in which the partial derivative
\begin{align}
    \frac{\partial \bm{D}_{\text{bs},k}(\tilde{\bm{\phi}})}{\partial \tilde{\bm{\phi}}} = \Big[ \frac{\partial \bm{D}_{\text{bs},k}(\phi_1^{(i)})}{\partial {\phi}_1^{(i)}},\ldots,\frac{\partial \bm{D}_{\text{bs},k}({\phi}_{N_\phi}^{(i)})}{\partial {\phi}_{N_\phi}^{(i)}} \Big],\nonumber \\
    \frac{\partial \bm{D}_{\text{ms},k}(\tilde{\bm{\theta}})}{\partial \tilde{\bm{\theta}}} = \Big[ \frac{\partial \bm{D}_{\text{ms},k}({\theta}_1^{(i)})}{\partial {\theta}_1^{(i)}},\ldots,\frac{\partial \bm{D}_{\text{ms},k}({\theta}_{N_\theta}^{(i)})}{\partial {\theta}_{N_\theta}^{(i)}} \Big]\nonumber 
\end{align}
are fixed and can be pre-computed. In this way, the sensing matrix $\bm{D}_k(\bm{\delta}_\phi,\bm{\delta}_\theta) $ becomes linear w.r.t. the unknown parameters $\bm{\delta}_\phi$ and $\bm{\delta}_\theta$.

\section{Probabilistic common sparsity modeling}
\label{sec:model}

In (\ref{eqn:optproblem}), the choice of $\epsilon$ has a great impact on the estimated number of paths, and consequently the accuracy of channel estimation. In particular, if the noise power at the receiver is small, $\epsilon$ should be set as a relatively small value for a smaller fitting square error. On the other hand, if the noise power is large, $\epsilon$ should be set relatively large to overcome noise overfitting. However, determining the best settings of $\epsilon$ under different SNRs requires heavy parameter tuning. Moreover, the noise power is generally unknown in practice, which makes parameter tuning not an option. Therefore, instead of directly solving (\ref{eqn:optproblem}), we propose to build a probabilistic model, which is able to introduce common sparsity into $\{\bm{X}_k\}_{k=1}^K$, while not requiring parameter tuning.

\subsection{Sparsity Inducing Probabilistic Model}
\label{subsec:probabilisticmodel}
To model the sparsity as shown in Fig. \ref{subfig:dualwidebandvirtualvec}, we propose to adopt the Gaussian-Gamma prior, which has been proved to promote sparsity in many machine learning tasks \cite{cheng2016probabilistic,chen2022bayesian}. Considering the shared sparsity pattern in different $\bm{x}_k$, the elements with the same position in $\bm{x}_k$ for $k=1$ to $K$ should be assigned the same prior distribution, all controlled by one common latent variable. Therefore, the augmented path loss $\{\bm{x}_k\}_{k=1}^K$ are modeled to be distributed as
\begin{align}
    p(\{\bm{x}_k\}_{k=1}^K) = \prod_{k=1}^{K}  \mathbb{CN} (\bm{x}_k|0,\text{diag}(\bm{\lambda})^{-1}), \text{ for } k = 1, 2, \ldots, K,
    \label{eqn:Xkmodel}
\end{align}
where $\bm\lambda \in \mathbb{R}^{N_\theta N_\phi}$ is Gamma distributed with
\begin{align}
    p(\bm{\lambda}) = \prod_{r=1}^{N_\theta N_\phi} \text{Gamma}(\bm{\lambda}_{r}|\bm{\gamma}_r,\bm{\beta}_r),
    \label{eqn:lambdamodel}
\end{align}
with $\bm{\lambda}_r$ standing for the $r$th element of $\bm{\lambda}$, $\bm{\gamma}_r$ and $\bm{\beta}_r$ are set to be very small values (e.g., $10^{-6}$) so that the prior distribution of $\bm{\lambda}_r$ is non-informative.

For the noise in the received signal, it follows independent and identically distributed (i.i.d.) complex Gaussian distribution with mean zero and precision $\xi$,

\begin{align}
    p(\bm{N}_{\text{w},k}) = \prod_{p=1}^{P} \mathbb{CN} (\bm{n}_{\text{w},k}(p) | 0, \xi^{-1} \bm{I}_{N_\text{s}}),
\end{align}
where $\bm{N}_{\text{w},k} = \bm{\Sigma}^{-\frac{1}{2}}\bm{U}^H \bm{N}_k$ is the noise after the whitening step, and $\bm{n}_{\text{w},k}(p)$ denotes the $p$-th column of $\bm{N}_{\text{w},k}$. Furthermore, as the noise precision (i.e., the inverse of the variance) is a positive value but unknown, we further model the noise precision $\xi$ as
\begin{align}
    p({\xi}) = \text{Gamma}({\xi}|\gamma_\xi,\beta_\xi),
    \label{eqn:ximodel}
\end{align}
where $\gamma_\xi$ and $\beta_\xi$ are also set to $10^{-6}$ to induce a non-informative prior on $\xi$.

\begin{figure}[!t]
    \centering
    \includegraphics[width=0.95\linewidth]{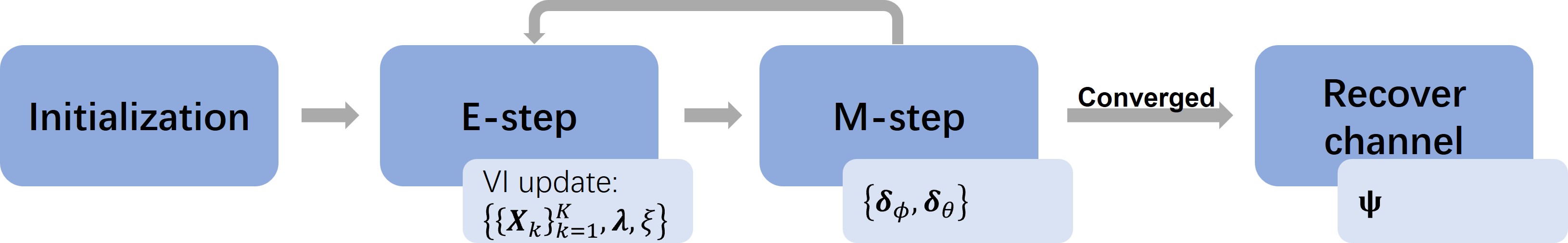}
    \caption{Outline of the proposed algorithm.}
    \label{fig:algorithmoutline}
\end{figure}

On the other hand, the likelihood of the $\bm{y}_{\text{w},k}$ given $\bm{x}_k$ and the noise precision $\xi$ is given by
\begin{align}
    p(\bm{y}_{\text{w},k}|\bm{x}_k,\xi) = &\mathbb{CN}\bigg(\bm{y}_{\text{w},k}\Big|\bm{D}_k(\bm{\delta}_\phi,\bm{\delta}_\theta) \bm{x}_k,\xi^{-1}\bm{I}_{PN_\text{s}}\bigg),\nonumber\\
    & \text{ for } k = 1, 2, \ldots, K.
    \label{eqn:yklikelihood}
\end{align}

\subsection{Variational EM Learning Framework}
To learn both the augmented path loss $\{\bm{x}_k\}_{k=1}^K$ and the grid offsets $\bm{\delta}_\phi$ and $\bm{\delta}_\theta$, an expectation-maximization (EM) algorithm \cite{mclachlan2007algorithm} is proposed (in the next section), which targets at minimizing the marginalized likelihood of $\{\bm{y}_{\text{w},k}\}_{k=1}^K$ w.r.t. $\bm{\delta}_\phi$ and $\bm{\delta}_\theta$, and treats the variables $\Big\{ \{\bm{x}_k\}_{k=1}^K,\bm{\lambda},\xi\Big\}$ in the last sub-section as latent variables. In each iteration of the algorithm, the latent variables are firstly updated with fixed channel parameters, which corresponds to the E-step. Specifically, as the exact expectation of variables cannot be analytically derived due to the complicated model proposed above, the variational inference (VI) is adopted. After the E-step, the channel parameters are updated with fixed latent variables, which corresponds to the M-step. Upon the convergence of the EM algorithm, the path loss and time delay are estimated. An outline of the proposed algorithm is depicted in Fig. \ref{fig:algorithmoutline}.

\section{Derivation of the algorithm}
\label{sec:algorithm}
In this section, our goal is to maximize the marginalized likelihood function $p(\{\bm{{y}}_{\text{w},k}\}_{k=1}^K|\bm{\delta}_{\phi},\bm{\delta}_{\theta}) = \int p(\{\bm{{y}}_{\text{w},k}\}_{k=1}^K,\bm{\Omega}|\bm{\delta}_{\phi},\bm{\delta}_{\theta}) d\bm{\Omega}$, in which $\bm{\Omega}$ is the shorthand notation for $\Big\{ \{\bm{x}\}_{k=1}^K,\bm{\lambda},\xi\Big\}$ and the joint likelihood is obtained by multiplying (\ref{eqn:yklikelihood}), (\ref{eqn:Xkmodel}), (\ref{eqn:lambdamodel}) and (\ref{eqn:ximodel}) as
\begin{align}
    \ln & p(\{\bm{{y}}_{\text{w},k}\}_{k=1}^K,\bm{\Omega}|\bm{\delta}_{\phi},\bm{\delta}_{\theta}) = KPN_\text{s} \ln\xi - \xi\sum_{k=1}^{K}\Big(\bm{y}_{\text{w},k} \nonumber \\
    & -\bm{D}_{k}(\bm{\delta}_{\phi},\bm{\delta}_{\theta})\bm{x}_k\Big)^{H}\Big(\bm{y}_{\text{w},k}-\bm{D}_{k}(\bm{\delta}_{\phi},\bm{\delta}_{\theta})\bm{x}_k\Big) \nonumber \\
    &+  K\sum_{r=1}^{N_\phi N_\theta} \ln \bm{\lambda}_{r} - \sum_{k=1}^{K} \bm{x}_k^H \text{diag}(\bm{\lambda})\bm{x}_k\nonumber \\
    & + \sum_{r=1}^{N_{\phi} N_\theta}(\bm{\gamma}_r-1)\ln \bm{\lambda}_{r} - \sum_{r=1}^{N_{\phi} N_\theta}\bm{\beta}_r \bm{\lambda}_r + (\gamma_\xi-1)\xi - \beta_\xi \xi.
    \label{eqn:yjointlikelihood}
\end{align}
Due to the complicated probabilistic model, the exact integration for getting rid of the latent variables $\bm{\Omega}$ is not tractable. To this end, we rewrite the maximization of the likelihood function by introducing variational distribution $q(\bm{\Omega})$, and the maximum likelihood estimation problem can be equivalently formulated as
\begin{align}
    &\max_{q( \bm{\Omega}),\bm{\delta}_\phi,\bm{\delta}_\theta} \ln p(\{\bm{{y}}_{\text{w},k}\}_{k=1}^K|\bm{\delta}_\phi,\bm{\delta}_\theta) \nonumber\\
    = & \underbrace{\int q(\bm{\Omega}) \ln{\frac{p(\bm{\Omega},\{\bm{{y}}_{\text{w},k}\}_{k=1}^K|\bm{\delta}_\phi,\bm{\delta}_\theta)}{q(\bm{\Omega})}}d\bm{\Omega}}_{L(\bm{\delta}_\phi,\bm{\delta}_\theta)} \nonumber \\
    \quad & \underbrace{- \int q(\bm{\Omega}) \ln{\frac{p(\bm{\Omega}|\bm{\delta}_\phi,\bm{\delta}_\theta,\{\bm{{y}}_{\text{w},k}\}_{k=1}^K)}{q(\bm{\Omega})}}d\bm{\Omega}}_{\text{KL}\Big(q(\bm{\Omega})||p(\bm{\Omega}|\bm{\delta}_\phi,\bm{\delta}_\theta,\{\bm{{y}}_{\text{w},k}\}_{k=1}^K)\Big)}.
    \label{eqn:EMformula}
\end{align}
As shown in (\ref{eqn:EMformula}), the marginalized log-likelihood is composed of two parts, i.e., the KL-divergence $\text{KL}\Big(q(\bm{\Omega})$ $||p(\bm{\Omega}|\bm{\delta}_\phi,\bm{\delta}_\theta,\{\bm{{y}}_{\text{w},k}\}_{k=1}^K)\Big)$ and the lower bound $L(\bm{\delta}_\phi,\bm{\delta}_\theta)$ \cite[pp.~450]{bishop2006pattern}. Since the KL-divergence is always larger than zero, (\ref{eqn:EMformula}) can be maximized by alternatively minimizing $\text{KL}\Big(q||p\Big)$ w.r.t. $q(\bm{\Omega})$ by fixing $\bm{\delta}_\phi$ and $\bm{\delta}_\theta$, and maximizing $L(\bm{\delta}_\phi,\bm{\delta}_\theta)$ w.r.t. $\bm{\delta}_\phi$ and $\bm{\delta}_\theta$ by fixing $q(\bm{\Omega})$.

\subsection{Minimization of KL Divergence (Variational E-step)}
In the KL divergence minimization step, the subproblem is
\begin{align}
    &\min_{q(\bm{\Omega})} \text{KL}\Big(q(\bm{\Omega})||p(\bm{\Omega}|\bm{\delta}_\phi,\bm{\delta}_\theta,\{\bm{{y}}_{\text{w},k}\}_{k=1}^K)\Big) 
\end{align}
If there is no restriction on $q(\bm{\Omega})$, the optimal solution is obviously $q(\bm{\Omega}) = p(\bm{\Omega}|\bm{\delta}_\phi,\bm{\delta}_\theta,\{\bm{{y}}_{\text{w},k}\}_{k=1}^K)$, which is the posterior distribution of the latent variables. However, this brings us back to the original challenge of intractable $\int p(\{\bm{{y}}_{\text{w},k}\}_{k=1}^K,\bm{\Omega}|\bm{\delta}_\phi,\bm{\delta}_\theta)d\bm{\Omega}$. Therefore, the mean-field approximation on $q(\bm{\Omega})$ is adopted, which assumes that the variable sets are independent with each other as follows,
\begin{align}
    q(\bm{\Omega}) = \prod_{k=1}^K q(\bm{x}_k) q(\bm{\lambda}) q(\xi).
\end{align}
Expressing $\bm{\Omega}$ using exhaustive but non-overlapping subsets $\{\bm{\Omega}_s\}_{s=1}^{S}$ (i.e., $\cup_{s=1}^S \bm{\Omega}_s = \bm{\Omega} $ and ${\bm{\Omega}_s} \cap {\bm{\Omega}_j}=\emptyset$ for $s\neq j$), the optimal distribution of $\bm{\Omega}_s$ can be derived as \cite[pp.~737]{murphy2012machine}
\begin{align}
    \ln q^\star(\bm{\Omega}_s) = \mathbb{E}_{\bm{\Omega}\setminus\bm{\Omega}_s} \Big\llbracket \ln p(\bm{\Omega},\{\bm{{y}}_{\text{w},k}\}_{k=1}^K|\bm{\delta}_\phi,\bm{\delta}_\theta)\Big\rrbracket + \text{const},
    \label{eqn:MFupdate}
\end{align}
where $\mathbb{E}_{\bm{\Omega}\setminus\bm{\Omega}_s}$ denotes the expectation of all variables except $\bm{\Omega}_s$.

Then by substituting (\ref{eqn:yjointlikelihood}) into (\ref{eqn:MFupdate}) and taking expectation on the corresponding variable sets, the variational distributions of $ \Big\{ \{\bm{x}_k\}_{k=1}^K,\bm{\lambda},\xi\Big\}$ can be derived accordingly. The results are summarized in the following and the detailed derivation can be found in Appendix A ().

\noindent \underline{Update for $\bm{x}_k$, for $k=1$ to $K$}

$q(\bm{x}_k)$ follows a complex Gaussian distribution $\mathbb{CN}(\bm{x}_k|\bm{m}_{\bm{x}_k},\bm{\Sigma}_{\bm{x}_k})$, with
\begin{align}
    & \bm{\Sigma}_{\bm{x}_k} = \Big( \mathbb{E}\llbracket \xi \rrbracket \bm{D}_{k}^H(\bm{\delta}_{\phi},\bm{\delta}_{\theta})\bm{D}_{k}(\bm{\delta}_{\phi},\bm{\delta}_{\theta}) + \text{diag}(\mathbb{E}\llbracket \bm\lambda \rrbracket) \Big)^{-1}, \nonumber \\
    & \bm{m}_{\bm{x}_k} = \mathbb{E}\llbracket\xi\rrbracket \bm{\Sigma}_{\bm{x}_k}\bm{D}_{k}^H(\bm{\delta}_{\phi},\bm{\delta}_{\theta}) \bm{y}_{\text{w},k}.
    \label{eqn:Xupdate}
\end{align}

\noindent \underline{Update for $\bm{\lambda}$}

Each element of $\bm{\lambda}$ follows a Gamma distribution, i.e., $q(\bm{\lambda}) = \prod_{r=1}^{N_\phi N_\theta} \text{Gamma}(\bm{\lambda}_r|\hat{\bm{\gamma}}_r,\hat{\bm{\beta}}_r)$ with
\begin{align}
    & \hat{\bm{\gamma}}_r = \bm{\gamma}_r + K, \nonumber \\
    & \hat{\bm{\beta}}_r = \bm{\beta}_r +  [\bm{\Sigma}_{\bm{x}_k} ]_{r,r} + [\bm{m}_{\bm{x}_k}]_r^*[\bm{m}_{\bm{x}_k}]_r,
    \label{eqn:lambdaupdate}
\end{align}
and expectation $\mathbb{E}\llbracket \bm{\lambda}_r \rrbracket = \hat{\bm{\gamma}}_r/\hat{\bm{\beta}}_r$.

\noindent \underline{Update for $\xi$}

The variational distribution of $\xi$ is a Gamma distribution $\text{Gamma}(\xi|\hat{\gamma}_\xi,\hat{\beta}_\xi)$ with
\begin{align}
    &\hat{\gamma}_\xi = KPN_\text{s} + \gamma_\xi, \nonumber \\
    &\hat{\beta}_\xi = \beta_\xi + \sum_{k=1}^{K} \bigg( \bm{y}_{\text{w},k}^H\bm{y}_{\text{w},k} +  \text{tr} \Big(\bm{D}_{k}^H(\bm{\delta}_{\phi},\bm{\delta}_{\theta})\bm{D}_{k}(\bm{\delta}_{\phi},\bm{\delta}_{\theta})  \nonumber \\
    &  \big( \bm{\Sigma}_{\bm{x}_k} + \bm{m}_{\bm{x}_k}\bm{m}_{\bm{x}_k}^H \big)\Big) - 2 \Re \Big\{ \bm{y}_{\text{w},k}^H \bm{D}_{k}(\bm{\delta}_{\phi},\bm{\delta}_{\theta}) \bm{m}_{\bm{x}_k} \Big\} \bigg).
    \label{eqn:xiupdate}
\end{align}

\subsection{Maximization of the Lower Bound (M-step)}
When $q(\bm{\Omega})$ is fixed, the maximization of the lower bound of (\ref{eqn:EMformula}) is
\begin{align}
    \max_{\bm{\delta}_\phi,\bm{\delta}_\theta} L(\bm{\delta}_\phi,\bm{\delta}_\theta) = \mathbb{E}_{q(\bm{\Omega})} \Big \llbracket \ln p(\bm{\Omega},\{\bm{{y}}_{\text{w},k}\}_{k=1}^K|\bm{\delta}_\phi,\bm{\delta}_\theta) \Big \rrbracket + \text{const}.
    \label{eqn:maxL}
\end{align}
Extracting the terms related to $\bm{\delta}_\phi$ and $\bm{\delta}_\theta$, it is shown in Appendix \ref{apdsec:mstepupdate} that (\ref{eqn:maxL}) can be simplified as
\begin{align}
    \min_{\bm{\delta}_\phi,\bm{\delta}_\theta} \text{ } &-2\Big(Re\{ \bm{v}_1  \}^T{\bm{\delta}_\theta} + Re\{ \bm{v}_2  \}^T{\bm{\delta}_\phi}\Big)  + {\bm{\delta}_\theta}^T \bm{E}_{11} {\bm{\delta}_\theta} \nonumber \\
    & + {\bm{\delta}_\phi}^T \bm{E}_{22} {\bm{\delta}_\phi}+ 2{\bm{\delta}_\theta}^T Re\{\bm{E}_{12}\} {\bm{\delta}_\phi},
    \label{eqn:emtoquadratic}
\end{align}
with
\begin{align}
    &\bm{v}_1 = \sum_{k=1}^K\Bigg( \bm{y}_{\text{w},k}^H  \Big(  \frac{\partial \bm{D}_{\text{bs},k}(\tilde{\bm{\phi}})}{\partial \tilde{\bm{\phi}}}  \otimes \bm{D}_{\text{ms},k}(\tilde{\bm{\theta}}) \Big) \text{diag}(\bm{m}_{\bm{x}_k}) {\bm{B}^{(1)}}\nonumber \\
    & - \text{diag}\bigg(
    \Big(   \frac{\partial \bm{D}_{\text{bs},k}(\tilde{\bm{\phi}})}{\partial \tilde{\bm{\phi}}} \otimes \bm{D}_{\text{ms},k}(\tilde{\bm{\theta}})   \Big)^H
    \Big(  \bm{D}_{\text{bs},k}(\tilde{\bm{\phi}})\otimes \bm{D}_{\text{ms},k}(\tilde{\bm{\theta}}) \Big)\nonumber\\
     &\times \big(   \bm{\Sigma}_{\bm{x}_k} + \bm{m}_{\bm{x}_k}\bm{m}_{\bm{x}_k}^H\big) \bigg)^T{\bm{B}^{(1)}}\Bigg)^T
    ,\nonumber \\
    &\bm{v}_2 = \sum_{k=1}^K\Bigg(\bm{y}_{\text{w},k}^H  \Big( \bm{D}_{\text{bs},k}(\tilde{\bm{\phi}})\otimes \frac{\partial \bm{D}_{\text{ms},k}(\tilde{\bm{\theta}})}{\partial \tilde{\bm{\theta}}}  \Big) \text{diag}(\bm{m}_{\bm{x}_k}) \bm{B}^{(2)}\nonumber\\
    & -\text{diag}\bigg(
    \Big(    \bm{D}_{\text{bs},k}(\tilde{\bm{\theta}}) \otimes  \frac{\partial\bm{D}_{\text{ms},k}(\tilde{\bm{\theta}})}{\partial \tilde{\bm{\theta}}}   \Big)^H
    \Big(  \bm{D}_{\text{bs},k}(\tilde{\bm{\phi}})\otimes \bm{D}_{\text{ms},k}(\tilde{\bm{\theta}}) \Big) \nonumber \\
     & \times \big(   \bm{\Sigma}_{\bm{x}_k} + \bm{m}_{\bm{x}_k}\bm{m}_{\bm{x}_k}^H\big) \bigg)^T
    {\bm{B}^{(2)}} \Bigg)^T,\nonumber \\
    &\bm{E}_{11} = \sum_{k=1}^K{\bm{B}^{(1)}}^H 
    \bigg(\Big(   \frac{\partial \bm{D}_{\text{bs},k}(\tilde{\bm{\phi}})}{\partial \tilde{\bm{\phi}}} \otimes \bm{D}_{\text{ms},k}(\tilde{\bm{\theta}})   \Big)^H  \nonumber \\
    & \times \Big(   \frac{\partial \bm{D}_{\text{bs},k}(\tilde{\bm{\phi}})}{\partial \tilde{\bm{\phi}}} \otimes \bm{D}_{\text{ms},k}(\tilde{\bm{\theta}})   \Big) \ast \big(   \bm{\Sigma}_{\bm{x}_k} + \bm{m}_{\bm{x}_k}\bm{m}_{\bm{x}_k}^H\big)^T \bigg) 
    {\bm{B}^{(1)}},\nonumber \\
    & \bm{E}_{22} =\sum_{k=1}^K {\bm{B}^{(2)}}^H
    \bigg(\Big(    \bm{D}_{\text{bs},k}(\tilde{\bm{\theta}}) \otimes  \frac{\partial\bm{D}_{\text{ms},k}(\tilde{\bm{\theta}})}{\partial \tilde{\bm{\theta}}}   \Big)^H  \nonumber\\
    &\times \Big(    \bm{D}_{\text{bs},k}(\tilde{\bm{\theta}}) \otimes  \frac{\partial\bm{D}_{\text{ms},k}(\tilde{\bm{\theta}})}{\partial \tilde{\bm{\theta}}}   \Big) \ast \big(   \bm{\Sigma}_{\bm{x}_k} + \bm{m}_{\bm{x}_k}\bm{m}_{\bm{x}_k}^H\big)^T \bigg) 
     \bm{B}^{(2)}, \nonumber \\
     & \bm{E}_{12} =\sum_{k=1}^K {\bm{B}^{(1)}}^H 
    \bigg(\Big(   \frac{\partial \bm{D}_{\text{bs},k}(\tilde{\bm{\phi}})}{\partial \tilde{\bm{\phi}}} \otimes \bm{D}_{\text{ms},k}(\tilde{\bm{\theta}})   \Big)^H  \nonumber \\
    & \times \Big(    \bm{D}_{\text{bs},k}(\tilde{\bm{\theta}}) \otimes  \frac{\partial\bm{D}_{\text{ms},k}(\tilde{\bm{\theta}})}{\partial \tilde{\bm{\theta}}}   \Big) \ast \big(   \bm{\Sigma}_{\bm{x}_k} + \bm{m}_{\bm{x}_k}\bm{m}_{\bm{x}_k}^H\big)^T \bigg)  
    \bm{B}^{(2)},
    \label{eqn:quad_formulation}
\end{align}
where $\bm{B}^{(1)} \in \mathbb{R}^{(N_\phi N_\theta) \times N_\phi}$ and $\bm{B}^{(2)} \in \mathbb{R}^{(N_\phi N_\theta )\times N_\theta}$ are defined as
\begin{align}
    &\bm{B}^{(1)} =  \bm{I}_{N_\phi} \odot \bm{1}_{N_\theta \times N_\phi},\nonumber \\
    &\bm{B}^{(2)} =  \bm{1}_{N_\phi \times N_\theta} \odot \bm{I}_{N_\theta},
    \label{eqn:Bformulation}
\end{align}
with $\odot$ denoting the Khatri-Rao product and $\bm{1}_{M\times N}$ denoting a $M \times N$ matrix with all elements being $1$.

Then the optimal solution to (\ref{eqn:emtoquadratic}) can be obtained by setting the differentiation of (\ref{eqn:emtoquadratic}) with respect to $\bm{\delta}_\phi$ and $\bm{\delta}_\theta$ to zero, and this results in
\begin{align}
    \begin{bmatrix}
    \bm{\delta}_\phi^\star \\
    \bm{\delta}_\theta^\star
    \end{bmatrix} = 
    \begin{bmatrix}
    \bm{E}_{11} & \bm{E}_{12} \\
    \bm{E}_{12}^H & \bm{E}_{22}
    \end{bmatrix}^{-1}
    \begin{bmatrix}
    \Re\{\bm{v}_1\}\\
    \Re\{\bm{v}_2\}
    \end{bmatrix}.
    \label{eqn:emupdate}
\end{align}

\subsection{Delay and Path Loss Estimation}
\label{subsec:delayalpha}
According to (\ref{eqn:dual_wideband_channel_model}) and (\ref{eqn:Xkrelation}), the relationship between $[\bm{X}_{k}]_{i,j}$ for $k=1$ to $K$ and the corresponding path loss and time delay is
\begin{align}
    \Big[ [\bm{X}_{1}]_{i,j},\ldots,[\bm{X}_{K}]_{i,j} \Big] =  \bm{\Upsilon}_{i,j} [e^{-j2\pi f_1\bm{\Psi}_{i,j}},\ldots,e^{-j2\pi f_K\bm{\Psi}_{i,j}}],
\end{align}
where $\bm{\Upsilon}_{i,j}$ is the path loss of the virtual path with AOA $\tilde{\bm{\theta}}_i$ and AOD $\tilde{\bm{\phi}}_j$. Denote $[\bm{X}_{1:K}]_{i,j} = \Big[ [\bm{X}_{1}]_{i,j},\ldots,[\bm{X}_{K}]_{i,j} \Big]^T$ and $\bm{e}(\bm{\Psi}_{i,j}) =[e^{-j2\pi f_1\bm{\Psi}_{i,j}},\ldots,e^{-j2\pi f_K\bm{\Psi}_{i,j}}]^T $, the problem of path loss and delay estimation is
\begin{align}
    \min_{\bm{\Upsilon}_{i,j},\bm{\Psi}_{i,j}} \Big\|[\bm{X}_{1:K}]_{i,j} - \bm{\Upsilon}_{i,j} \bm{e}(\bm{\Psi}_{i,j})\Big\|_2^2.
    \label{eqn:delayest}
\end{align}
With fixed $\bm{\Psi}_{i,j}$, the optimal solution for $\bm{\Upsilon}_{i,j}$ is
\begin{align}
    \bm{\Upsilon}_{i,j}^\star = \frac{\bm{e}^H(\bm{\Psi}_{i,j})[\bm{X}_{1:K}]_{i,j}}{\bm{e}^H(\bm{\Psi}_{i,j})\bm{e}(\bm{\Psi}_{i,j})} = \frac{1}{K}\bm{e}^H(\bm{\Psi}_{i,j})[\bm{X}_{1:K}]_{i,j}.
    \label{eqn:alphaupdate}
\end{align}
Then by plugging (\ref{eqn:alphaupdate}) back to (\ref{eqn:delayest}), $\bm{\Upsilon}_{i,j}^\star$ is eliminated, and the problem becomes
\begin{align}
    & \min_{\bm{\Psi}_{i,j}} \Big\|[\bm{X}_{1:K}]_{i,j} - \frac{1}{K} \bm{e}^H(\bm{\Psi}_{i,j})[\bm{X}_{1:K}]_{i,j}\bm{e}(\bm{\Psi}_{i,j})\Big\|_2^2 \nonumber \\
    =&  \max_{\bm{\Psi}_{i,j}} \Big( \bm{e}^H(\bm{\Psi}_{i,j})[\bm{X}_{1:K}]_{i,j}\Big)^H \Big( \bm{e}^H(\bm{\Psi}_{i,j})[\bm{X}_{1:K}]_{i,j}\Big),
    \label{eqn:tauupdate}
\end{align}
in which the best solution can be found by one-dimensional grid search. Notice that for each $(i,j)$, (\ref{eqn:tauupdate}) is solved independently. Furthermore, we only need to solve for the $(i,j)$ where $[\bm{X}_{1:K}]_{i,j}$ is non-zero. Since $\bm{X}_k$ is sparse, only a small number of delay estimation problems in the form of (35) need to be solved.

\begin{algorithm}[!tb]
\SetAlgoLined
 \textbf{initialization:} Input the observed signal $\{\bm{y}_{\text{w},k}\}_{k=1}^K$ and system information $N_t$, $N_r$, $P$, $N_\text{s}$, $K$
 
 \While{Not Converged}{
    \textbf{E-step}
    
    \quad Update $\bm{X}$ according to (\ref{eqn:Xupdate}),
    
    \quad Update $\bm{\lambda}$ according to (\ref{eqn:lambdaupdate}),
    
    \quad Update $\xi$ according to (\ref{eqn:xiupdate}),

    \textbf{M-step}
    
    \quad Update $\bm{\delta}_\phi$ and $\bm{\delta}_\phi$ according to (\ref{eqn:emupdate}),
    
    Combine collapsed paths,
 }
 
 Update $\bm{\tau}$ according to (\ref{eqn:tauupdate}),
 
 Update $\bm{\bar\alpha}$ according to (\ref{eqn:alphaupdate}).
 
 \caption{Variational EM algorithm for dual-wideband channel estimation.}
 \label{alg:graphTTC_opt}
\end{algorithm}

\subsection{Summary and Further Discussion}
The algorithm can be summarized as Algorithm \ref{alg:graphTTC_opt}. After estimating the channel parameters, the channel is recovered using (\ref{eqn:dual_wideband_channel_model}), and the signals are recovered by (\ref{eqn:Y_k_received}) but without the noise term. According to \cite{tzikas2008variational}, since the algorithm is in the EM framework, the convergence of Algorithm \ref{alg:graphTTC_opt} is guaranteed. Specifically, in each iteration, $L(\bm{\delta}_\phi,\bm{\delta}_\theta)$, which is the lower bound for $\ln p(\{\bm{{y}}_{\text{w},k}\}_{k=1}^K|\bm{\delta}_\phi,\bm{\delta}_\theta)$, is non-decreasing in both the E-step and M-step. On the other hand, the likelihood will not change in the E-step, but gains a larger increase than the lower bound in the M-step due to a larger KL-divergence, so is also non-decreasing.

\subsubsection{Complexity Analysis}

For each KL-divergence minimization step, the computational complexity is $\bm{O}\Big( K (N_\phi N_\theta)^3 + K (N_\phi N_\theta )^2 P N_{\text{s}} \Big)$, which mainly comes from the computation of (\ref{eqn:Xupdate}) and (\ref{eqn:xiupdate}). For the maximization of the lower bound, it takes $\bm{O}\Big( K(N_\phi N_\theta)^2(P N_\text{s} + N_\phi) + (N_\phi + N_\theta)^3 \Big)$.

\subsubsection{Initilization}

For the probabilistic model, the hyper-parameters $\{\bm{\gamma}_r,\bm{\beta}_r\}_{r=1}^{N_\phi N_\theta}$ and $\{\gamma_\xi,\beta_\xi\}$ are all set as $10^{-6}$ so that the prior distributions are non-informative \cite{babacan2014bayesian}. For the initialization of $\mathbb{E} \llbracket \bm{\lambda}_r \rrbracket$, $\mathbb{E} \llbracket \xi \rrbracket$ and $\bm{\Sigma}_{\bm{x}_k}$, they are accordingly calculated as $\mathbb{E} \llbracket \bm{\lambda}_r \rrbracket = \bm{\gamma}_r/\bm{\beta}_r = 1$, $\mathbb{E} \llbracket \xi \rrbracket=\gamma_\xi/\beta_\xi=1$ and $\bm{\Sigma}_{\bm{x}_k} = \text{diag}(\mathbb{E} \llbracket \bm{\lambda} \rrbracket)^{-1} = \bm{I}_{N_\phi N_\theta}$, respectively. For the initialization of $\bm{m}_{\bm{x}_k}$, it is initialized by the results of OMP in each sub-channel.

\subsubsection{Grid update and pruning}

In Algorithm \ref{alg:graphTTC_opt}, we only consider fixed grids $\tilde{\bm{\phi}}$ and $\tilde{\bm{\theta}}$ with offsets $\bm{\delta}_\phi$ and $\bm{\delta}_\theta$. While such a strategy allows off-grid AOD/AOAs, the accuracy still highly depends on the accuracy of the first-order Taylor approximation (\ref{eqn:taylorapprox}). In practice, we could allow the update of the grids in each iteration, as $\tilde{\bm{\phi}}^{(i+1)} = \tilde{\bm{\phi}}^{(i)} + \bm{\delta}_\phi$ and $\tilde{\bm{\theta}}^{(i+1)} = \tilde{\bm{\theta}}^{(i)} + \bm{\delta}_\theta$, where the superscript denotes the $i$-th iteration. By substituting the updated $\tilde{\bm{\phi}}^{(i+1)}$ and $\tilde{\bm{\theta}}^{(i+1)}$ into (\ref{eqn:EMformula}), none of the equations in Algorithm \ref{alg:graphTTC_opt} would change, except that the defined grids are different in each iteration. Moreover, at the end of one iteration, such a substitution will not change the value of the likelihood, KL-divergence, or the lower bound. Since the lower bound and the likelihood will take a larger value in the next iteration, the convergence of the algorithm is still guaranteed.

On the other hand, because of the sparsity-promoting property of the probabilistic model, most of the estimated $\bm{x}_k$ are with values very close to zero. Therefore, during the execution of the algorithm, we can discard columns of $\bm{D}_k(\bm{\delta}_\phi,\bm{\delta}_\theta)$ if the corresponding values of $\bm{x}_k$ are all smaller than a threshold, e.g., $10^{-2}$, for $k=1$ to $K$. Furthermore, if we allow the update of grids as discussed above, there may be a situation that two different sets of AOD/AOAs become the same. Under this condition, we combine these two AOD/AOA sets, by pruning the one with a smaller augmented path loss. While the grid pruning strategy changes the lower bound, as well as the likelihood function, it is worth noting that after grid pruning, it can be seen as solving (\ref{eqn:EMformula}) with a smaller problem size, with the latest solution acting as the initialization of the new problem, and thus the convergence of the algorithm is still guaranteed. Furthermore, the pruning strategy greatly reduces the computational complexity because of the reduction of dictionary size in the probabilistic model.

\section{Simulation Results and Discussions}
\label{sec:experiments}
In the simulations, the system configuration is set as follows unless specified otherwise. The downlink channel has $N_t = 64$ transmitting antennas, $N_r= 32$ receiving antennas. OFDM with $K_0 = 256$ sub-channels is considered, with $K = 16$ sub-channels used for training (located at the $16^\text{th}$, $32^\text{nd}$, ..., $256^\text{th}$ subcarriers). The number of observed OFDM symbols is $P = 16$, and $N_\text{s} = 6$. The bandwidth is $f_s = 1.76$GHz, which is in accordance with the IEEE 802.11ad standard \cite{perahia2010ieee}, and the carrier frequency is $f_{\text{c}} = 60$GHz. All experiments are implemented in Matlab R2018b on a desktop with an Intel i7 6-core computer with $32$GB RAM.

In the simulations, the following methods are compared:

\noindent    1) \hspace{0.1cm} ESPRIT with the true path number assumed to be known, which would be denoted as 'ESPRIT' in the simulations;
    
\noindent    2) \hspace{0.1cm} ESPRIT with the path number estimated by minimum description length (MDL), denoted as 'ESPRIT-MDL';
    
\noindent    3) \hspace{0.1cm} OMP operated on the dictionaries in the angular domain (i.e., ${\bm D}_{\text{bs},k}(\tilde{\bm{\phi}}) \otimes \bm{D}_{\text{ms},k}(\tilde{\bm{\theta}})$ in (\ref{eqn:optproblem})), with the grid sizes $N_\phi$ and $N_\theta$ all set as $256$, denoted as 'OMP (angular)';

\noindent    4) \hspace{0.1cm} Off-grid OMP in the angular domain, in which the off-grid dictionary (\ref{eqn:taylorapprox}) is adopted, with the grid sizes set as $128$ for both $N_\phi$ and $N_\theta$. The algorithm is implemented by iteratively updating $\bm{x}_k$ through OMP and updating the dictionary by minimizing the target function in (\ref{eqn:emtoquadratic}) but without the expectation. This scheme is denoted as ‘Off-grid OMP (angular)’;

\noindent    5) \hspace{0.1cm} OMP operated on the dictionary in the angular-delay domain \cite{jian2019angle,wang2019block,wang2018spatial}, with the grid sizes set as $64$, $64$ and $50$ for the AOD, AOA and delay, respectively. It is denoted as 'OMP (angular-delay)'. Note that the grid size for each parameter is set much smaller compared to methods 3) and 4) due to the fact that the dictionary in the angular-delay domain is intrinsically much larger than that in the angular domain when the same grid density is chosen;

\noindent   6) \hspace{0.1cm} The proposed method, with the grid sizes set as $128$ for both $N_\phi$ and $N_\theta$, and is denoted as 'Proposed'.

\noindent For the first four algorithms (i.e., ESPRIT-based methods and OMP-based methods in the angular domain), they can only be performed independently on each sub-carrier, with each sub-problem in the form
\begin{align}
    &\min_{ \bm{x}_k} \Big\| \bm{y}_{\text{w},k}-\Big({\bm D}_{\text{bs},k}(\tilde{\bm{\phi}}) \otimes \bm{D}_{\text{ms},k}(\tilde{\bm{\theta}})\Big)\bm{x}_k \Big\|_2^2, \text{ s.t. }  \Big\| \bm{x}_k \Big\|_0 \le \hat{L},
    \label{eqn:optproblem_narrowband}
\end{align}
separately from $k=1$ to $K$, where $\hat{L}$ is the maximum of the possible path number. To combine these estimates from different sub-carriers, we first pair the estimated AOD/AOA's with the true one by minimizing the square error, then compute the average of the paired AOD/AOA's for evaluating the accuracy of AOD/AOA estimation. Any spurious estimated components not matched with the true AOD/AOA's are discarded. Note that this strategy is not achievable in practice as the true AOD/AOA’s are not available.


In the subsections \ref{subsec:EXP1} and \ref{subsec:EXP2} below, the precoding and combining matrices are all set to $[\bm{I},\bm{0}]^T$ as in \cite{lin2020tensor} so that the ESPRIT method is applicable. In this case, methods 1) to 4) described above are compared with the proposed algorithm. In the subsection \ref{subsec:EXP3} below, we consider cases where $\bm{W}$ and $\bm{F}$ are randomly generated to simulate the general operating scenarios. In this case, the ESPRIT-based methods are not applicable, and the methods 3) to 5) are compared to the proposed algorithm.
\begin{figure*}[!t]
    \centering
    \begin{subfigure}[b]{0.42\textwidth}
         \centering
         \includegraphics[width=\textwidth]{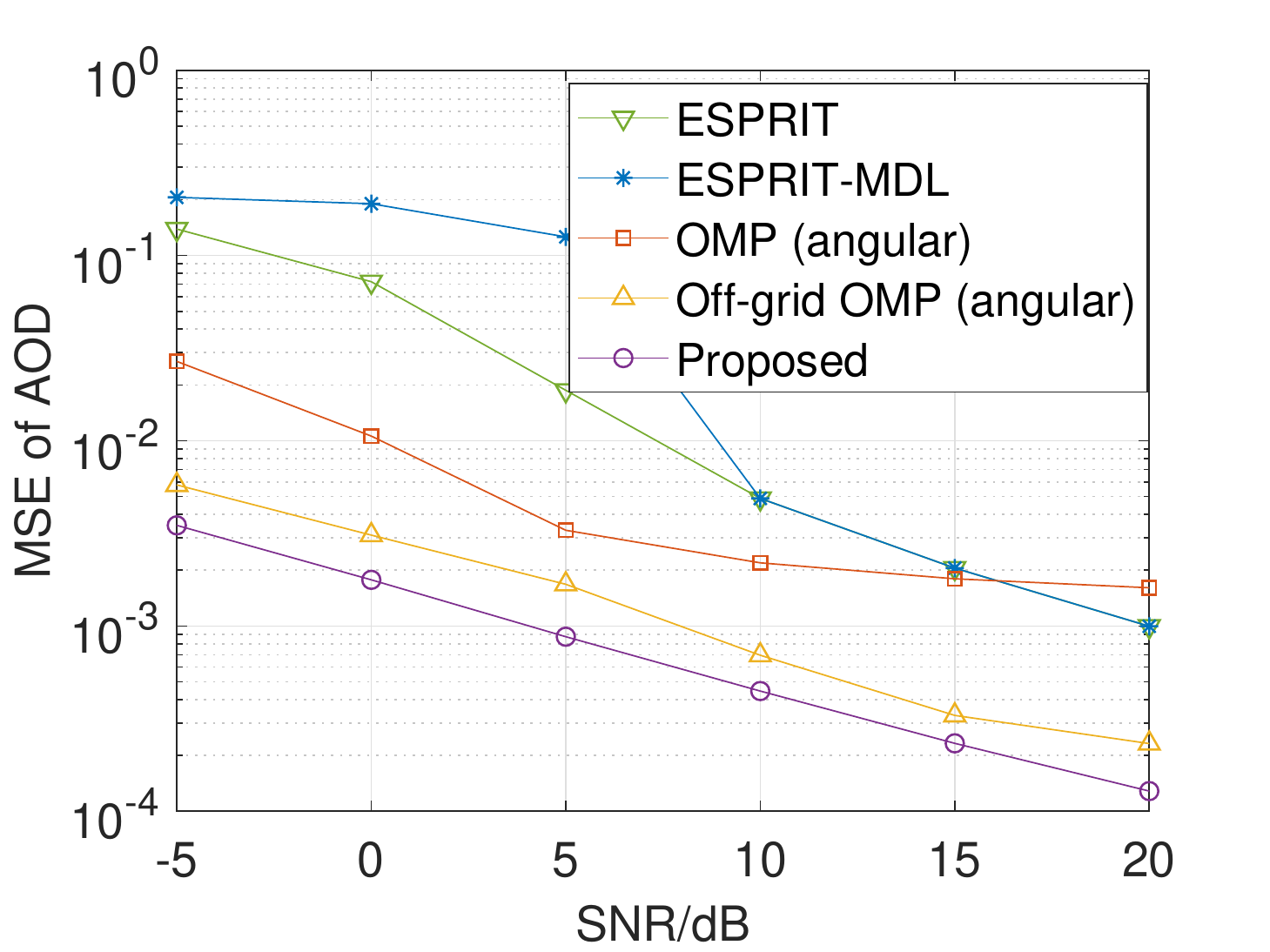}
         \caption{MSE of AOD estimation}
         \label{subfig:phimse}
     \end{subfigure}
     \begin{subfigure}[b]{0.42\textwidth}
         \centering
         \includegraphics[width=\textwidth]{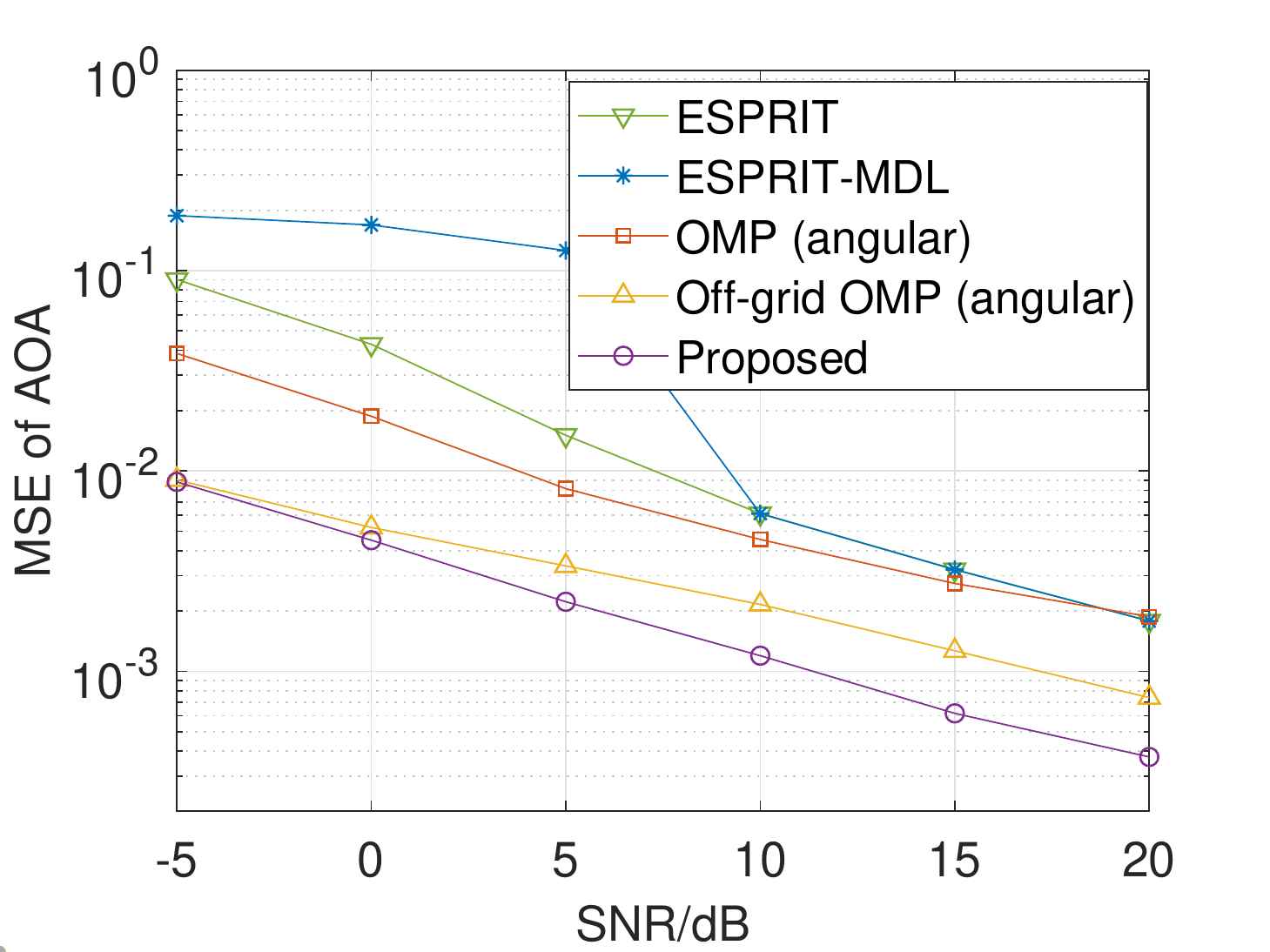}
         \caption{MSE of AOA estimation}
         \label{subfig:thetamse}
     \end{subfigure}
     \begin{subfigure}[b]{0.42\textwidth}
         \centering
         \includegraphics[width=\textwidth]{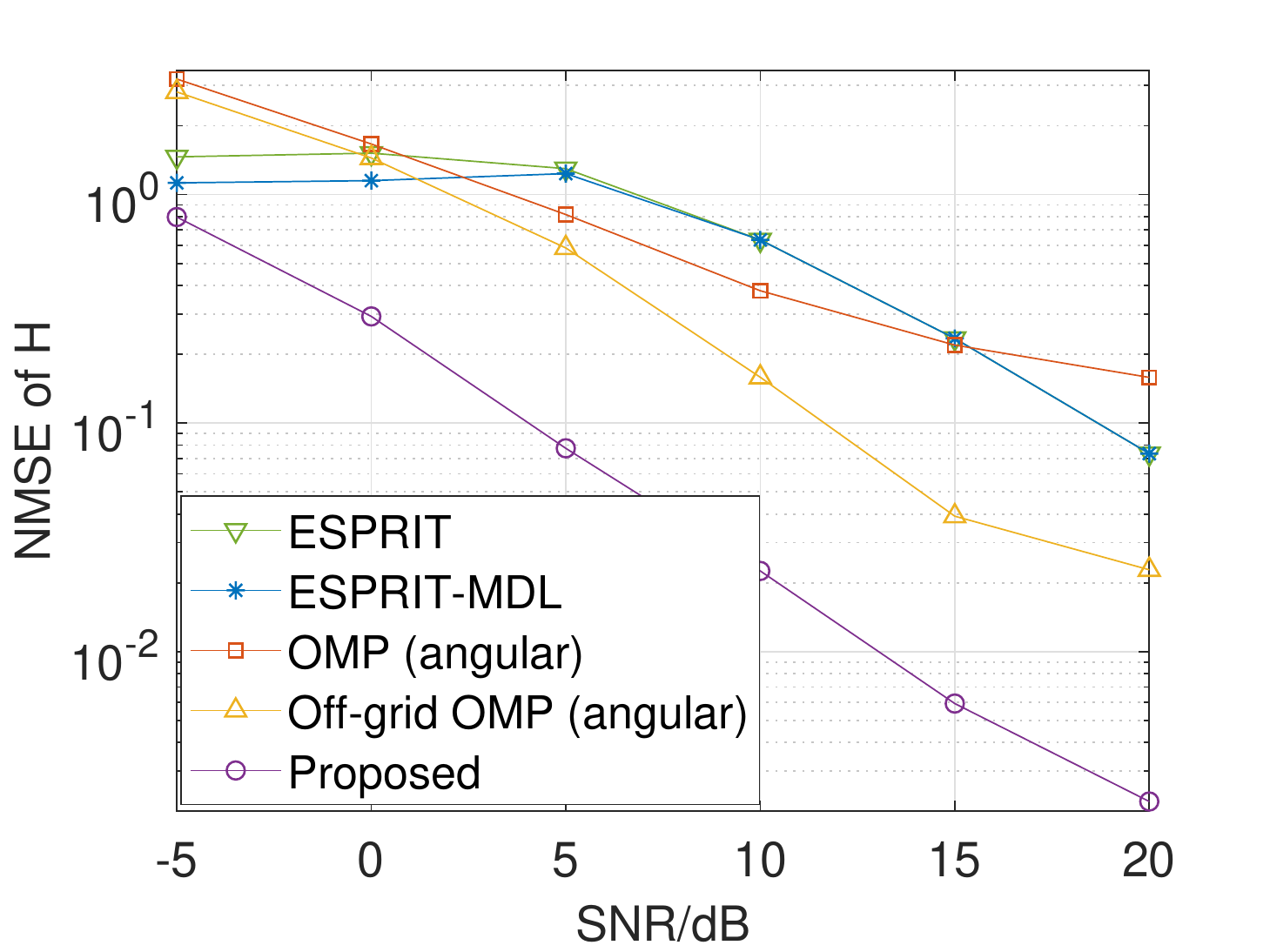}
         \caption{NMSE of channel estimation}
         \label{subfig:hNMSE_snr}
     \end{subfigure}
     \begin{subfigure}[b]{0.42\textwidth}
         \centering
         \includegraphics[width=\textwidth]{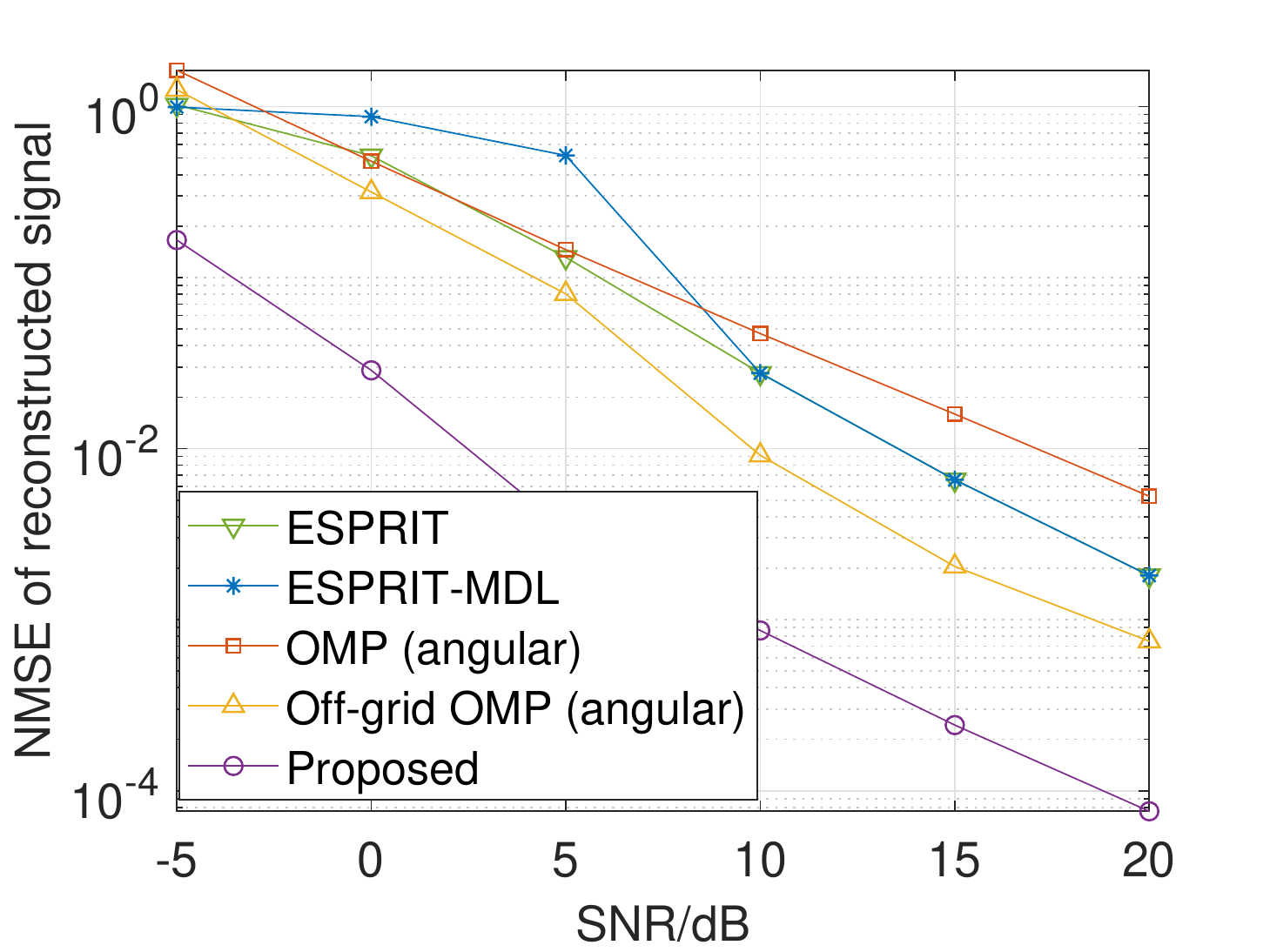}
         \caption{NMSE of signal estimation}
         \label{subfig:yNMSE_sne}
     \end{subfigure}
    \caption{Performance of channel estimation w.r.t. noise power.}
    \label{fig:anglemse}
\end{figure*}
\subsection{Estimation Results w.r.t. Different SNRs}
\label{subsec:EXP1}
We first test the accuracy of the AOD/AOA estimation of various algorithms. The number of paths is set as $L = 4$, and for clear visualization of the estimated AOD/AOA's under different levels of noise, the AOD/AOA's are fixed in this experiment as $(-32.14^\circ,60.66^\circ)$, $(-4.02^\circ,7.22^\circ)$, $(24.10^\circ,-39.17^\circ)$, $(66.29^\circ,26.91^\circ)$, leading to the normalized angle pairs $(-0.26,0.43)$, $(-0.03, 0.06)$, $(0.20, -0.31)$, $(0.45, 0.22)$. The absolute values of the path loss are also fixed as $0.87$, $0.58$, $0.32$, $0.7$. Under this setting, noise are generated for SNR from $-5$dB to $20$dB, and for each noise level, each compared algorithm is tested for $100$ Monte-Carlo trials. The mean square error (MSE) of AOD/AOA and the normalized mean square error (NMSE) of the estimated channel and recovered signal are recorded. 

The performance of the compared algorithms is presented in Fig. \ref{fig:anglemse}. Under different noise levels, the proposed algorithm obviously outperforms the competitors, in terms of the accuracy of AOD/AOA estimation, channel estimation, and signal recovery. Specifically, since ESPRIT-MDL does not have prior knowledge of the number of paths, its AOD/AOA estimation degrades as the SNR decreases. For Off-grid OMP (angular), while its performance is improved by updating the grid, it is still worse than that of the proposed method.

\begin{figure*}[!t]
    \centering
    \begin{subfigure}[b]{0.19\textwidth}
         \centering
         \includegraphics[width=\textwidth]{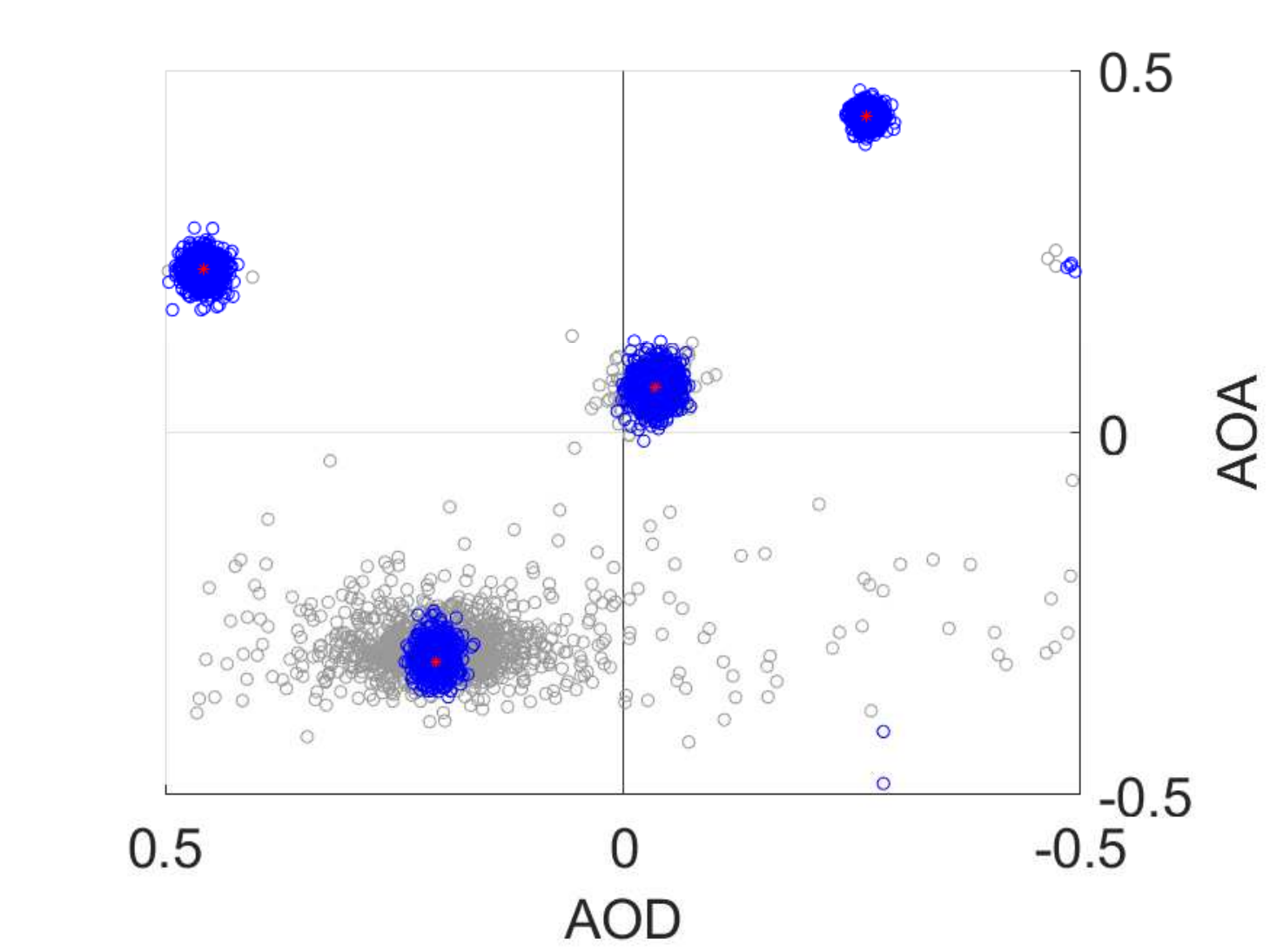}
     \end{subfigure}
     \begin{subfigure}[b]{0.19\textwidth}
         \centering
         \includegraphics[width=\textwidth]{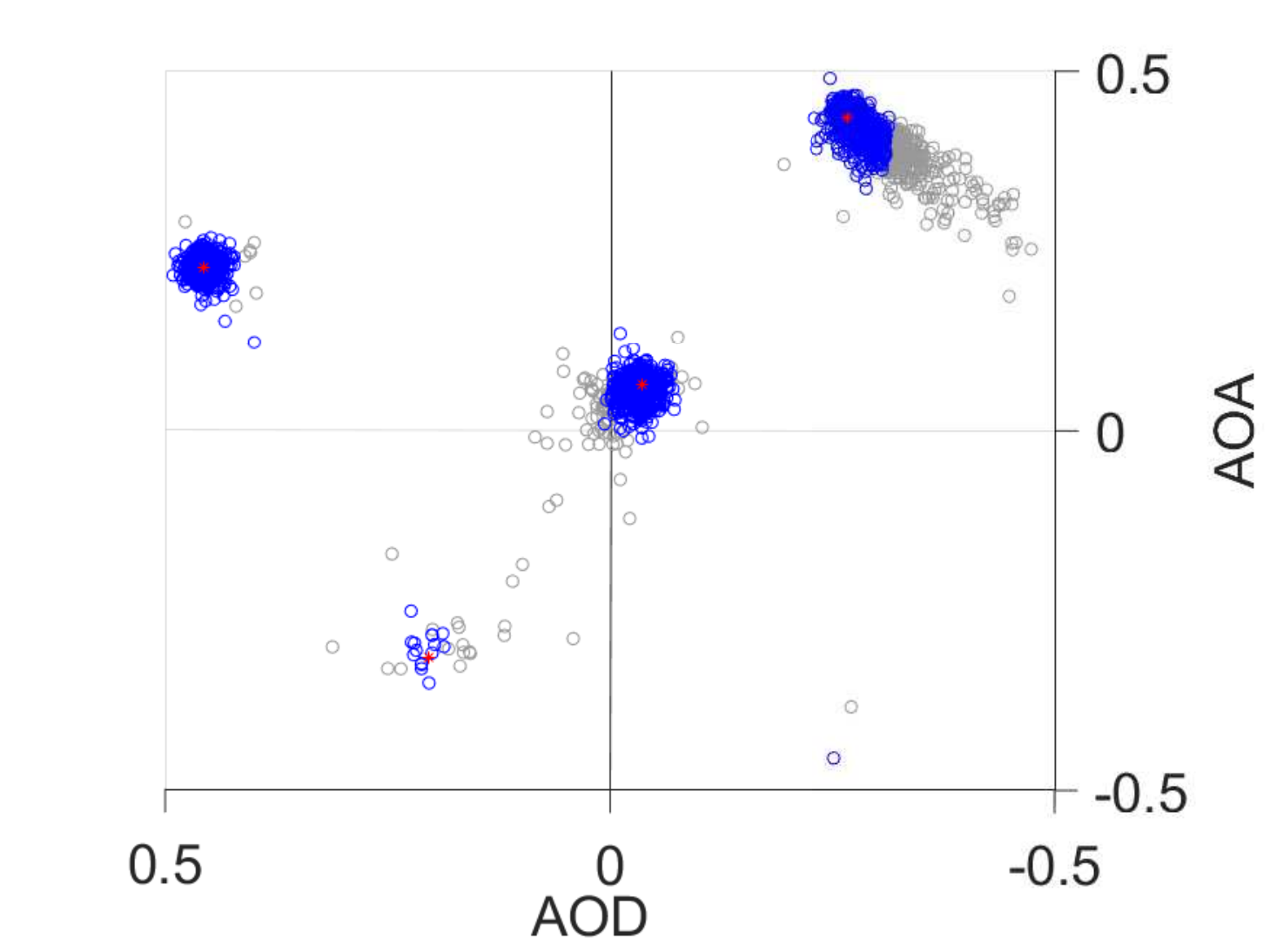}
     \end{subfigure}
     \begin{subfigure}[b]{0.19\textwidth}
         \centering
         \includegraphics[width=\textwidth]{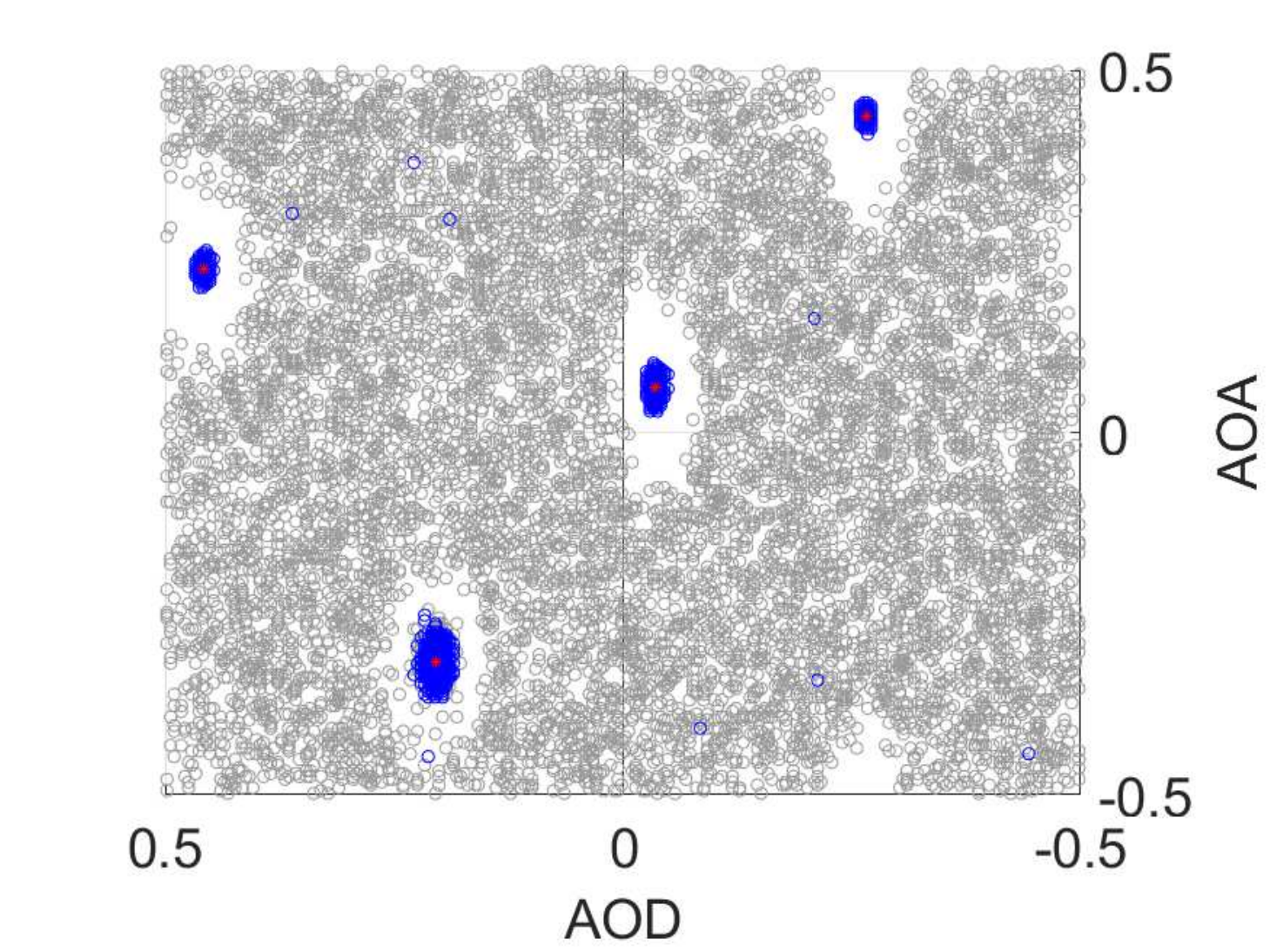}
     \end{subfigure}
     \begin{subfigure}[b]{0.19\textwidth}
         \centering
         \includegraphics[width=\textwidth]{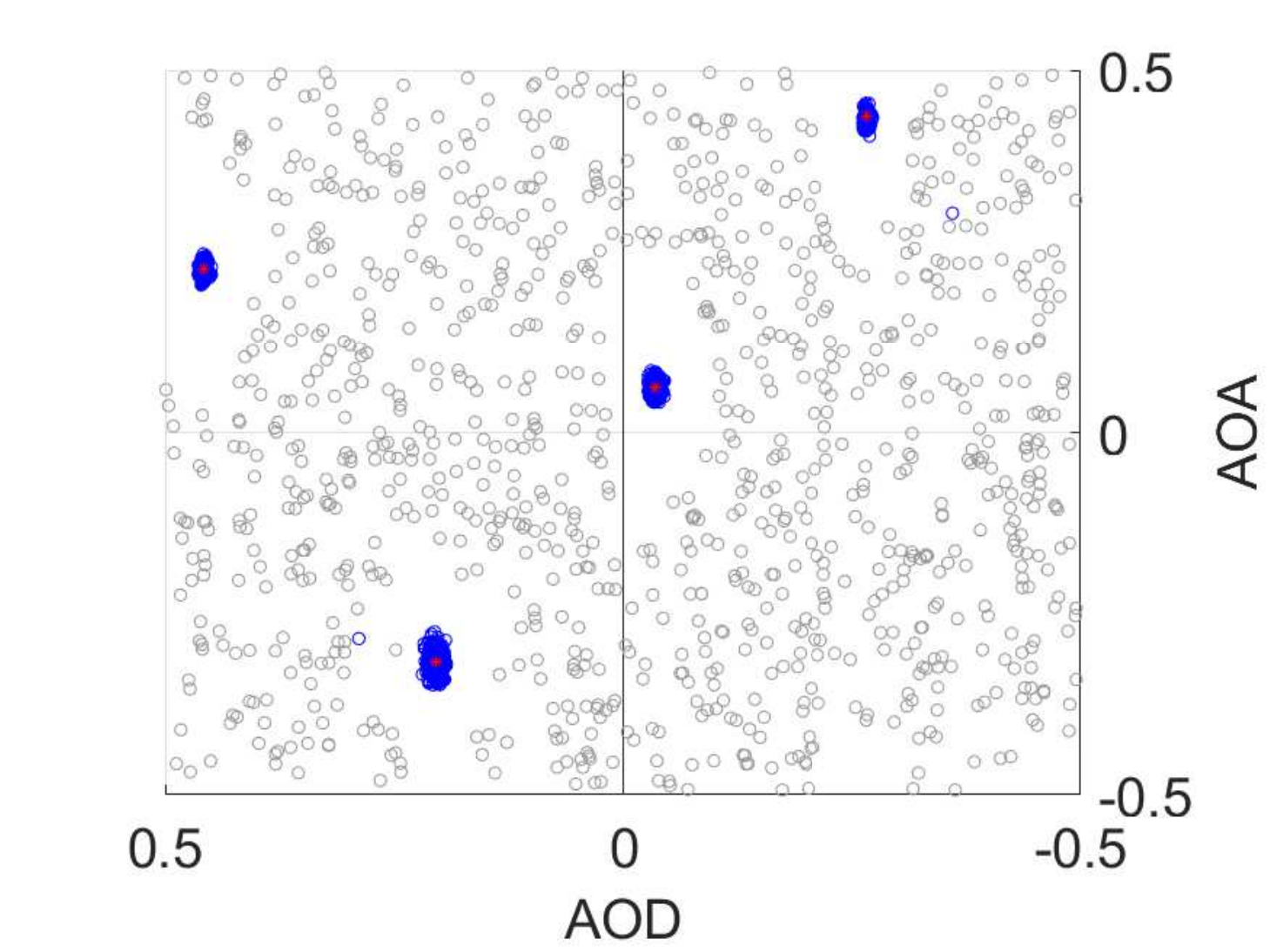}
     \end{subfigure}
     \begin{subfigure}[b]{0.19\textwidth}
         \centering
         \includegraphics[width=\textwidth]{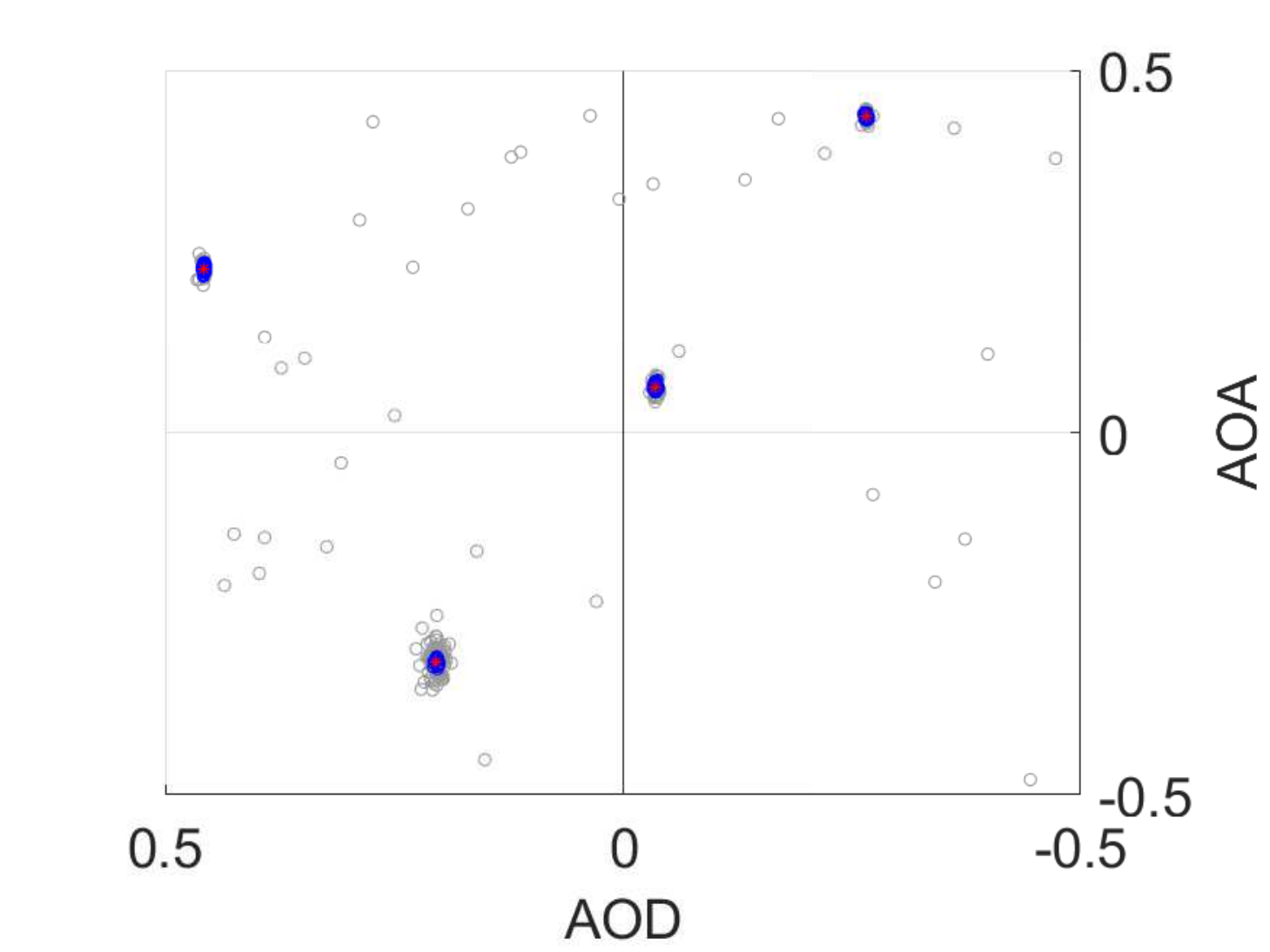}
     \end{subfigure}
     
    \begin{subfigure}[b]{0.19\textwidth}
         \centering
         \includegraphics[width=\textwidth]{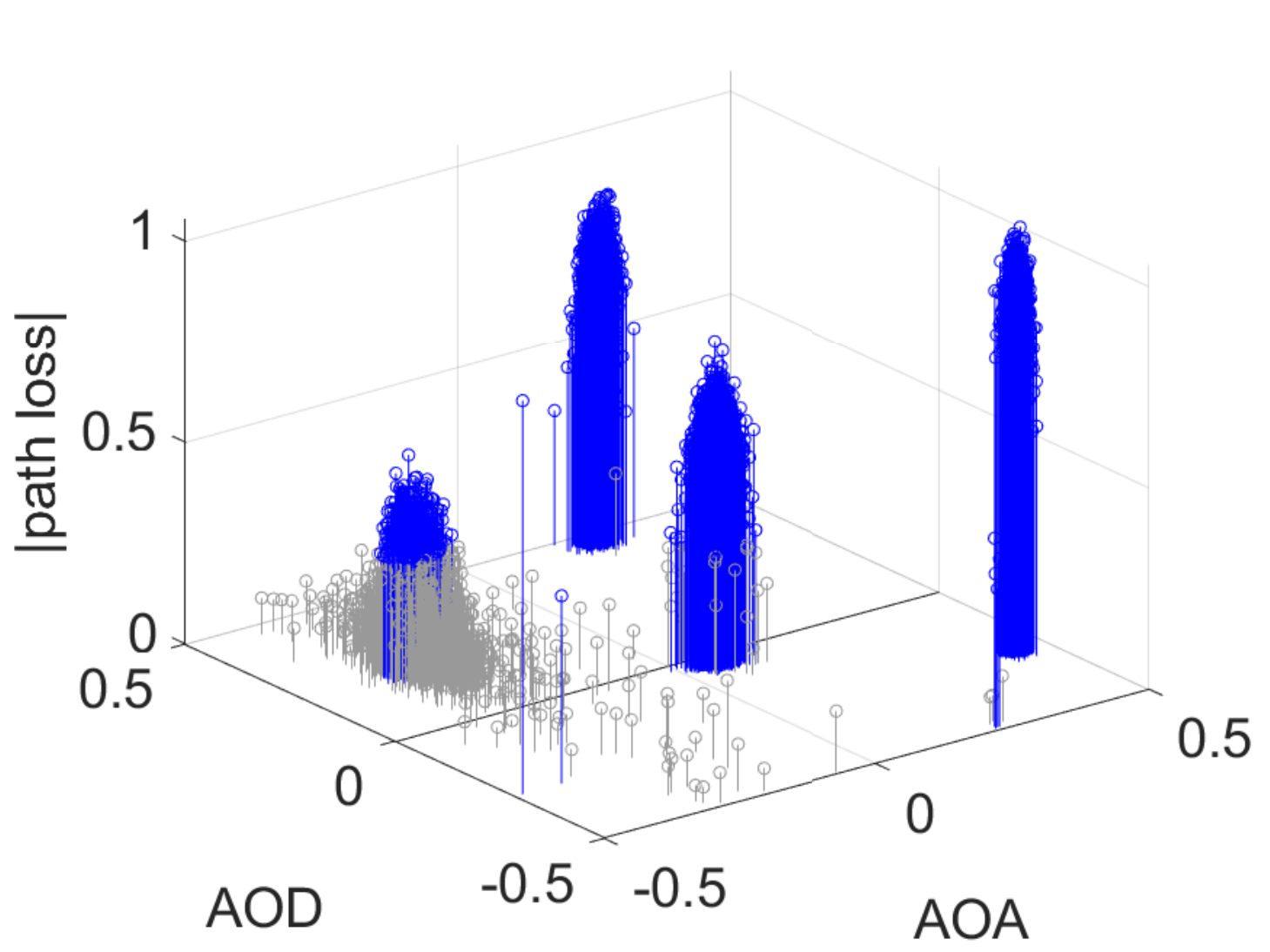}
         \caption{ESPRIT with known number of path}
         \label{subfig:angleESPRIT3D}
     \end{subfigure}
     \begin{subfigure}[b]{0.19\textwidth}
         \centering
         \includegraphics[width=\textwidth]{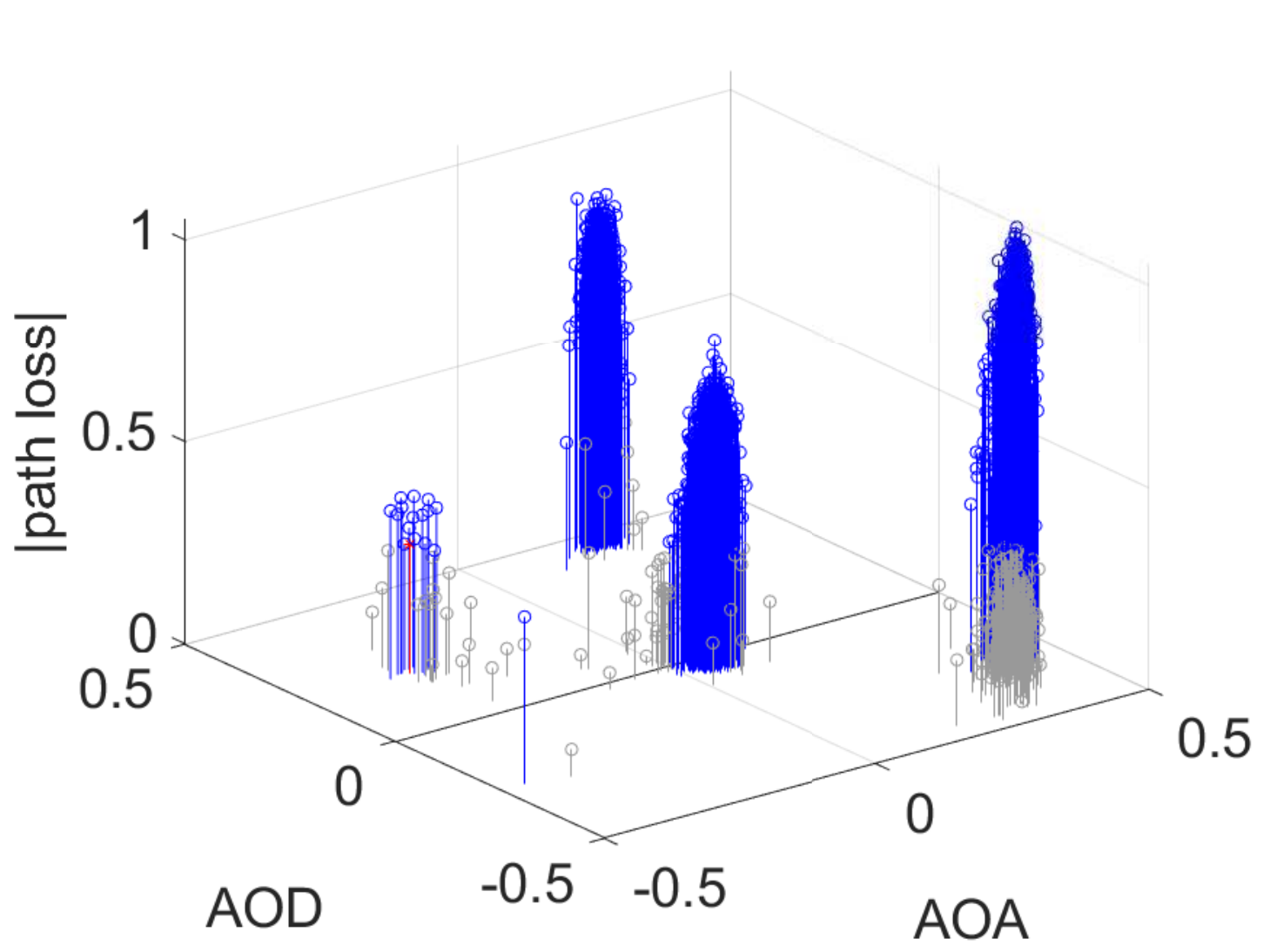}
         \caption{ESPRIT with path number estimated by MDL}
         \label{subfig:angleESPRITmdl3D}
     \end{subfigure}
     \begin{subfigure}[b]{0.19\textwidth}
         \centering
         \includegraphics[width=\textwidth]{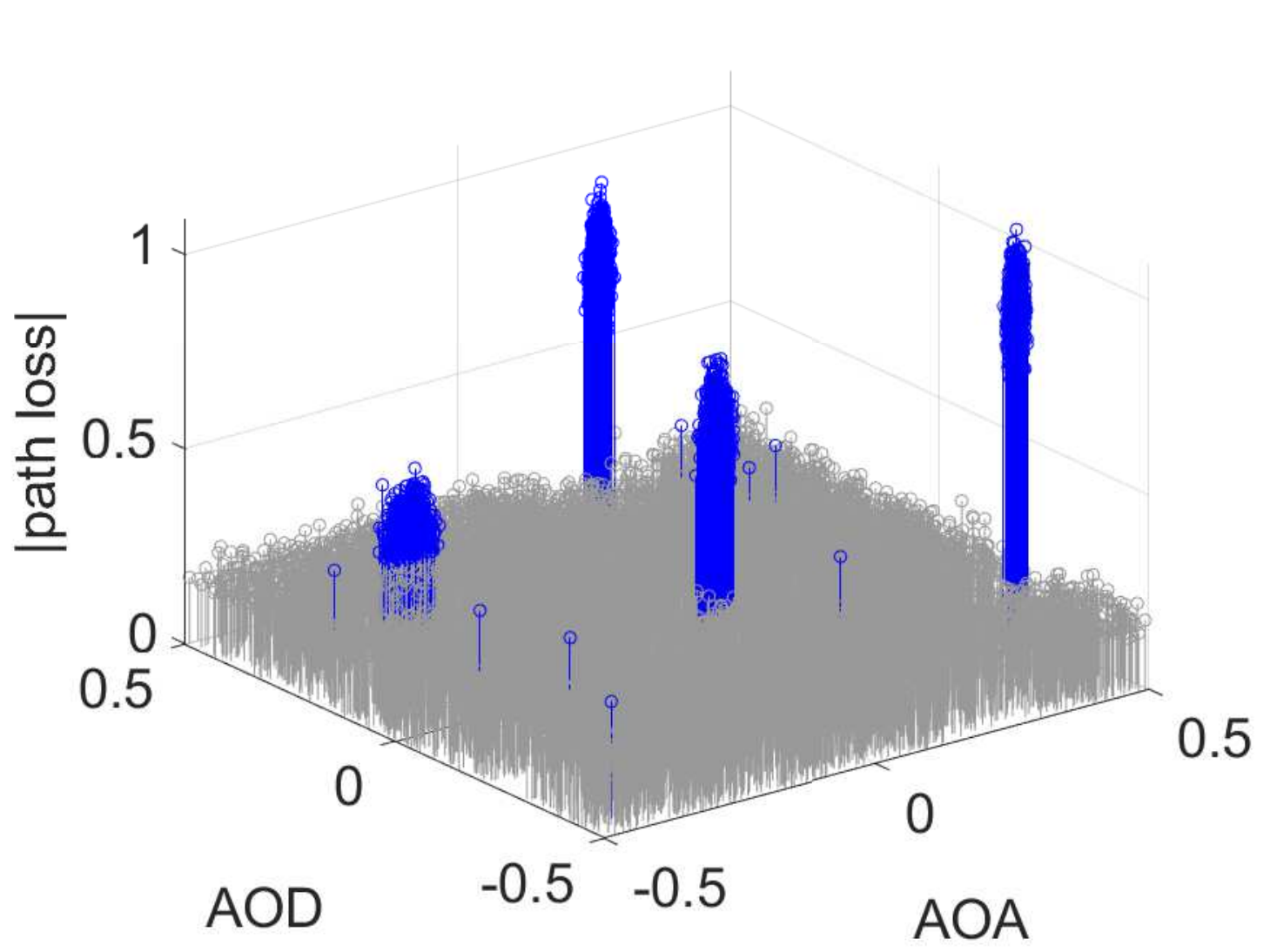}
         \caption{OMP (angular)\\\text{ } }
         \label{subfig:angleOMP3D}
     \end{subfigure}
     \begin{subfigure}[b]{0.19\textwidth}
         \centering
         \includegraphics[width=\textwidth]{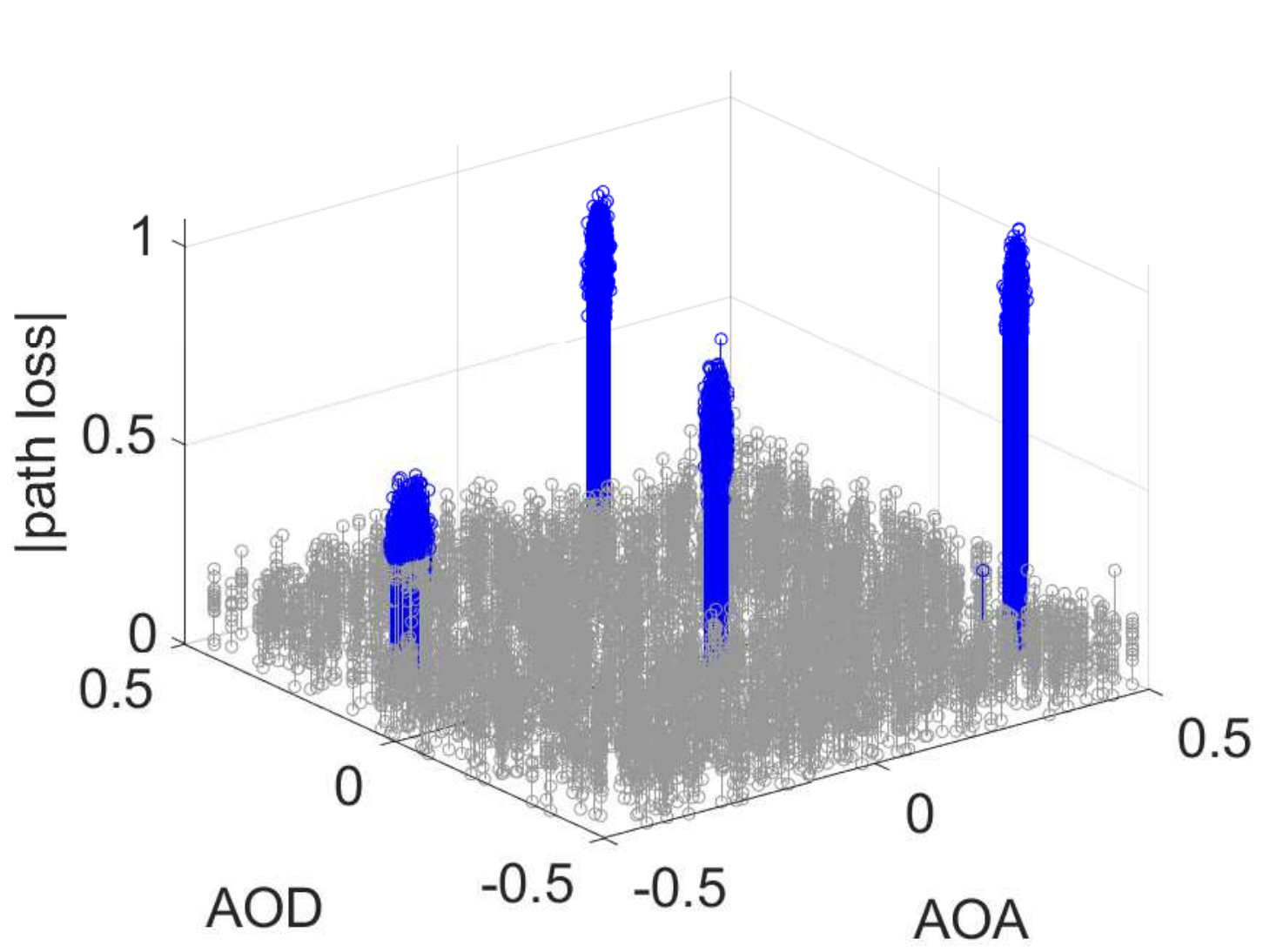}
         \caption{Off-grid OMP\\ (angular)}
         \label{subfig:angleoffgridOMP3D}
     \end{subfigure}
     \begin{subfigure}[b]{0.19\textwidth}
         \centering
         \includegraphics[width=\textwidth]{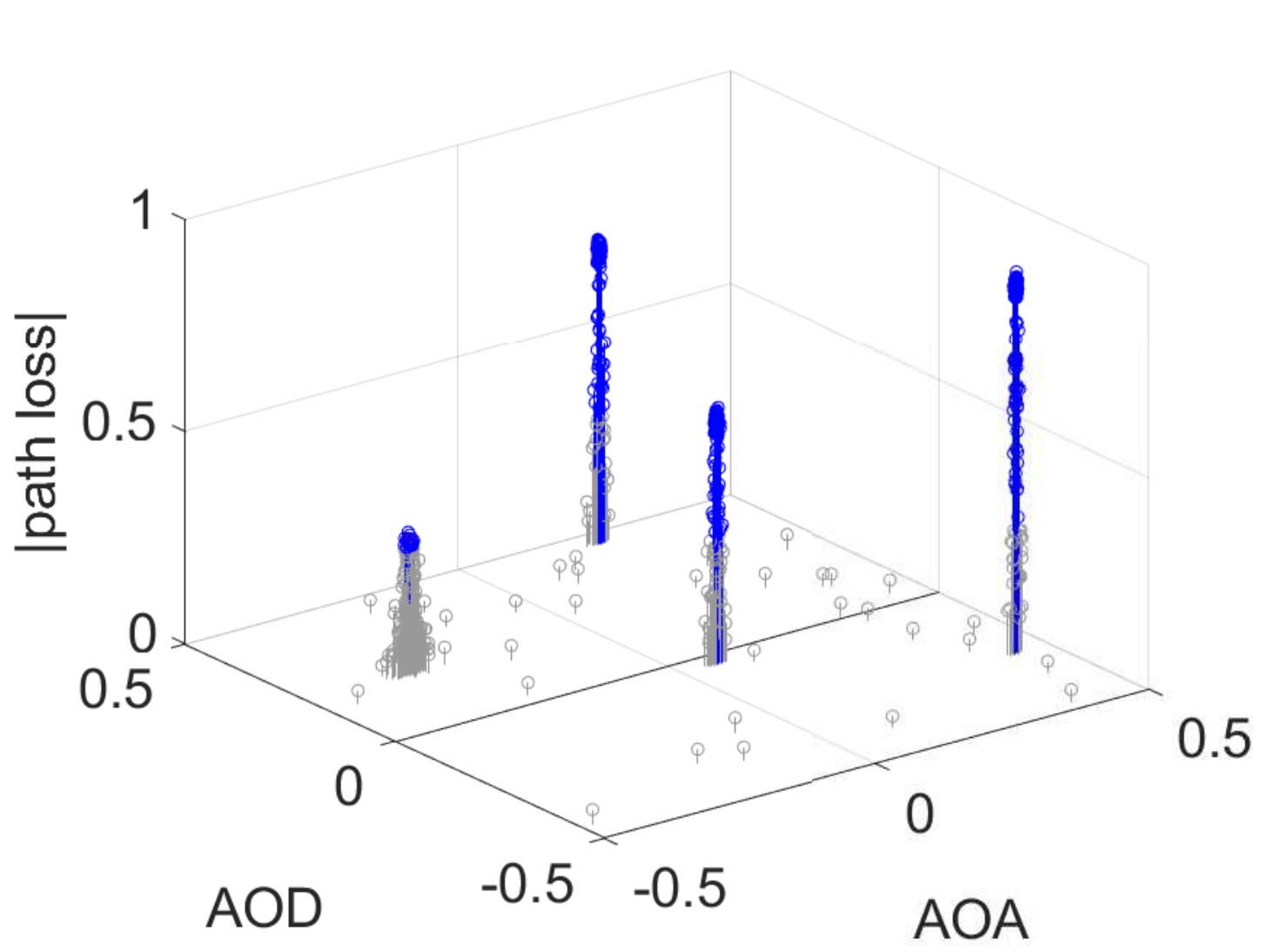}
         \caption{The proposed\\\text{ }}
         \label{subfig:angleEM3D}
     \end{subfigure}
    \caption{All estimated AODs and AOAs in 100 trials under 5dB SNR, first row: 2D illustration, second row: 3D illustration, with z-axis denoting the absolute value of the corresponding path loss.}
    \label{fig:angleplot}
\end{figure*}

To visualize the good results of the proposed algorithm, Fig. \ref{fig:angleplot} shows the AOD/AOA/path loss estimation of the compared algorithms for all $100$ trials under SNR$ = 5$dB. For ESPRIT-based and OMP-based methods, as they cannot adopt the common sparsity among different sub-bands, for each trial they would give $K$ estimations of the parameters, and all the estimations are shown. For better visualization, path losses with an absolute value smaller than $0.3$ are painted in gray, while those larger than $0.3$ are plotted in blue. As can be seen, the AOD/AOA estimation from the proposed algorithm is the most accurate, as all the $100$ sets of estimated angle pairs tightly concentrate around the ground truth. The drawback of the ESPRIT-MDL is obviously shown in Fig. \ref{subfig:angleESPRITmdl3D}, in which the success rate of detecting the angle pair $(0.20, -0.31)$ is very low as the blue points around this angle pair are few. On the other hand, as shown in Fig. \ref{subfig:angleOMP3D} and \ref{subfig:angleoffgridOMP3D}, both OMP and Off-grid OMP mistakenly detect more paths (shown as gray bars spread all around the AOD-AOA map). The reason is that MDL underestimates the number of paths, while OMP overfits the noise by using more path numbers.

\subsection{Estimation Results under Different System Settings}
\label{subsec:EXP2}
In this subsection, we test the performance of the compared methods under different system settings. Especially, the number of training subcarriers $K$, the number of time frames $P$, the number of transmitting symbols $N_\text{s}$ in each sub-channel in a frame, the carrier frequency $f_{\text{c}}$ and the bandwidth $f_s$ are varied.

\begin{figure*}[!h]
    \centering
    \begin{subfigure}[b]{0.42\textwidth}
         \centering
         \includegraphics[width=\textwidth]{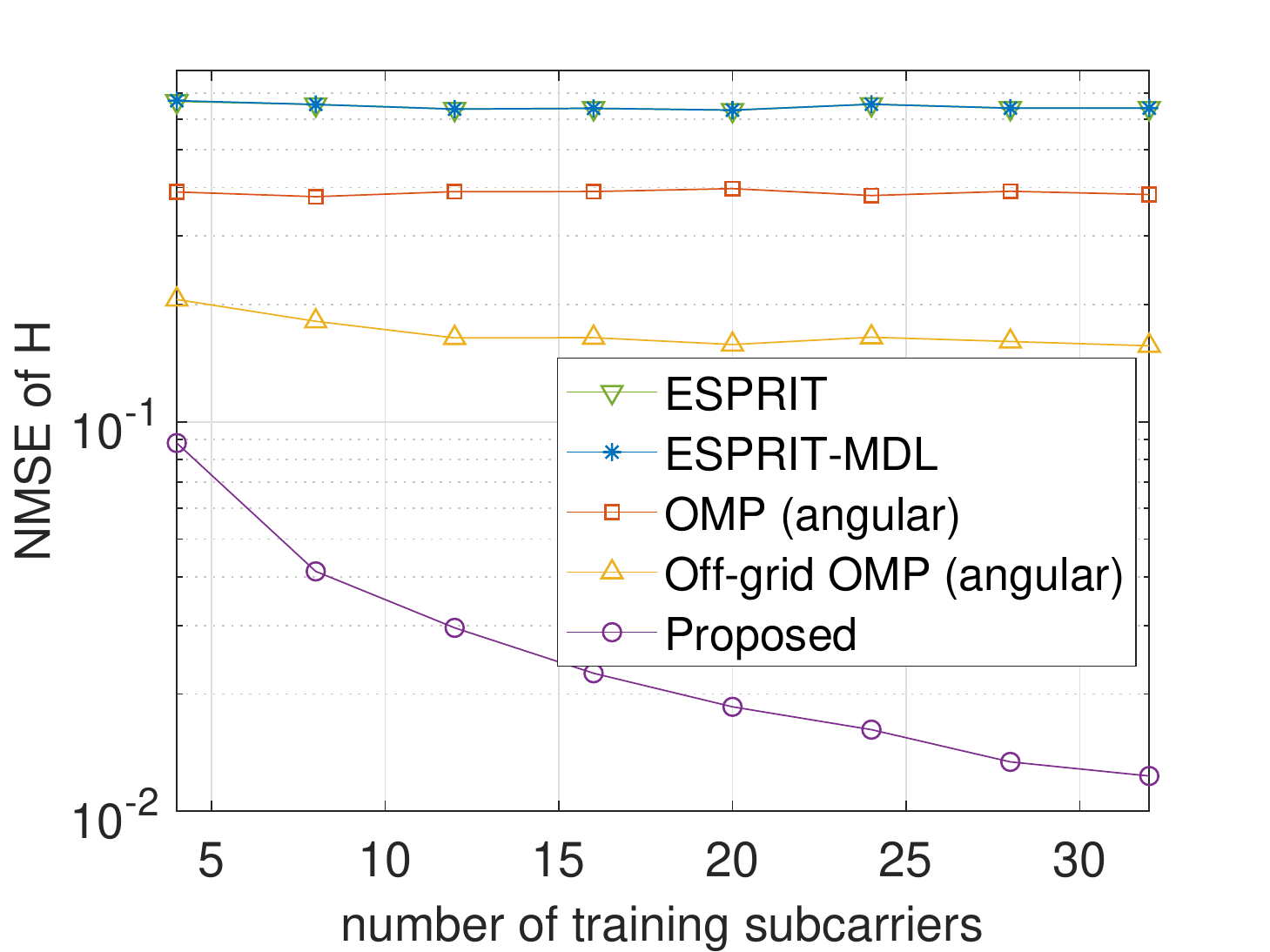}
         \caption{w.r.t. $K$}
         \label{subfig:H2K}
     \end{subfigure}
     \begin{subfigure}[b]{0.42\textwidth}
         \centering
         \includegraphics[width=\textwidth]{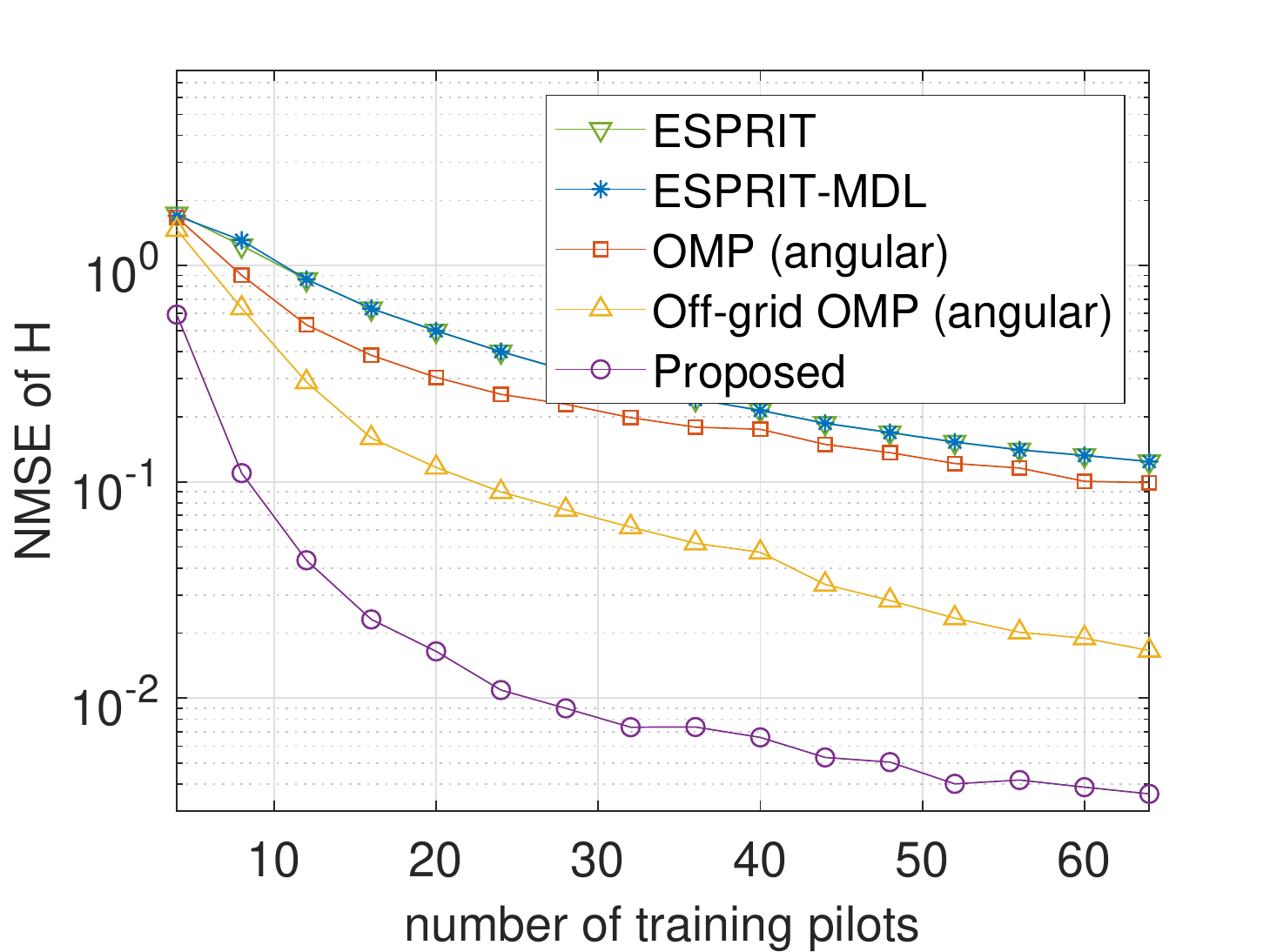}
         \caption{w.r.t. $P$}
         \label{subfig:H2P}
     \end{subfigure}
     \begin{subfigure}[b]{0.42\textwidth}
         \centering
         \includegraphics[width=\textwidth]{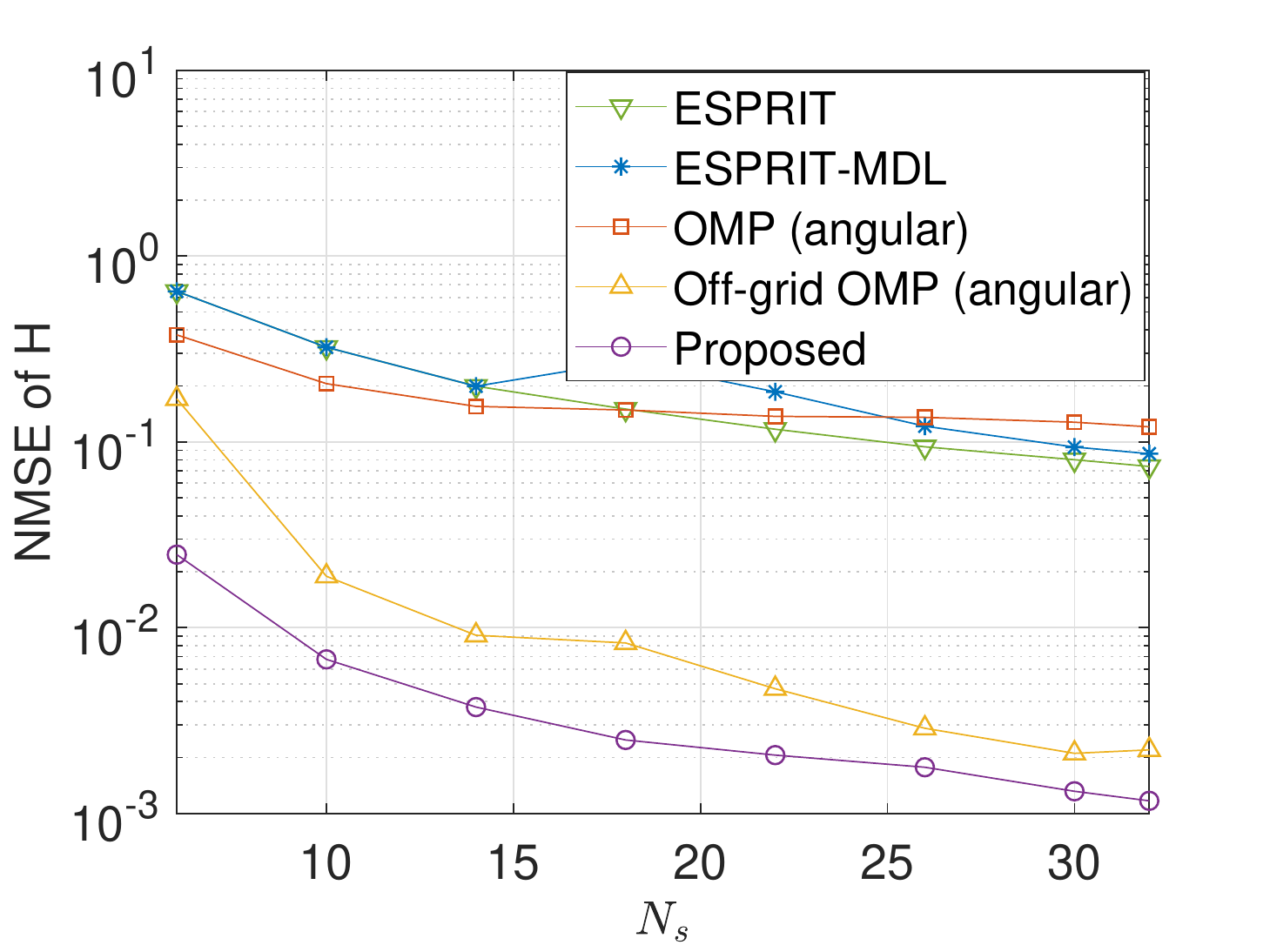}
         \caption{w.r.t. $N_\text{s}$}
         \label{subfig:H2Q}
     \end{subfigure}
     \begin{subfigure}[b]{0.42\textwidth}
         \centering
         \includegraphics[width=\textwidth]{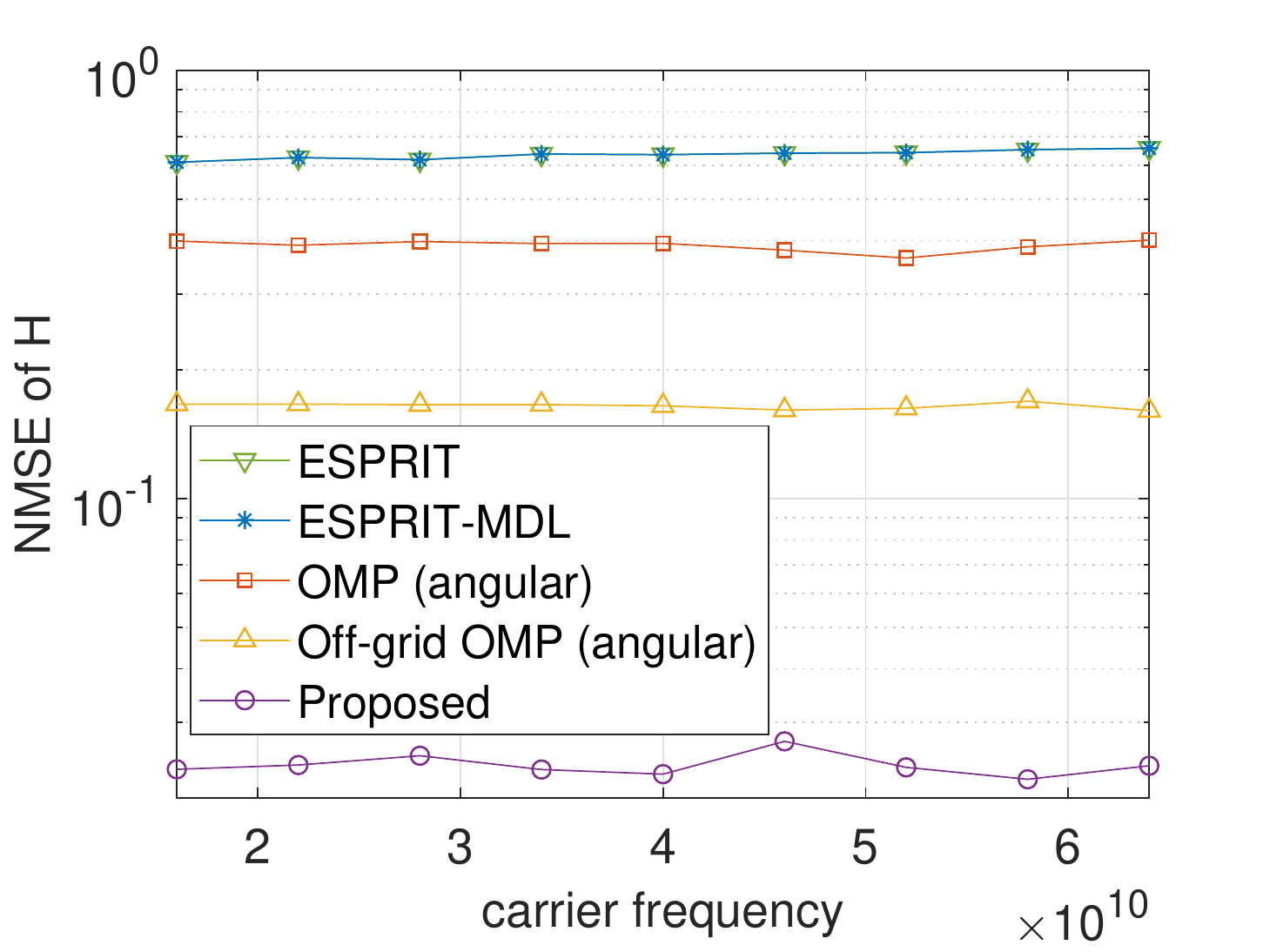}
         \caption{w.r.t. $f_{\text{c}}$}
         \label{subfig:H2fc}
     \end{subfigure}
     \begin{subfigure}[b]{0.42\textwidth}
         \centering
         \includegraphics[width=\textwidth]{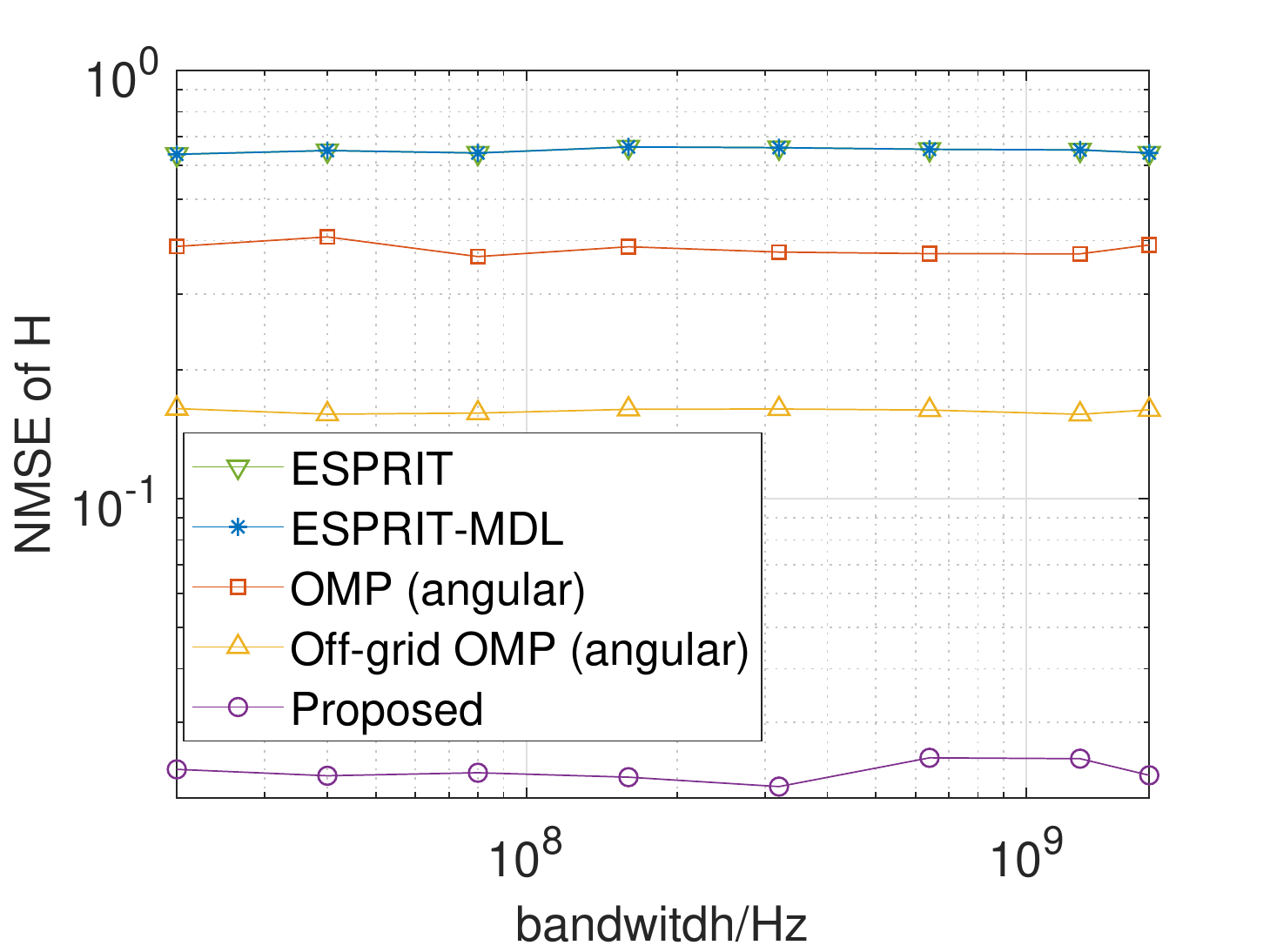}
         \caption{w.r.t. $f_s$}
         \label{subfig:H2fs}
     \end{subfigure}
    \caption{Performance of channel estimation w.r.t. different system configuration under 10dB SNR.}
    \label{fig:H2system}
\end{figure*}

As can be seen from Fig. \ref{fig:H2system}, the proposed algorithm achieves the best performance under all of the settings. Especially, when the number of training subcarriers increases, ESPRIT-based and OMP-based  methods do not perform better, while the accuracy of the proposed method improves. That is because the compared algorithms can only estimate the channel for each OFDM sub-band and it clearly shows that simply averaging the results from different sub-bands does not work. In contrast, the proposed method combines information from all bands and estimates the channel as a whole, thus achieving superior performance.

\begin{figure*}[!t]
    \centering
    \begin{subfigure}[b]{0.42\textwidth}
         \centering
         \includegraphics[width=\textwidth]{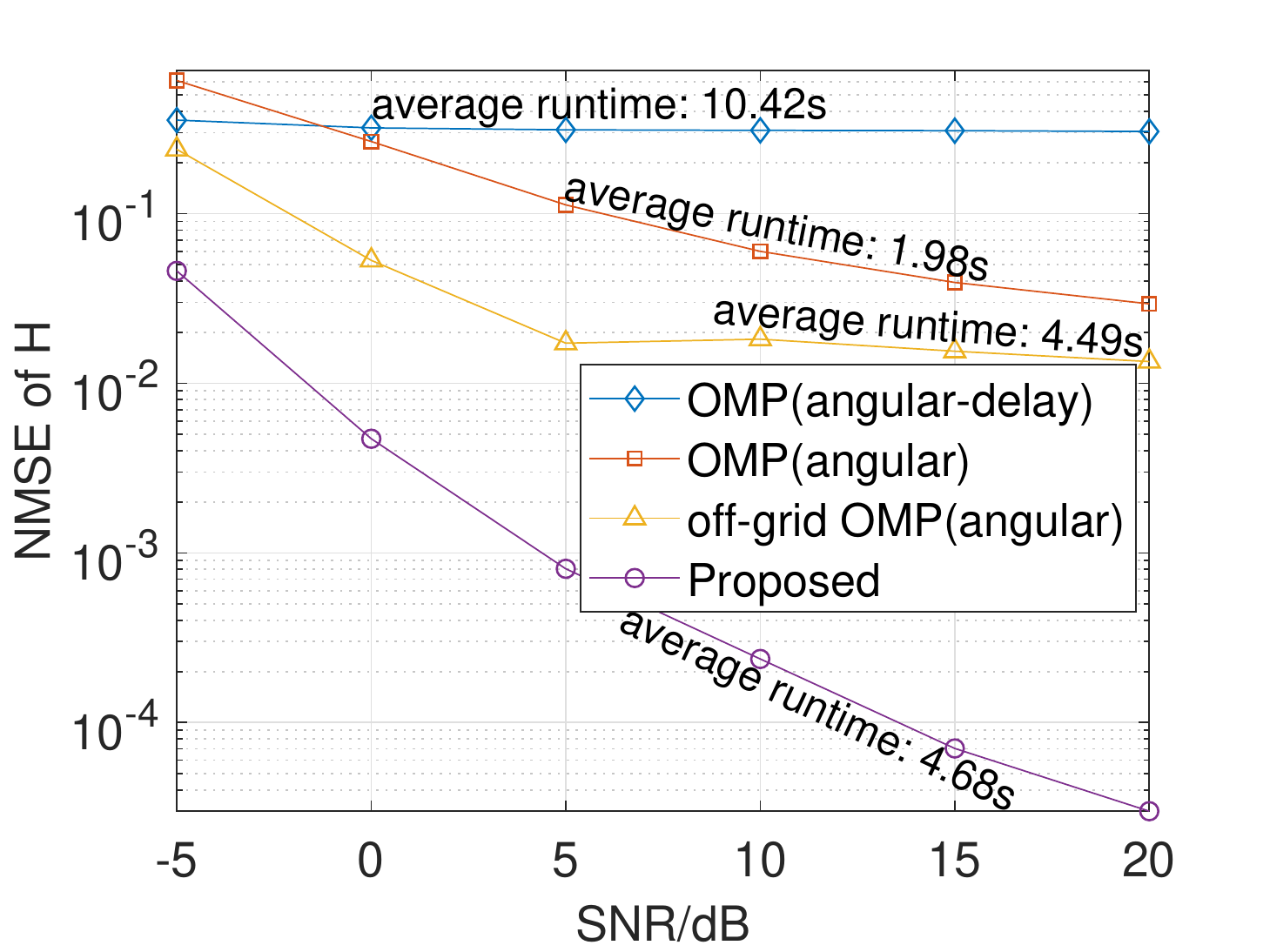}
         \caption{w.r.t. SNR}
         \label{subfig:H2snr_WF}
     \end{subfigure}
    \begin{subfigure}[b]{0.42\textwidth}
         \centering
         \includegraphics[width=\textwidth]{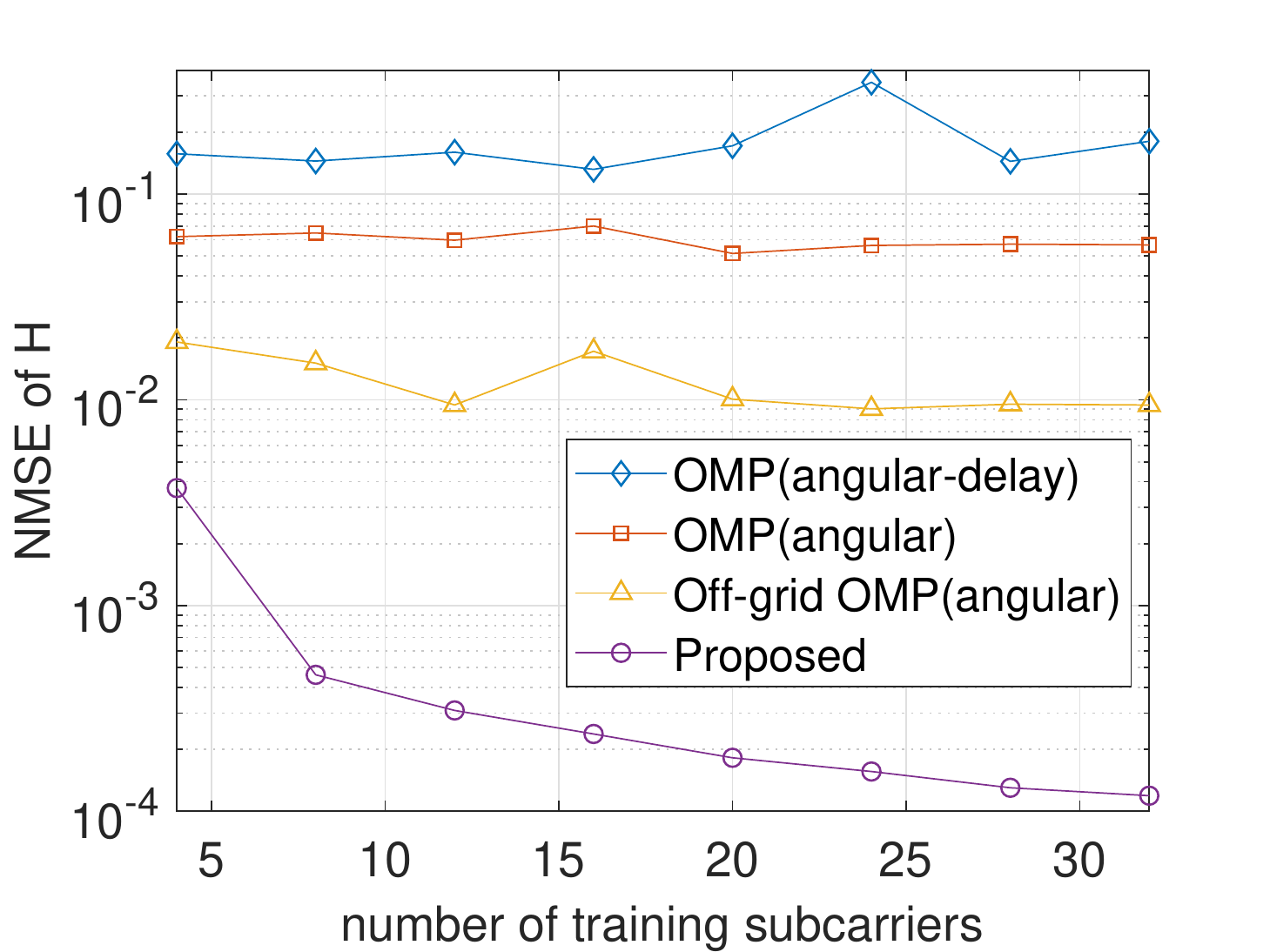}
         \caption{w.r.t. $K$}
         \label{subfig:H2K_WF}
     \end{subfigure}
     \begin{subfigure}[b]{0.42\textwidth}
         \centering
         \includegraphics[width=\textwidth]{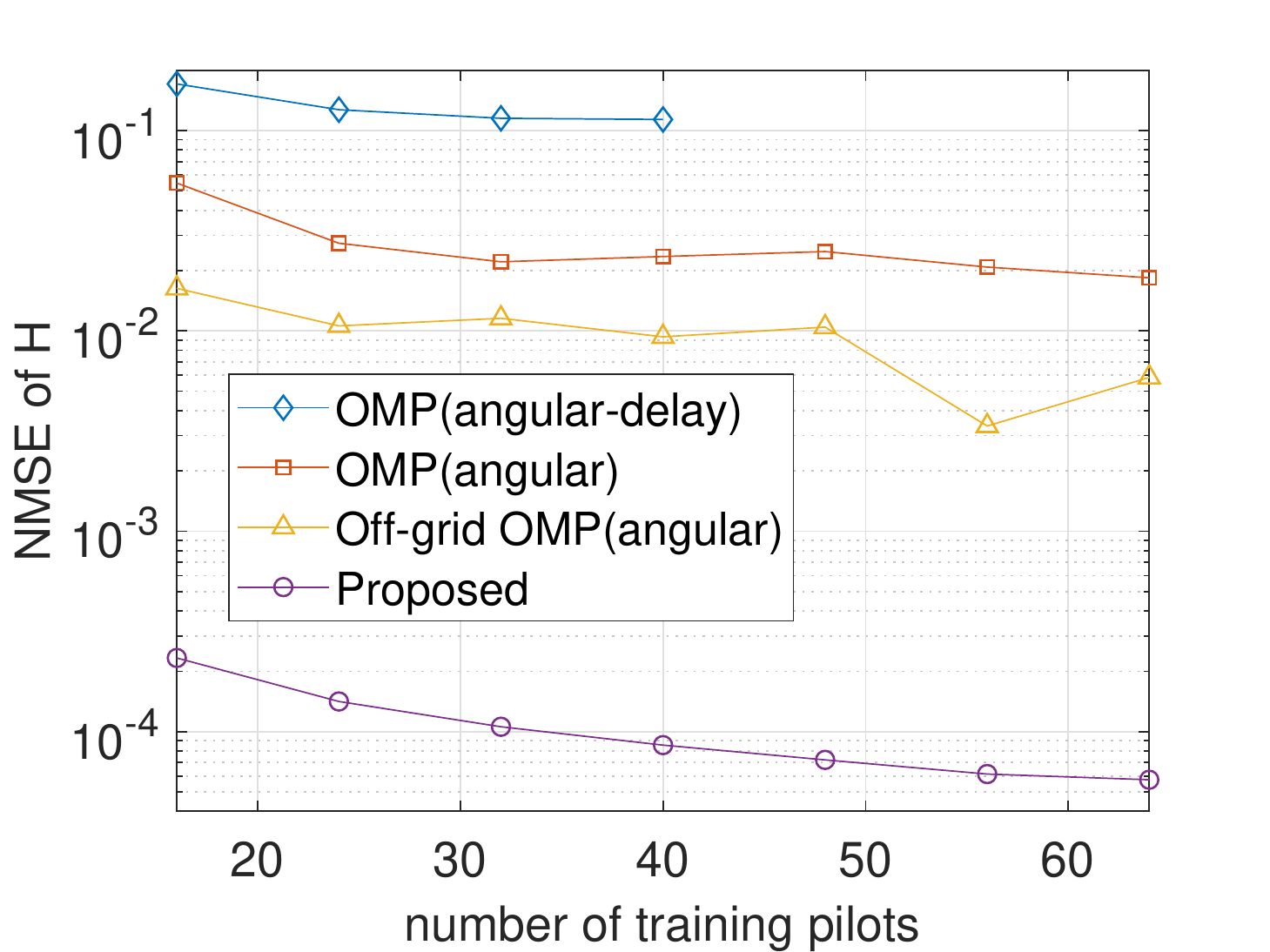}
         \caption{w.r.t. $P$}
         \label{subfig:H2P_WF}
     \end{subfigure}
     \begin{subfigure}[b]{0.42\textwidth}
         \centering
         \includegraphics[width=\textwidth]{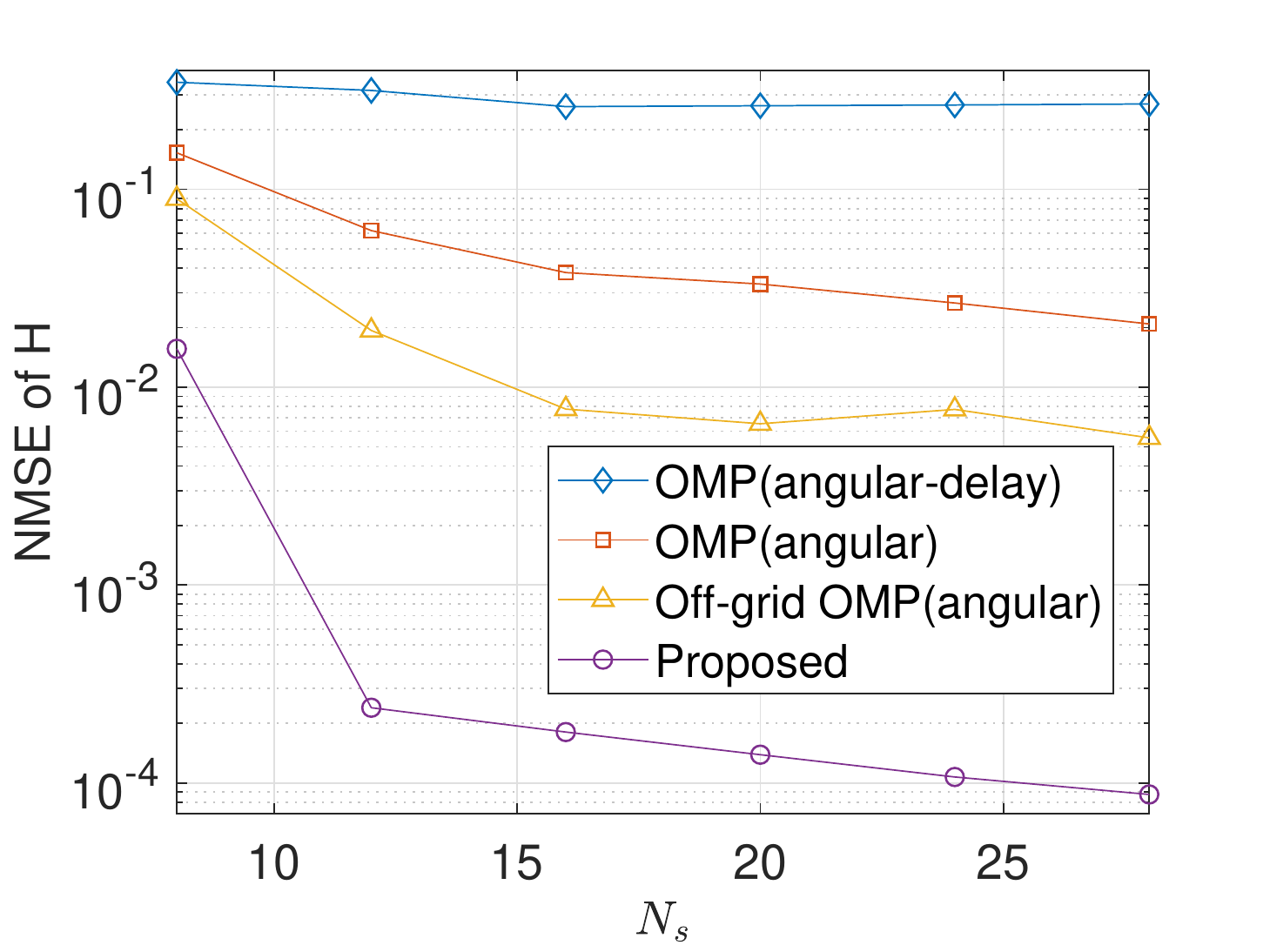}
         \caption{w.r.t. $N_\text{s}$}
         \label{subfig:H2Q_WF}
     \end{subfigure}
    \caption{Performance of channel estimation w.r.t. different system configuration under SNR$=10$dB, with random $\bm{W}$ and $\bm{F}$.}
    \label{fig:H2system_WF}
\end{figure*}

\subsection{Random Precoding and Combining Matrices}
\label{subsec:EXP3}
In this subsection, the case where $\bm{W}$ and $\bm{F}$ cannot take the form of an identity matrix is considered. Consequently, ESPRIT-based methods cannot be adopted anymore. 
On the other hand, the methods exploiting sparsity pattern in the angular-delay domain \cite{jian2019angle,wang2019block,wang2018spatial} is considered for comparison. 
In this experiment, the default number of RF chains at the receiver $N_\text{s}$ is set as $12$ instead of $6$, as the random $\bm{W}$ and $\bm{F}$ make it harder to find the correct basis from the dictionaries. Different noise levels, the number of training sub-carriers $K$, the number of training frames $P$, and the number of symbols of a frame on one sub-carrier $N_\text{s}$ are considered.

The performance of the compared methods is shown in Fig. \ref{fig:H2system_WF}.
It can be seen that under all circumstances the proposed method outperforms OMP in the angular domain (both on-grid and off-grid) and OMP (angular-delay) significantly. 
Specifically, as can be seen from Fig. \ref{subfig:H2snr_WF}-\ref{subfig:H2Q_WF}, the accuracy of the estimated channel from OMP (angular-delay) does not get higher under higher SNR, or with more training sub-carriers, training pilot frames or longer length of training symbols. This may be because the dictionary is with fixed grids, and even with more training resources, the grid mismatch restricts the performance. In addition, as recorded in Fig. \ref{subfig:H2snr_WF}, the average runtime of OMP (angular-delay) is much longer than the other three methods. 
Even though off-grid methods could be adopted to improve the performance of OMP (angular-delay), the runtime is still unacceptable. Furthermore, as shown in Fig. \ref{subfig:H2P_WF}, when $P$ is equal to or larger than $48$, the performance of OMP (angular-delay) cannot be successfully recorded since it exceeds the RAM limitation ($32$GB).

In addition, it can be seen from Fig. \ref{subfig:H2K_WF} that with more training sub-carriers, the performance of the proposed method becomes better, while that of OMP (angular) does not. This is because OMP (angular) estimates the channel in different sub-bands separately. As shown in Fig. \ref{subfig:H2K_WF} and \ref{subfig:H2Q_WF}, when $K = 4$ or $N_{\text{s}}=8$, the performance of the proposed method is not particularly impressive, but still outperforms other schemes even with larger $K$ or $N_s$. However, with $K$ equal to $8$ or above, or $N_s$ equal to $12$ or above, the performance of the proposed method is unmistakably better than the competing algorithms.

\section{Conclusion}
\label{sec:conclusion}
In this paper, to estimate the dual-wideband channel in general system settings (with precoding and combining matrices not restricted to identity matrices), the common sparsity among different subcarriers in the virtual channel model has been exploited, and then a probabilistic model has been built to induce the desired form of sparsity. To overcome the grid mismatch in the virtual channel model, grid offset parameters have been introduced by first-order Taylor approximation of the sensing matrix. A variational EM algorithm has been derived to estimate the variables in the probabilistic model as well as the grid offsets. Simulation results have shown that the proposed algorithm estimates the path number more accurately than MDL, and the AOD/AOA's more accurately than ESPRIT- and OMP-based methods. Furthermore, owning to its superior performance on channel parameter estimation, the proposed method typically provides $10$ to $100$ times smaller NMSE compared with ESPRIT- and OMP-based methods, under different system configuration.















\bibliographystyle{IEEEtran}
\bibliography{IEEEtran}

\appendices

\clearpage
\newpage

\section{Derivation of (\ref{eqn:Xupdate}), (\ref{eqn:lambdaupdate}) and (\ref{eqn:xiupdate})}
\label{apdsec:viupdate}
According to (\ref{eqn:MFupdate}), the variational distribution $q(\bm{x}_k)$ is obtained by taking expectation on (\ref{eqn:yjointlikelihood}) except $\bm{x}_k$, which gives
\begin{align}
    & \ln q(\bm{x}_k) = \mathbb{E}_{\bm{\Omega}\setminus \{\bm{x}_k\}} \Big\llbracket - \xi \Big(\bm{y}_{\text{w},k}-\bm{D}_{k}(\bm{\delta}_{\phi},\bm{\delta}_{\theta})\bm{x}_k\Big)^{H}\nonumber \\
    & \quad \times \Big(\bm{y}_{\text{w},k}-\bm{D}_{k}(\bm{\delta}_{\phi},\bm{\delta}_{\theta})\bm{x}_k\Big) - \bm{x}_k^H \text{diag}(\bm{\lambda})\bm{x}_k \Big \rrbracket + \text{const} \nonumber \\
    & = - \bm{x}_k^H \underbrace{\Big( \mathbb{E}\llbracket \xi \rrbracket \bm{D}_{k}^H(\bm{\delta}_{\phi},\bm{\delta}_{\theta})\bm{D}_{k}(\bm{\delta}_{\phi},\bm{\delta}_{\theta}) + \text{diag}(\mathbb{E}\llbracket \bm\lambda \rrbracket) \Big)}_{\bm{\Sigma}_{\bm{x}_k}^{-1}} \bm{x}_k \nonumber \\
    & \quad + 2 \Re \Big(  \mathbb{E}\llbracket \xi \rrbracket  \bm{y}_{\text{w},k}^H \bm{D}_{k}(\bm{\delta}_{\phi},\bm{\delta}_{\theta})  \bm{x}_k\Big) + \text{const}.
    \label{eqn:apd_qx}
\end{align}
It is observed that (\ref{eqn:apd_qx}) is quadratic w.r.t. $\bm{x}_k$, and therefore $q(\bm{x}_k)$ follows a complex Gaussian distribution with its covariance given by $\bm{\Sigma}_{\bm{x}_k}$ and mean
\begin{align}
    \bm{m}_{\bm{x}_k} = \mathbb{E}\llbracket \xi \rrbracket \bm{\Sigma}_{\bm{x}_k}\bm{D}_{k}^H(\bm{\delta}_{\phi},\bm{\delta}_{\theta}) \bm{y}_{\text{w},k}.
    \label{eqn:apdXupdate}
\end{align}

For $\bm{\lambda}$, the variational distribution is derived by taking expectation on (\ref{eqn:yjointlikelihood}) except $\bm{\lambda}$:
\begin{align}
    &\ln q(\bm{\lambda}) = \sum_{r=1}^{N_\phi N_\theta} (\bm{\gamma}_r +K-1)\ln \bm{\lambda}_{r} \nonumber \\
    & \quad - \sum_{r=1}^{N_\phi N_\theta} \Big( \bm{\beta}_r + \sum_{k=1}^K \mathbb{E} \llbracket (\bm{x}_k)_r^*(\bm{x}_k)_r\rrbracket \Big) \bm{\lambda}_r + \text{const}.
    \label{eqnapx:ln_q_lambda}
\end{align}
Comparing (\ref{eqnapx:ln_q_lambda}) with the form of Gamma distribution, it is found that each $\bm{\lambda}_r$ follows a Gamma distribution with $\hat{\bm{\gamma}}_r = \bm{\gamma}_r + K$ and $\hat{\bm{\beta}}_r = \bm{\beta}_r + \mathbb{E} \llbracket (\bm{x}_k)_r^*(\bm{x}_k)_r\rrbracket$. Since $\mathbb{E} \llbracket (\bm{x}_k)_r^*(\bm{x}_k)_r\rrbracket = [\bm{\Sigma}_{\bm{x}_k} ]_{r,r} + [\bm{m}_{\bm{x}_k}]_r^*[\bm{m}_{\bm{x}_k}]_r$, we obtain (\ref{eqn:lambdaupdate}).

Similarly, taking expectation on (\ref{eqn:yjointlikelihood}) w.r.t. all variables except $\xi$, the variational distribution of $\xi$ is obtained as
\begin{align}
    \ln q(\xi) &= (KPN_\text{s} + \gamma_\xi -1)\ln \xi - \bigg( \beta_\xi + \sum_{k=1}^{K} \Big( \bm{y}_{\text{w},k}^H\bm{y}_{\text{w},k} \nonumber \\
    &  +  \mathbb{E}\big \llbracket \bm{x}_k^H  \bm{D}_{k}^H(\bm{\delta}_{\phi},\bm{\delta}_{\theta})\bm{D}_{k}(\bm{\delta}_{\phi},\bm{\delta}_{\theta}) \bm{x}_k\big\rrbracket \nonumber \\
    &- 2 \Re \big( \bm{y}_{\text{w},k}^H \bm{D}_{k}(\bm{\delta}_{\phi},\bm{\delta}_{\theta}) \mathbb{E}\big \llbracket\bm{x}_k \big \rrbracket \big) \Big) \bigg)  \xi + \text{const},
\end{align}
which indicates that $\xi$ follows a Gamma distribution $\text{Gamma}(\xi | \hat{\gamma}_\xi,\hat{\beta}_\xi)$ with
\begin{align}
    & \hat{\gamma}_\xi = KPN_\text{s} + \gamma_\xi, \nonumber \\
    & \hat{\beta}_\xi = \beta_\xi + \sum_{k=1}^{K} \Big( \bm{y}_{\text{w},k}^H\bm{y}_{\text{w},k} +  \mathbb{E}\big \llbracket \bm{x}_k^H  \bm{D}_{k}^H(\bm{\delta}_{\phi},\bm{\delta}_{\theta}) \nonumber \\
    & \times \bm{D}_{k}(\bm{\delta}_{\phi},\bm{\delta}_{\theta}) \bm{x}_k\big\rrbracket - 2 \Re \big( \bm{y}_{\text{w},k}^H \bm{D}_{k}(\bm{\delta}_{\phi},\bm{\delta}_{\theta}) \mathbb{E}\big \llbracket\bm{x}_k \big \rrbracket \big) \Big).
\end{align}
Since $\mathbb{E}\big \llbracket \bm{x}_k^H  \bm{D}_{k}^H(\bm{\delta}_{\phi},\bm{\delta}_{\theta})\bm{D}_{k}(\bm{\delta}_{\phi},\bm{\delta}_{\theta}) \bm{x}_k\big\rrbracket = \text{tr} \Big(\bm{D}_{k}^H(\bm{\delta}_{\phi},\bm{\delta}_{\theta}) $ $\times \bm{D}_{k}(\bm{\delta}_{\phi},\bm{\delta}_{\theta})  \big( \bm{\Sigma}_{\bm{x}_k} + \bm{m}_{\bm{x}_k}\bm{m}_{\bm{x}_k}^H \big)\Big)$, we obtain (\ref{eqn:xiupdate}).
\section{Derivation of (\ref{eqn:emtoquadratic})}
\label{apdsec:mstepupdate}
Taking out the terms that are only related with $\bm{\delta}_\phi$ and $\bm{\delta}_\theta$, (\ref{eqn:maxL}) becomes
\begin{align}
    &\min_{\bm{\delta}_\phi,\bm{\delta}_\theta} \mathbb{E}_{\{q(\bm{x}_k)\}_{k=1}^K}\Big\llbracket\sum_{k=1}^{K} \Big\|\bm{y}_{\text{w},k} - {\bm{D}_{k}(\bm{\delta}_\phi,\bm{\delta}_\theta)} \bm{x}_k\Big\|_2^2\Big \rrbracket \nonumber \\
    =& \min_{\bm{\delta}_\phi,\bm{\delta}_\theta} \sum_{k=1}^{K} \bigg(  \text{tr}\Big( {\bm{D}_{k}^H(\bm{\delta}_\phi,\bm{\delta}_\theta)} {\bm{D}_{k}(\bm{\delta}_\phi,\bm{\delta}_\theta)} \big(   \bm{\Sigma}_{\bm{x}_k} + \bm{m}_{\bm{x}_k}\bm{m}_{\bm{x}_k}^H\big) \Big) \nonumber \\
    & \quad - 2\Re \Big( \bm{y}_{\text{w},k}^H \bm{D}_{k}(\bm{\delta}_{\phi},\bm{\delta}_{\theta}) \bm{m}_{\bm{x}_k} \Big) \bigg).
    \label{eqn:apdmin_2_delta}
\end{align}
By applying (\ref{eqn:taylorapprox}) to the trace term of (\ref{eqn:apdmin_2_delta}),  it can be derived that
\begin{align}
    &\text{tr}\Big( {\bm{D}_{k}^H(\bm{\delta}_\phi,\bm{\delta}_\theta)} {\bm{D}_{k}(\bm{\delta}_\phi,\bm{\delta}_\theta)} \big(   \bm{\Sigma}_{\bm{x}_k} + \bm{m}_{\bm{x}_k}\bm{m}_{\bm{x}_k}^H\big) \Big) \nonumber \\
    =&\text{tr} \Bigg( \bigg(  \Big( \big(\frac{\partial \bm{D}_{\text{bs},k}(\tilde{\bm{\phi}})}{\partial \tilde{\bm{\phi}}} \text{diag}(\bm{\delta}_\phi) \big)\otimes \bm{D}_{\text{ms},k}(\tilde{\bm{\theta}})  \Big)^H 
    \Big(  \big(\frac{\partial \bm{D}_{\text{bs},k}(\tilde{\bm{\phi}})}{\partial \tilde{\bm{\phi}}}\nonumber \\
    & \times \text{diag}(\bm{\delta}_\phi) \big)\otimes \bm{D}_{\text{ms},k}(\tilde{\bm{\theta}}) \Big) + \Big(  \big(\frac{\partial \bm{D}_{\text{bs},k}(\tilde{\bm{\phi}})}{\partial \tilde{\bm{\phi}}} \text{diag}(\bm{\delta}_\phi) \big)\nonumber \\
    &\otimes \bm{D}_{\text{ms},k}(\tilde{\bm{\theta}}) \Big)^H \Big( \bm{D}_{\text{bs},k}(\tilde{\bm{\phi}}) 
    \otimes \big(\frac{\partial \bm{D}_{\text{ms},k}(\tilde{\bm{\theta}})}{\partial \tilde{\bm{\theta}}} \text{diag}(\bm{\delta}_\theta)\big)  \Big)  \nonumber \\
    &+  \Big( \bm{D}_{\text{bs},k}(\tilde{\bm{\phi}})\otimes \big(\frac{\partial \bm{D}_{\text{ms},k}(\tilde{\bm{\theta}})}{\partial \tilde{\bm{\theta}}} \text{diag}(\bm{\delta}_\theta)\big)  \Big)^H
    \Big( \big(\frac{\partial \bm{D}_{\text{bs},k}(\tilde{\bm{\phi}})}{\partial \tilde{\bm{\phi}}}\nonumber\\
    & \times \text{diag}(\bm{\delta}_\phi) \big)\otimes \bm{D}_{\text{ms},k}(\tilde{\bm{\theta}})  \Big)+ \Big(  \bm{D}_{\text{bs},k}(\tilde{\bm{\phi}})\otimes \big(\frac{\partial \bm{D}_{\text{ms},k}(\tilde{\bm{\theta}})}{\partial \tilde{\bm{\theta}}} \nonumber \\
    & \times \text{diag}(\bm{\delta}_\theta)\big) \Big)^H 
    \Big(  \bm{D}_{\text{bs},k}(\tilde{\bm{\phi}})\otimes \big(\frac{\partial \bm{D}_{\text{ms},k}(\tilde{\bm{\theta}})}{\partial \tilde{\bm{\theta}}} \text{diag}(\bm{\delta}_\theta)\big) \Big)  \nonumber\\
     &+ 2\Re \bigg\{\Big(  \bm{D}_{\text{bs},k}(\tilde{\bm{\phi}})\otimes \bm{D}_{\text{ms},k}(\tilde{\bm{\theta}}) \Big)^H \Big( \big(\frac{\partial \bm{D}_{\text{bs},k}(\tilde{\bm{\phi}})}{\partial \tilde{\bm{\phi}}} \text{diag}(\bm{\delta}_\phi) \big) \nonumber \\
     &\otimes \bm{D}_{\text{ms},k}(\tilde{\bm{\theta}})  \Big) \bigg\} + 2\Re \bigg\{ \Big( \bm{D}_{\text{bs},k}(\tilde{\bm{\phi}})\otimes \bm{D}_{\text{ms},k}(\tilde{\bm{\theta}})  \Big)^H \nonumber \\
     & \times \Big( \bm{D}_{\text{bs},k}(\tilde{\bm{\phi}})\otimes \big(\frac{\partial \bm{D}_{\text{ms},k}(\tilde{\bm{\theta}})}{\partial \tilde{\bm{\theta}}} \text{diag}(\bm{\delta}_\theta)\big)  \Big) \bigg\} \bigg) \nonumber \\
     & \times \big(   \bm{\Sigma}_{\bm{x}_k} + \bm{m}_{\bm{x}_k}\bm{m}_{\bm{x}_k}^H\big) \Bigg).
    \label{eqn:apdtraceterm}
\end{align}
While (\ref{eqn:apdtraceterm}) looks complicated, most of its components are with similar forms and can be put into standard linear or quadratic form in term of $\bm{\delta}_\phi$ and $\bm{\delta}_\theta$. For example, the second-order term w.r.t. $\bm{\delta}_\phi$ in (\ref{eqn:apdtraceterm}) can be reformulated as
\begin{align}
    &\text{tr} \Bigg( \Big( \big(\frac{\partial \bm{D}_{\text{bs},k}(\tilde{\bm{\phi}})}{\partial \tilde{\bm{\phi}}} \text{diag}(\bm{\delta}_\phi) \big)\otimes \bm{D}_{\text{ms},k}(\tilde{\bm{\theta}})  \Big)^H \nonumber \\
    &\times \Big(  \big(\frac{\partial \bm{D}_{\text{bs},k}(\tilde{\bm{\phi}})}{\partial \tilde{\bm{\phi}}} \text{diag}(\bm{\delta}_\phi) \big)\otimes \bm{D}_{\text{ms},k}(\tilde{\bm{\theta}}) \Big)  \big(   \bm{\Sigma}_{\bm{x}_k} + \bm{m}_{\bm{x}_k}\bm{m}_{\bm{x}_k}^H\big) \Bigg)\nonumber \\
    &= \text{tr} \Bigg( \Big( \text{diag}(\bm{\delta}_\phi) \otimes \bm{I}_{N_\theta}\Big)^H 
    \Big(   \frac{\partial \bm{D}_{\text{bs},k}(\tilde{\bm{\phi}})}{\partial \tilde{\bm{\phi}}} \otimes \bm{D}_{\text{ms},k}(\tilde{\bm{\theta}})   \Big)^H \nonumber \\
    &\times \Big(   \frac{\partial \bm{D}_{\text{bs},k}(\tilde{\bm{\phi}})}{\partial \tilde{\bm{\phi}}} \otimes \bm{D}_{\text{ms},k}(\tilde{\bm{\theta}})   \Big)
    \Big( \text{diag}(\bm{\delta}_\phi) \otimes \bm{I}_{N_\theta}\Big) \nonumber \\
    &\times \big(   \bm{\Sigma}_{\bm{x}_k} + \bm{m}_{\bm{x}_k}\bm{m}_{\bm{x}_k}^H\big) \Bigg)\nonumber \\
    &= \Big( \bm{\delta}_\phi \odot \bm{1}_{N_\theta} \Big)^H \bigg(\Big(   \frac{\partial \bm{D}_{\text{bs},k}(\tilde{\bm{\phi}})}{\partial \tilde{\bm{\phi}}} \otimes  \bm{D}_{\text{ms},k}(\tilde{\bm{\theta}})   \Big)^H  
    \Big(   \frac{\partial \bm{D}_{\text{bs},k}(\tilde{\bm{\phi}})}{\partial \tilde{\bm{\phi}}} \nonumber \\
    &\otimes \bm{D}_{\text{ms},k}(\tilde{\bm{\theta}})   \Big) \ast \big(   \bm{\Sigma}_{\bm{x}_k} + \bm{m}_{\bm{x}_k}\bm{m}_{\bm{x}_k}^H\big)^T \bigg)  \Big( \bm{\delta}_\phi \odot \bm{1}_{N_\theta} \Big) \nonumber \\
    &=  \bm{\delta}_\phi^H {\bm{B}^{(1)}}^H 
    \bigg(\Big(   \frac{\partial \bm{D}_{\text{bs},k}(\tilde{\bm{\phi}})}{\partial \tilde{\bm{\phi}}} \otimes  \bm{D}_{\text{ms},k}(\tilde{\bm{\theta}})   \Big)^H  
    \Big(   \frac{\partial \bm{D}_{\text{bs},k}(\tilde{\bm{\phi}})}{\partial \tilde{\bm{\phi}}} \nonumber \\
    &\otimes \bm{D}_{\text{ms},k}(\tilde{\bm{\theta}})   \Big) \ast \big(   \bm{\Sigma}_{\bm{x}_k} + \bm{m}_{\bm{x}_k}\bm{m}_{\bm{x}_k}^H\big)^T \bigg)
    {\bm{B}^{(1)}} \bm{\delta}_\phi,
\end{align}
where $\bm{1}_{N_\theta}$ denotes an all-ones vector with length $N_\theta$, and $\bm{B}^{(1)}$ is defined in (\ref{eqn:Bformulation}).

With similar derivations applied to various terms of (\ref{eqn:apdtraceterm}), (\ref{eqn:apdtraceterm}) can be written as
    \begin{align}
    &\text{tr}\Big( {\bm{D}_{k}^H(\bm{\delta}_\phi,\bm{\delta}_\theta)} {\bm{D}_{k}(\bm{\delta}_\phi,\bm{\delta}_\theta)} \big(   \bm{\Sigma}_{\bm{x}_k} + \bm{m}_{\bm{x}_k}\bm{m}_{\bm{x}_k}^H\big) \Big) \nonumber \\
    =& \bm{\delta}_\phi^H {\bm{B}^{(1)}}^H 
    \bigg(\Big(   \frac{\partial \bm{D}_{\text{bs},k}(\tilde{\bm{\phi}})}{\partial \tilde{\bm{\phi}}} \otimes \bm{D}_{\text{ms},k}(\tilde{\bm{\theta}})   \Big)^H  
    \Big(   \frac{\partial \bm{D}_{\text{bs},k}(\tilde{\bm{\phi}})}{\partial \tilde{\bm{\phi}}} \nonumber \\
    & \otimes \bm{D}_{\text{ms},k}(\tilde{\bm{\theta}})   \Big)\ast \big(   \bm{\Sigma}_{\bm{x}_k} + \bm{m}_{\bm{x}_k}\bm{m}_{\bm{x}_k}^H\big)^T \bigg)  
    {\bm{B}^{(1)}} \bm{\delta}_\phi  \nonumber \\
    +&  \bm{\delta}_\phi^H {\bm{B}^{(1)}}^H 
    \bigg(\Big(   \frac{\partial \bm{D}_{\text{bs},k}(\tilde{\bm{\phi}})}{\partial \tilde{\bm{\phi}}} \otimes \bm{D}_{\text{ms},k}(\tilde{\bm{\theta}})   \Big)^H  
    \Big(    \bm{D}_{\text{bs},k}(\tilde{\bm{\theta}}) \nonumber \\
    & \otimes  \frac{\partial\bm{D}_{\text{ms},k}(\tilde{\bm{\theta}})}{\partial \tilde{\bm{\theta}}}   \Big) \ast \big(   \bm{\Sigma}_{\bm{x}_k} + \bm{m}_{\bm{x}_k}\bm{m}_{\bm{x}_k}^H\big)^T \bigg)  
    \bm{B}^{(2)} \bm{\delta}_\theta  \nonumber \\
    +&  \bm{\delta}_\theta^H {\bm{B}^{(2)}}^H
    \bigg(\Big(    \bm{D}_{\text{bs},k}(\tilde{\bm{\theta}}) \otimes  \frac{\partial\bm{D}_{\text{ms},k}(\tilde{\bm{\theta}})}{\partial \tilde{\bm{\theta}}}   \Big)^H  
    \Big(   \frac{\partial \bm{D}_{\text{bs},k}(\tilde{\bm{\phi}})}{\partial \tilde{\bm{\phi}}} \nonumber \\
    & \otimes \bm{D}_{\text{ms},k}(\tilde{\bm{\theta}})   \Big) \ast \big(   \bm{\Sigma}_{\bm{x}_k} + \bm{m}_{\bm{x}_k}\bm{m}_{\bm{x}_k}^H\big)^T \bigg) 
    {\bm{B}^{(1)}} \bm{\delta}_\phi \nonumber \\
    +&  \bm{\delta}_\theta^H {\bm{B}^{(2)}}^H
    \bigg(\Big(    \bm{D}_{\text{bs},k}(\tilde{\bm{\theta}}) \otimes  \frac{\partial\bm{D}_{\text{ms},k}(\tilde{\bm{\theta}})}{\partial \tilde{\bm{\theta}}}   \Big)^H  
    \Big(    \bm{D}_{\text{bs},k}(\tilde{\bm{\theta}})\nonumber \\
    &  \otimes  \frac{\partial\bm{D}_{\text{ms},k}(\tilde{\bm{\theta}})}{\partial \tilde{\bm{\theta}}}   \Big) \ast \big(   \bm{\Sigma}_{\bm{x}_k} + \bm{m}_{\bm{x}_k}\bm{m}_{\bm{x}_k}^H\big)^T \bigg) 
     \bm{B}^{(2)} \bm{\delta}_\theta \nonumber \\
    +& 2\Re \bigg\{ \text{diag}\bigg(
    \Big(   \frac{\partial \bm{D}_{\text{bs},k}(\tilde{\bm{\phi}})}{\partial \tilde{\bm{\phi}}} \otimes \bm{D}_{\text{ms},k}(\tilde{\bm{\theta}})   \Big)^H
    \Big(  \bm{D}_{\text{bs},k}(\tilde{\bm{\phi}})\nonumber \\
    & \otimes \bm{D}_{\text{ms},k}(\tilde{\bm{\theta}}) \Big)
    \big(   \bm{\Sigma}_{\bm{x}_k} + \bm{m}_{\bm{x}_k}\bm{m}_{\bm{x}_k}^H\big) \bigg)^T
    {\bm{B}^{(1)}} \bm{\delta}_\phi
    \bigg\}\nonumber \\
    +& 2\Re \bigg\{ \text{diag}\bigg(
    \Big(    \bm{D}_{\text{bs},k}(\tilde{\bm{\theta}}) \otimes  \frac{\partial\bm{D}_{\text{ms},k}(\tilde{\bm{\theta}})}{\partial \tilde{\bm{\theta}}}   \Big)^H
    \Big(  \bm{D}_{\text{bs},k}(\tilde{\bm{\phi}})\nonumber \\
    & \otimes \bm{D}_{\text{ms},k}(\tilde{\bm{\theta}}) \Big)
    \big(   \bm{\Sigma}_{\bm{x}_k} + \bm{m}_{\bm{x}_k}\bm{m}_{\bm{x}_k}^H\big) \bigg)^T
    {\bm{B}^{(2)}} \bm{\delta}_\theta
    \bigg\}.
    \label{eqn:apdtrace}
\end{align}

On the other hand, by applying (\ref{eqn:taylorapprox}) to the second term of (\ref{eqn:apdmin_2_delta}), it can be reformulated as
\begin{align}
    &2\Re \Big( \bm{y}_{\text{w},k}^H \bm{D}_{k}(\bm{\delta}_{\phi},\bm{\delta}_{\theta}) \bm{m}_{\bm{x}_k} \Big) \nonumber \\
    =& 2\Re \Bigg( \bm{y}_{\text{w},k}^H \bigg( \Big(  \big(\frac{\partial \bm{D}_{\text{bs},k}(\tilde{\bm{\phi}})}{\partial \tilde{\bm{\phi}}} \text{diag}(\bm{\delta}_\phi) \big)\otimes \bm{D}_{\text{ms},k}(\tilde{\bm{\theta}}) \Big) \nonumber \\
    &+ \Big( \bm{D}_{\text{bs},k}(\tilde{\bm{\phi}})\otimes \big(\frac{\partial \bm{D}_{\text{ms},k}(\tilde{\bm{\theta}})}{\partial \tilde{\bm{\theta}}} \text{diag}(\bm{\delta}_\theta)\big)  \Big)  \bigg) \bm{m}_{\bm{x}_k} \Bigg)+ \text{const} \nonumber \\
    =& 2\Re \Bigg( \bm{y}_{\text{w},k}^H  \Big(  \frac{\partial \bm{D}_{\text{bs},k}(\tilde{\bm{\phi}})}{\partial \tilde{\bm{\phi}}}  \otimes \bm{D}_{\text{ms},k}(\tilde{\bm{\theta}}) \Big) \text{diag}(\bm{m}_{\bm{x}_k}) \bm{B}^{(1)} \bm{\delta}_\phi \nonumber \\
    & + \bm{y}_{\text{w},k}^H  \Big( \bm{D}_{\text{bs},k}(\tilde{\bm{\phi}})\otimes \frac{\partial \bm{D}_{\text{ms},k}(\tilde{\bm{\theta}})}{\partial \tilde{\bm{\theta}}}  \Big) \text{diag}(\bm{m}_{\bm{x}_k}) \bm{B}^{(2)} \bm{\delta}_\theta
    \Bigg) \nonumber \\
    & + \text{const}.
    \label{eqn:apdsecondterm}
\end{align}
By collecting first- and second-order coefficients w.r.t. $\bm{\delta}_\phi$ and $\bm{\delta}_\theta$ from (\ref{eqn:apdtrace}) and (\ref{eqn:apdsecondterm}), (\ref{eqn:emtoquadratic}) can be obtained.
\end{document}